\documentclass[12pt]{article}
\usepackage[toc,page]{appendix}
\usepackage{xcolor}
\usepackage{tikz}
\usepackage{tikz-feynman}
\tikzfeynmanset{compat=1.1.0}
\usetikzlibrary{arrows.meta}
\usepackage{mathtools}
\usepackage{amsthm,amsmath,latexsym,amssymb,amsfonts,amscd}
\usepackage{graphics,lscape,fancyhdr,array,stmaryrd,euscript,wrapfig}
\usepackage{empheq,slashed}
\usepackage{verbatim,slashed}
\numberwithin{equation}{section}
\usepackage{hyperref,setspace}
\usepackage{dsfont}
\usepackage{mathrsfs}

\usepackage[numbers,sort&compress]{natbib}
\usepackage[nottoc]{tocbibind}
\usepackage{float} 

\newcommand{\pl}{\partial}

\newcommand{\be}{\begin{equation}}
\newcommand{\ee}{\end{equation}}

\newcommand{\fdu}[2]{{}_{#1}{}^{#2}\,}

\newcommand{\besubeqs}{\begin{subequations}}
\newcommand{\esubeqs}{\end{subequations}}

\newcommand{\zb}{{\bar{z}}}

\newcommand{\pb}{{\bar{p}}}
\newcommand{\qb}{{\bar{q}}}

\newcommand{\jb}{{\bar{j}}}

\newcommand{\pfrac}[1]{{\frac{\pl}{\pl #1}}}

\newcommand{\deltas}[1]{{\delta\left(#1\right)}}
\newcommand{\PP}{{\mathbb{P}}}

\newcommand{\PPb}{{\bar{\mathbb{P}}}}
\newcommand{\NPP}{\mathbb{N}_{\PP}}
\newcommand{\NPPb}{\mathbb{\bar N}_{\PPb}}

\newcommand{\mA}{\mathcal{A}}
\newcommand{\mtA}{\tilde{\mathcal{A}}}
\newcommand{\fA}{\mathfrak{f}}

\newcommand{\definition}{\mathrel{\mathop:}=}

\usepackage{cancel}

\begin{document}
\pagenumbering{gobble}
\hfill
\vskip 0.01\textheight
\begin{center}
{\Large\bfseries Massless spinning fields on the Light-Front\\
\vspace{2mm}
(quartic vertices and amplitudes)}

\vspace{0.4cm}

\vskip 0.03\textheight
\renewcommand{\thefootnote}{\fnsymbol{footnote}} Mattia Serrani${}^{\pi}$
\renewcommand{\thefootnote}{\arabic{footnote}}
\vskip 0.03\textheight
\centering
\href{mailto:mattia.serrani@umons.ac.be}{\texttt{mattia.serrani@umons.ac.be}}
\vskip 0.03\textheight

{\em ${}^{\pi}$ Service de Physique de l'Univers, Champs et Gravitation, \\ Universit\'e de Mons, 20 place du Parc, 7000 Mons, 
Belgium}
\end{center}

\vskip 0.02\textheight

\begin{abstract}
Within the light-front approach in flat space, we study the closure of the Poincaré algebra at the quartic order, specifically the non-holomorphic constraint involving both MHV and anti-MHV vertices. We first recover some well-established results: the existence of Yang-Mills theory and gravity, as well as the inconsistency of interacting multi-graviton theories. We explicitly construct several lower-derivative and lower-spin quartic vertices. We then turn to theories involving massless higher-spin fields. It becomes evident that the quartic constraint does not allow many cubic interactions to survive, in accordance with the well-known no-go results. Nevertheless, once higher-derivative cubic vertices are included, we find nontrivial solutions to the full quartic constraint and determine the corresponding quartic vertices. On this basis, we conjecture the complete set of quartic vertices that solve the light-cone consistency conditions. Exploiting this, we find all allowed unitary local higher-spin theories and identify new families of local quasi-chiral higher-spin theories. We then determine all local higher-spin four-point amplitudes using the spinor-helicity formalism together with locality in the form of consistent factorization. We conclude with a short discussion on non-locality.
\end{abstract}

\newpage
\tableofcontents
\newpage
\section{Introduction}
\label{section1}
\pagenumbering{arabic}
\setcounter{page}{2}

The light-cone or light-front approach to dynamics is perhaps the most powerful tool to search for new theories within the (perturbative) field theory approach or for proving no-go theorems against particular types of theories. For example, string theory was first quantised in the light-cone gauge \cite{Goddard:1973qh}, and the finiteness of $\mathcal{N}=4$ super-Yang-Mills theory was established in the light-cone gauge \cite{Mandelstam:1982cb,Brink:1982wv}. The first cubic interactions of massless higher-spin fields were also constructed in the light-cone gauge \cite{Bengtsson:1983pd}. The complete classification of cubic interactions of massless higher-spin fields was obtained in the light-cone gauge \cite{Bengtsson:1983pg,Bengtsson:1983pd,Bengtsson:1986kh,Metsaev:1991nb,Metsaev:1991mt,Fradkin:1991iy,Metsaev:1993ap,Metsaev:2005ar,Metsaev:2007rn}, as well as the construction of the first example of a perturbatively local higher-spin theory \cite{Metsaev:1991mt,Metsaev:1991nb,Ponomarev:2016lrm}.\footnote{... with massless propagating fields; otherwise, there are plenty of ``topological'' theories in $3d$ \cite{Blencowe:1988gj,Bergshoeff:1989ns,Campoleoni:2010zq,Henneaux:2010xg,Grigoriev:2020lzu,Pope:1989vj,Fradkin:1989xt,Grigoriev:2019xmp} and the conformal higher-spin gravity \cite{Segal:2002gd,Tseytlin:2002gz,Bekaert:2010ky, Basile:2022nou}. The only nontopological theory with massless fields that has not been constructed via the light-cone approach is \cite{Sharapov:2024euk}, but it owes its existence to the covariant form \cite{Sharapov:2022faa,Sharapov:2022wpz,Sharapov:2022awp,Sharapov:2022nps,Sharapov:2023erv} of the chiral higher-spin gravity \cite{Metsaev:1991mt,Metsaev:1991nb,Ponomarev:2016lrm}, which was first found in the light-cone gauge. } It is beyond the scope of the paper, but the light-cone gauge also allows one to construct off-shell formulations of supersymmetric theories.

There are several advantages to the light-cone gauge. Firstly, it is a unitary gauge, and all redundant degrees of freedom associated with gauge symmetries are absent (not even present to begin with). Secondly, the light-cone approach operates with physical degrees of freedom only and, hence, allows one to find unambiguous results concerning the (non)existence of particular types of interactions. By contrast, within any covariant approach, some results can depend on the type of Lorentz covariant field chosen to contain the physical degrees of freedom, e.g. the dual graviton does not exhibit gravitational interactions \cite{Bekaert:2002uh}. Thirdly, within the perturbative field-theory approach, the aim is to construct the charges of the Poincaré algebra, including the Hamiltonian, which is the minimal requirement for obtaining a Poincaré-invariant S-matrix. The light-cone approach does precisely this and requires nothing more.

In $4d$, the structure of interactions exhibits additional features that are absent in higher dimensions. There are significantly more cubic interactions (for a given set of three spins) in the light-cone gauge \cite{Metsaev:1991mt,Metsaev:1991nb,Bengtsson:2014qza, Conde:2016izb} or in the spinor-helicity language \cite{Benincasa:2011pg,Benincasa:2007xk} than the standard description in terms of a totally-symmetric (Fronsdal) tensor field can give, see e.g. \cite{Boulanger:2006gr,Zinoviev:2008ck,Manvelyan:2010je}. The light-cone form itself exhibits several peculiar features. Firstly, the cubic vertices split into holomorphic (h or MHV) and anti-holomorphic (ah or anti-MHV), which is difficult to see in a covariant approach. Secondly, the closure of the Poincaré algebra at the quartic order has a remarkable structure: there are two sectors that contain the h--h and ah--ah vertices and receive no contribution from quartic generators (we call them holomorphic constraints), but these equations are already powerful enough to fix the spectrum and coupling constants \cite{Metsaev:1991mt,Metsaev:1991nb,Ponomarev:2017nrr,Serrani:2025owx}. In this way, upon setting the ah-vertices to zero, one obtains consistent chiral/self-dual theories.  

The present paper is divided into two parts, which can be read largely independently. The first part, up to Section \ref{section6}, is devoted to the study of the non-holomorphic quartic constraint in the light-cone gauge. This completes the analysis of higher-spin interactions at quartic order within the light-cone framework. The second part, presented in Section \ref{section7}, adopts a complementary perspective by studying four-point amplitudes in the spinor-helicity formalism that satisfy the factorisation constraints. To systematically explore all possible cases, we employ an efficient version of the factorisation procedure. The two approaches, based respectively on the light-cone formulation and on spinor-helicity methods, lead to fully consistent results. This agreement is expected, since it was shown in \cite{Ponomarev:2016cwi} that the light-cone deformation procedure can be reformulated entirely in terms of the spinor-helicity formalism at all orders in the interactions.\footnote{For earlier works exploring the relation between the spinor-helicity formalism and the light-cone formulation, see \cite{Ananth:2012un,Bengtsson:2016alt,Bengtsson:2016hss}.} While the light-cone approach provides explicit off-shell expressions for the local quartic vertices, the spinor-helicity approach allows us to compute all local higher-spin four-point amplitudes explicitly.

The first part of the paper is a systematic study of the quartic light-cone non-holomorphic constraint, i.e. of type h--ah. In a previous paper, we derived a general solution to the (anti)-holomorphic constraints \cite{Serrani:2025owx} and explicitly classified theories with gauge and gravitational interactions (one- and two-derivative, respectively). However, parity and unitarity require both h- and ah-vertices to be present at the same time and to have equal couplings. Therefore, the non-holomorphic constraint becomes essential, and this is also where most of the no-go theorems hide. The first important steps were already taken in \cite{Metsaev:1991nb}. 

We will warm up with Yang-Mills theory and gravity and proceed to numerous cases involving higher-spin fields. We strengthen some no-go results and find some yes-go options. We also consider quasi-chiral cases, a term coined in \cite{Adamo:2022lah}, where we allow for both h- and ah-vertices, but in an asymmetric fashion. 

In the second part, we use the on-shell spinor-helicity formalism for massless particles to determine all local higher-spin four-point amplitudes consistent with factorisation and compute them explicitly. Similar ideas have been explored in several previous works. The pioneering study of higher-spin amplitudes from the perspective of factorisation is \cite{Benincasa:2007xk} (see also \cite{Schuster:2008nh,Fotopoulos:2010ay,Dempster:2012vw}),\footnote{For related results in the covariant approach, see \cite{Taronna:2017wbx,Roiban:2017iqg}.} which initiated a systematic search for constructible theories using the BCFW recursion relations \cite{Britto:2004ap,Britto:2005fq} through a four-particle test starting from cubic amplitudes. However, the BCFW construction requires amplitudes to exhibit sufficiently good behaviour under large complex deformations of the external momenta. In particular, it requires the amplitude to vanish in the limit $z\to \infty$, where $z$ is the complex deformation parameter, so that the contribution from the contour integral at infinity can be neglected. While this condition is satisfied in Yang--Mills theory and gravity, it generally fails for higher-spin amplitudes, making the BCFW approach less suitable for studying higher-spin interactions.

In subsequent works \cite{Benincasa:2011kn,Benincasa:2011pg}, a generalised notion of constructibility was introduced, allowing for amplitudes that do not vanish in the limit $z\to\infty$ and, consequently, for the presence of boundary contributions. These contributions are directly related to self-consistent quartic vertices, which correspond to solutions of the homogeneous quartic constraint in the light-cone formulation. The resulting generalised four-particle test can therefore be consistently applied to higher-spin amplitudes. These works provided important partial results, including the identification of several local four-point amplitudes constructed from cubic higher-spin amplitudes. However, a complete and systematic classification of local higher-spin four-point amplitudes was still lacking.

Another very interesting work addressing higher-spin four-point amplitudes consistent with factorisation is \cite{McGady:2013sga}. The philosophy adopted there is closer to the one pursued in the present work, as it does not rely on a BCFW complex deformation but instead derives the constraints directly from four-point factorisation.

The main difference is that the analysis of \cite{McGady:2013sga} assumes the four-point amplitude to be constructed from a given three-point amplitude $\mA_3$ together with its parity-conjugate amplitude $\bar{\mA}_3$, then assuming unitarity. In contrast, we do not make this assumption and derive the factorisation constraints in full generality. Their analysis therefore corresponds to the unitary case that will be discussed in Section \ref{section6}. A second difference is that we do not just derive the conditions for the existence of local four-point amplitudes, but determine their explicit form in all cases. Our motivation is to classify all local four-point amplitudes, rather than only those arising in local unitary theories. Indeed, assuming parity invariance would, for example, exclude quasi-chiral higher-spin theories, which are instead local higher-spin theories at the quartic order and may be consistent to all orders. 

Lastly, we will also observe the emergence of color-kinematics duality and the double-copy structure at the level of non-holomorphic higher-spin four-point amplitudes. For similar results in the chiral sector, see \cite{Ponomarev:2017nrr,Ponomarev:2024jyg}.

\subsection*{Summary of the main results}
We summarise here the main results of the paper. 
The first results are presented in Section \ref{section5}, where we study the quartic light-cone constraint in detail for the cubic vertices of Yang–Mills theory and gravity. We find that self-dual Yang–Mills theory is described by a complex Lie algebra with a structure constant $\fA^c_{[ab]}$; while self-dual gravity is described by a complex commutative and associative algebra with a structure constant $g^c_{(ab)}$. Moreover, upon assuming CPT symmetry or unitarity, we show that the full Yang–Mills theory is governed by a Lie algebra with imaginary and fully antisymmetric structure constants $\fA_{[abc]}$. Gravity, instead, is governed by a commutative, symmetric, and associative algebra with a real structure constant $g_{(abc)}$. We also derive the explicit form of the local quartic vertices for Yang–Mills theory, given in \eqref{h4_1_YM} and \eqref{h4_2_YM}, and for gravity in \eqref{GR}. We verified the results by matching the four-point amplitudes obtained by summing the exchange diagrams and the quartic vertex. An analogous analysis can be readily extended to other classes of lower-spin vertices, as well as to vertices involving higher-spin fields.

\begin{figure}[H]
    \centering
    \begin{tikzpicture}
        \begin{feynman}
            \vertex (i1) at (-6, 1) {\(\lambda_2\)};
            \vertex (i2) at (-6,-1) {\(\lambda_1\)};
            \vertex (i3) at (-2, 1) {\(\lambda_3\)};
            \vertex (i4) at (-2,-1) {\(\lambda_4\)};

            \vertex (ii1) at (1, 1) {\(\lambda_2\)};
            \vertex (ii2) at (1,-1) {\(\lambda_1\)};
            \vertex (ii3) at (3, 1) {\(\lambda_3\)};
            \vertex (ii4) at (3,-1) {\(\lambda_4\)};

            \vertex (v1) at (-5, 0);
            \vertex (v3) at (-3, 0);

            \node at (-5.8, 0) {\(C\)};
            \node at (-2.2, 0) {\(\bar{C}\)};
            \node at (-0.5,0) {\Large $+$};
            
            \vertex at (-5, 0.5) {\(\omega\)};
            \vertex at (-3.2, 0.5) {\(-\omega\)};
            
            \diagram* {
                (i1) -- (v1),
                (i2) -- (v1),
                (v1) -- [plain] (v3),
                (v3) -- (i3),
                (v3) -- (i4),
            };
            \diagram* {
                (ii1) -- (ii4),
                (ii2) -- (ii3),
            };

        \end{feynman}
    \end{tikzpicture}
    \caption{Generic $C\bar{C}$ exchange with contact term $C^{\lambda_1,\lambda_2,\lambda_3,\lambda_4}$.}
    \label{fig1}
\end{figure}
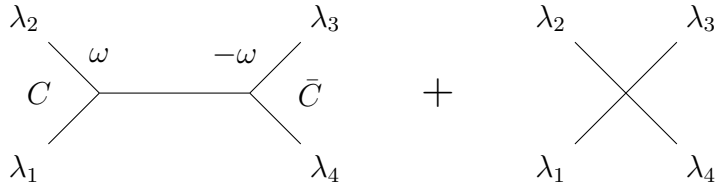

In Section \ref{section6}, all local quartic vertices, including higher-spin fields, that solve the quartic light-cone constraint are determined. Given two cubic vertices, one holomorphic $C^{\lambda_1,\lambda_2,\omega}$ ($\lambda_{12}+\omega>0$) and the other anti-holomorphic $\bar{C}^{-\omega,\lambda_3,\lambda_4}$ ($\lambda_{34}-\omega<0$), forming the exchange as shown in Figure \ref{fig1}, a local quartic vertex $C^{\lambda_1,\lambda_2,\lambda_3,\lambda_4}$ exists if the following sets of inequalities are satisfied:\footnote{We adopt the notation $\lambda_{ij}=\lambda_i+\lambda_j$ and $\lambda_{ijk}=\lambda_i+\lambda_j+\lambda_k$.}
\begin{equation}\label{conditionIntro}
\begin{aligned}
&\lambda_1 \leq \lambda_2+\omega+k-1\,,\qquad &
&\lambda_2 \leq \lambda_1+\omega+k-1\,,\qquad &
&\lambda_3 \leq \lambda_4+\omega+k-1\,,\\
&\lambda_4 \leq \lambda_3+\omega+k-1\,,\qquad &
&\lambda_{12} \geq \lambda_{34}\,,\qquad &
&k=0,1,2,3\,.
\end{aligned}
\end{equation}
\noindent
When \eqref{conditionIntro} is satisfied only for $k=1,2,3$,\footnote{By ``satisfied only for'' we mean that at least one of the inequalities appearing in \eqref{conditionIntro} is saturated, i.e. it holds as an equality.} additional exchange diagrams are required, as discussed in the main text. Accordingly, local quartic vertices fall into four classes: self-consistent quartic vertices ($k=0$);\footnote{In this case, $\omega$ no longer represents an exchange field; instead, it parametrizes the number of derivatives carried by the quartic vertex, $D=\lambda_{12}-\lambda_{34}+2\omega-2$.} vertices requiring a single-channel exchange ($k=1$); vertices requiring exchange in two channels ($k=2$), as in Yang–Mills theory; and vertices requiring $s$-, $t$-, and $u$-channel exchanges ($k=3$), as in gravity.

We used these results to determine all local unitary theories up to quartic order. We found the following possibilities:
\begin{itemize}
    \item Theories consisting exclusively of abelian couplings, both holomorphic (\(C\)) and anti-holomorphic (\(\bar{C}\)), that satisfy the triangular inequalities. These form unitary theories that can involve higher-spin fields.

    \item Theories that contain at least a parity-related pair of lower-spin non-abelian couplings. In this case, only additional lower-spin couplings are allowed, while all higher-spin couplings, whether abelian or non-abelian, are ruled out.
\end{itemize}
We further identify a class of local ``quasi-chiral'' higher-spin theories that extend both self-dual Yang–Mills theory and self-dual gravity. These theories allow for both holomorphic and anti-holomorphic cubic couplings and are therefore distinct from the chiral HS-SDYM and HS-SDGR theories \cite{Ponomarev:2017nrr, Krasnov:2021nsq}, while remaining non-parity-invariant and non-unitary.

In Section~\ref{section7}, we determine all local four-point higher-spin amplitudes in flat space by exploiting spinor-helicity properties such as little-group scaling and the locality of the amplitudes. We also explicitly check the results in several cases by summing the exchange diagrams together with the local quartic contact vertex. When the conditions \eqref{conditionIntro} are satisfied for $k=0$, self-consistent local quartic vertices are allowed. These correspond to solutions of the homogeneous quartic constraint and lead to the following amplitudes
\begin{equation}
    \mA^{(D)}_{\text{homo}}(1_{\lambda_1}2_{\lambda_2}3_{\lambda_3}4_{\lambda_4})=\sum_{i=0}^{\frac{D-d}{2}}\left(c_i\,s^it^{\frac{D-d}{2}-i}\right)[12]^{2\lambda_2}\langle 34\rangle^{\lambda_1-\lambda_{234}}[13]^{\lambda_{13}-\lambda_{24}}[14]^{\lambda_{14}-\lambda_{23}}\,,
\end{equation}
where $D$ denotes the total number of derivatives carried by the quartic vertex, or equivalently the mass dimension of the amplitude, while $d=3\lambda_1-\lambda_{234}$ is the total sum of the powers of spinor brackets, and $c_i$ are $\frac{D-d}{2}+1$ free coefficients. In particular, we need to associate the helicities $\lambda_1,\lambda_2,\lambda_3,\lambda_4$ with the fields $1,2,3,4$ in such a way as to maximise $d$, namely $d=\max\limits_{i\neq j\neq k\neq \ell}\{3\lambda_i-\lambda_{jk\ell},-3\lambda_i+\lambda_{jk\ell}\}$, where $i,j,k,\ell=1,2,3,4$. We can also exchange $\langle\;\rangle$ and $[\;]$ when necessary.

When the conditions \eqref{conditionIntro} are satisfied for $k=1$, amplitudes that include single-channel exchange diagrams are allowed and are given by
\begin{subequations}
\begin{align}
    \mA^{(D)}_{s}(1_{\lambda_1}2_{\lambda_2}3_{\lambda_3}4_{\lambda_4})&=k_s\frac{t^{\frac{D-d+2}{2}}}{s}[12]^{2\lambda_2}\langle 34\rangle^{\lambda_1-\lambda_{234}}[13]^{\lambda_{13}-\lambda_{24}}[14]^{\lambda_{14}-\lambda_{23}}+\mA^{(D)}_{\text{homo}}\,,\\
    \mA^{(D)}_{t}(1_{\lambda_1}2_{\lambda_2}3_{\lambda_3}4_{\lambda_4})&=k_t\frac{u^{\frac{D-d+2}{2}}}{t}[12]^{2\lambda_2}\langle 34\rangle^{\lambda_1-\lambda_{234}}[13]^{\lambda_{13}-\lambda_{24}}[14]^{\lambda_{14}-\lambda_{23}}+\mA^{(D)}_{\text{homo}}\,,\\
    \mA^{(D)}_{u}(1_{\lambda_1}2_{\lambda_2}3_{\lambda_3}4_{\lambda_4})&=k_u\frac{s^{\frac{D-d+2}{2}}}{u}[12]^{2\lambda_2}\langle 34\rangle^{\lambda_1-\lambda_{234}}[13]^{\lambda_{13}-\lambda_{24}}[14]^{\lambda_{14}-\lambda_{23}}+\mA^{(D)}_{\text{homo}}\,,
\end{align}
\end{subequations}
where $k_{\bullet}\sim C\bar{C}$ is the product of cubic couplings in the $s$-, $t$-, and $u$-channels. In particular, when $D=d-2$, we have a unique solution corresponding to
\begin{subequations}
\begin{align}
    \mA^{(d-2)}_{s}(1_{\lambda_1}2_{\lambda_2}3_{\lambda_3}4_{\lambda_4})&=\frac{k_s}{s}[12]^{2\lambda_2}\langle 34\rangle^{\lambda_1-\lambda_{234}}[13]^{\lambda_{13}-\lambda_{24}}[14]^{\lambda_{14}-\lambda_{23}}\,,\\
    \mA^{(d-2)}_{t}(1_{\lambda_1}2_{\lambda_2}3_{\lambda_3}4_{\lambda_4})&=\frac{k_t}{t}[12]^{2\lambda_2}\langle 34\rangle^{\lambda_1-\lambda_{234}}[13]^{\lambda_{13}-\lambda_{24}}[14]^{\lambda_{14}-\lambda_{23}}\,,\\
    \mA^{(d-2)}_{u}(1_{\lambda_1}2_{\lambda_2}3_{\lambda_3}4_{\lambda_4})&=\frac{k_u}{u}[12]^{2\lambda_2}\langle 34\rangle^{\lambda_1-\lambda_{234}}[13]^{\lambda_{13}-\lambda_{24}}[14]^{\lambda_{14}-\lambda_{23}}\,.
\end{align}
\end{subequations}
When the conditions \eqref{conditionIntro} are satisfied only for $k=2$, and we have $D=d-4$, Yang-Mills-like (YM-like) amplitudes are allowed and are given by
\begin{subequations}
\begin{align}
    \mA^{(d-4)}_{st}(1_{\lambda_1}2_{\lambda_2}3_{\lambda_3}4_{\lambda_4})&=\frac{k_{st}}{st}[12]^{2\lambda_2}\langle 34\rangle^{\lambda_1-\lambda_{234}}[13]^{\lambda_{13}-\lambda_{24}}[14]^{\lambda_{14}-\lambda_{23}}\,,\\
    \mA^{(d-4)}_{us}(1_{\lambda_1}2_{\lambda_2}3_{\lambda_3}4_{\lambda_4})&=\frac{k_{us}}{us}[12]^{2\lambda_2}\langle 34\rangle^{\lambda_1-\lambda_{234}}[13]^{\lambda_{13}-\lambda_{24}}[14]^{\lambda_{14}-\lambda_{23}}\,,\\
    \mA^{(d-4)}_{tu}(1_{\lambda_1}2_{\lambda_2}3_{\lambda_3}4_{\lambda_4})&=\frac{k_{tu}}{tu}[12]^{2\lambda_2}\langle 34\rangle^{\lambda_1-\lambda_{234}}[13]^{\lambda_{13}-\lambda_{24}}[14]^{\lambda_{14}-\lambda_{23}}\,.
\end{align}
\end{subequations}
These are not independent since $\mA^{(d-4)}_{st}=\frac{u}{t}\mA^{(d-4)}_{us}=\frac{u}{s}\mA^{(d-4)}_{tu}$. As we will show in the main text, this reflects the existence of a Jacobi identity, which implies the existence of two color-ordered amplitudes and the BCJ amplitude relations for non-holomorphic higher-spin amplitudes. When the conditions \eqref{conditionIntro} are satisfied only for $k=3$, and we have $D=d-6$, gravity-like (GR-like) amplitudes are allowed and are given by
\begin{align}
    &\mA^{(d-6)}_{stu}(1_{\lambda_1}2_{\lambda_2}3_{\lambda_3}4_{\lambda_4})=\frac{k_{stu}}{stu}[12]^{2\lambda_2}\langle 34\rangle^{\lambda_1-\lambda_{234}}[13]^{\lambda_{13}-\lambda_{24}}[14]^{\lambda_{14}-\lambda_{23}}\,,
\end{align}
with $k_{stu}=k_s=k_t=k_u$. In the case of gravity, this reproduces the MHV four-point graviton amplitude. As we will show in the main text, GR-like amplitudes can be obtained through a ``double copy'' construction from YM-like ones.

\subsection*{Outline of the paper}

The outline of the paper is as follows:
In Section \ref{section2}, we review various no-go theorems to appreciate the no-/yes-go results in the paper.
In Section \ref{section3}, we briefly recall the light-front approach to massless higher-spin interactions through the study of the light-cone constraints.
In Section \ref{section4}, we focus on the quartic constraint, highlighting some of its features and explaining our strategy for studying it. We describe the two main methods employed to study the system of PDEs arising from it.
In Section \ref{section5}, we solve the quartic constraint for lower-derivative quartic vertices, recovering known results such as the existence of Yang–Mills and gravity vertices. We compute the four-point amplitudes by summing exchange contributions and quartic contact terms.
In Section \ref{section6}, we extend the analysis to higher-derivative quartic vertices. We conjecture the complete set of quartic vertices that solve the light-cone quartic constraints. We use them to find all unitary local higher-spin theories in flat space and identify new families of local quasi-chiral higher-spin theories. In Section \ref{section7}, we find all local higher-spin four-point amplitudes.
In Section \ref{section8}, we discuss the possibility of quartic vertices with exchange-type non-localities.
Finally, in Section \ref{section8}, we summarise our conclusions and comment on possible future directions.

We also include six appendices. 
In Appendix \ref{AppendixA}, we explain how unitarity and parity are implemented in the light-cone formalism and, in particular, how they act on the quartic densities. 
In Appendix \ref{AppendixB}, we collect some useful formulas and relations.  
In Appendix \ref{AppendixC}, we comment on the quartic constraint in the presence of a cubic scalar vertex. 
Appendix \ref{AppendixD} collects some explicit solutions for quartic vertices.
Appendix \ref{AppendixE} contains a review of the on-shell relation between the light-cone and spinor-helicity formalisms. 
Appendix \ref{AppendixF} collects some of the tables we used to conjecture the complete set of quartic vertices that solve the light-cone quartic constraints.

\section{No-go's}\label{section2}

To better understand the role of the no-go and, more importantly, the yes-go results in this paper and \cite{Metsaev:1991mt,Metsaev:1991nb,Ponomarev:2016lrm,Ponomarev:2017nrr,Serrani:2025owx}, it is necessary to review the most well-known no-go-type arguments against nontrivial interactions of massless higher-spin fields. 

Various no-go theorems or other constraints implied by higher spin symmetry can be grouped into several categories: (A) global vs. (B) local; flat space (I) vs. de Sitter or anti-de Sitter (II). In the latter category, one may also add results for general gravitational background (III), or backgrounds that are less restrictive than (I-II), e.g. pp-wave or self-dual. Under (A), we group the results that constrain observables at `infinity', such as the S-matrix or the holographic S-matrix. Under (B), we collect results that impose restrictions on interactions within the perturbative approach to field theory, which is usually understood as some form of the Noether procedure.

Some important distinctions between (A) and (B) include: (1) simplicity or even triviality of the asymptotic observables from (A) does not necessarily imply that (B) is empty, e.g. interactions can conspire as to produce a simple (or trivial) S-matrix \cite{Ponomarev:2016lrm,Ponomarev:2017nrr,Skvortsov:2018jea,Skvortsov:2020wtf,Skvortsov:2020gpn}; (2) any results of (A)-type do not imply that there is a local quantum field theory that can deliver them. In fact, in some cases, it is quite clear that a given $S$-matrix that is a solution of the higher-spin symmetry constraints cannot result from a field theory. Lastly, it is never too late to rethink some of the explicit and hidden assumptions underlying the no-go results. We do not attempt to do this below and, instead, just state what every result means in ``plain terms'', i.e. what kind of conclusions are usually drawn.

\paragraph{A-category: global constraints of higher spin symmetry.} Global constraints can be derived by requiring $S$-matrix type observables to be invariant under higher spin transformations. Depending on the situation, these can be just constraints to ensure the masslessness of external states or an assumption of extended global symmetries.

{\bf AI.} In flat space, the canonical observable is the $S$-matrix and free massless fields (which are the only ones relevant for the asymptotic states) can be described, for instance, by (gauge-fixed) Fronsdal tensor $\Phi_{a_1...a_s} $ \cite{Fronsdal:1978rb} with gauge symmetry of the form
\begin{align}
    \delta \Phi_{a_1...a_s} &= \pl_{(a_1}\xi_{a_2...a_s)}\,.
\end{align}
The most powerful result in this category is Weinberg's low-energy theorem \cite{Weinberg:1964ew}. Roughly speaking, it implies that the S-matrix for any theory with massless higher-spin fields has to be the trivial one, $S=1$. There are at least two important remarks. Firstly, the conclusion is true once the higher-spin interactions survive in the low energy limit, which corresponds to the most interesting nonabelian interactions. Such interaction vertices are not trivially gauge-invariant and are the ones to produce the low-energy constraints. If one takes the abelian interactions only, no interesting constraints arise. Secondly, Weinberg's theorem is about the ``number of derivatives in the nonabelian interactions''. For one- and two-derivative interactions (gauge and gravitational ones), one gets the charge conservation and the universality of the gravitational coupling. For higher-derivative interactions, the constraints are just too strong. However, the number of derivatives depends on how the physical degrees of freedom are embedded into a covariant Lorentz tensor. In the Fronsdal formulation ``gauge'' and ``gravitational'' interactions turn out to have a higher-derivative form. By contrast, in the chiral formulation \cite{Krasnov:2021nsq}, which follows from twistor theory \cite{Penrose:1965am,Hughston:1979tq,Eastwood:1981jy,Woodhouse:1985id}, the gauge and gravitational interactions have the usual number of derivatives. This allows for an interesting loophole \cite{Tran:2022amg}. 

Another powerful constraint is the Coleman-Mandula theorem \cite{Coleman:1967ad} that implies that $S$-matrix cannot have ``extended symmetries'', where the latter means the existence of some charges $Q_{a_1...a_{s-1}}$ that transform in a higher-spin representation of the Lorentz group, i.e. with $s>2$.  

An obvious way to avoid all of the above arguments is to have a theory with nontrivial interactions that conspire to give a trivial S-matrix, as it happens in chiral theory \cite{Ponomarev:2016lrm,Ponomarev:2017nrr,Skvortsov:2018jea,Skvortsov:2020wtf,Skvortsov:2020gpn,Krasnov:2021nsq} and, more generally, in all self-dual theories. The nontriviality of interactions can still reveal itself in more nontrivial backgrounds.

{\bf AII.} In the case of anti-de Sitter space, the requirement of masslessness of external states is indistinguishable from the requirement to have extended symmetries, realized by some higher spin charges. Massless higher-spin fields are AdS/CFT dual to conserved higher-spin currents
\begin{align}
    \delta \Phi_{a_1...a_s} &= \nabla_{(a_1}\xi_{a_2...a_s)} && \longleftrightarrow && \pl^k J_{k n_2...n_s}=0\,.
\end{align}
There is an extension of the Weinberg-Coleman theorems to the AdS/CFT setting: all CFTs in $d\geq3$ that have at least one higher-spin current are free CFTs (possibly, in disguise) that have, in fact, infinitely many higher-spin currents \cite{Maldacena:2011jn,Boulanger:2013zza,Alba:2013yda,Alba:2015upa}. Therefore, the AdS/CFT analogue of $S=1$ is $S=$ free CFT on the boundary. 

An interesting deviation from this ``trivial'' S-matrix is that one can consider different boundary conditions and, eventually, a higher-spin gravity can be dual to something interesting, e.g. to the critical vector model \cite{Sezgin:2002rt,Klebanov:2002ja,Sezgin:2003pt,Leigh:2003gk} and, more generally, to Chern-Simons matter theories, see e.g. \cite{Giombi:2011kc}. A recent development \cite{Skvortsov:2018uru,Sharapov:2022awp,Jain:2024bza,Aharony:2024nqs} is the duality between chiral higher-spin gravity and a subsector of Chern-Simons matter theories.

\paragraph{B-category: local constraints of higher spin symmetry.} What falls into this category are various attempts to construct higher-spin theories as (reasonably) local field theories. Here, the main players are the Noether procedure, the light-front approach, and various AdS/CFT-related ideas. 

{\bf BI. Flat. } The Noether procedure is a way to search for theories by fixing some free field content and then deforming the action and gauge symmetries with higher-order terms while using gauge invariance and locality to constrain interactions. 
\begin{align}
    \delta \Phi_{a_1...a_s} &= \pl_{(a_1}\xi_{a_2...a_s)} +\mathcal{O}(\Phi\xi)\,.
\end{align}
Gauge symmetry is important in maintaining the correct number of degrees of freedom. While some cubic interactions are available \cite{Bengtsson:1986kh,Metsaev:1991mt,Metsaev:1991nb,Berends:1984rq,Boulanger:2006gr,Zinoviev:2008ck,Manvelyan:2010je}, there is an obstruction at the quartic order \cite{Bekaert:2010hp,Roiban:2017iqg,Ponomarev:2017nrr}. The obstruction is that one cannot find a local quartic vertex that is compatible with any ``interesting'' cubic interaction (note that abelian cubic interactions are consistent, but they do not induce any deformation of gauge symmetries and do not contain interactions such as gravitational and gauge). One might think of relaxing the standard definition of locality, but it is not clear how (if one completely abandons locality, then anything can be made into a theory \cite{Barnich:1993vg}). Nevertheless, the best one can do at present is to construct a chiral/self-dual subsector of a hypothetical theory that has some degree of nonlocality. 

{\bf BII. (A)dS. } Not surprisingly, the same problems persist in $(A)dS$. Thanks to the AdS/CFT duality, one can get a better handle on the problem since one can just take a free CFT on the boundary and try to ``reconstruct'' the interactions in $AdS$ that would give the same correlation functions back. It happens that such interactions are too nonlocal \cite{Bekaert:2015tva,Sleight:2017pcz,Ponomarev:2017qab}: the degree of nonlocality is the same as that of an exchange diagram, i.e. one cannot separate classical from quantum, roughly speaking.   

Note that the initial excitement about massless higher-spin fields in AdS was based on \cite{Vasiliev:1986bq} where ``gravitational interactions'' of higher-spin fields were constructed in $AdS_4$, while no such result was available in flat space in terms of Fronsdal fields. However, the complete classification of the cubic interactions in flat space \cite{Bengtsson:1986kh} has both gauge and gravitational interactions on the list, but in the light-cone gauge. Therefore, \cite{Vasiliev:1986bq} revealed that, while gauge and gravitational interactions cannot be constructed in terms of Fronsdal fields in flat space, one can do so in $AdS_4$, i.e. it revealed some singularity of the Fronsdal description. Objectively, there is a one-to-one correspondence between cubic interactions in flat space and $AdS_4$, \cite{Metsaev:2018xip,Nagaraj:2018nxq}.

{\bf BIII. More general backgrounds. } It is easy to show that the Fronsdal equations, with partial derivatives replaced by covariant ones, cease to be gauge invariant on general gravitational backgrounds \cite{Aragone:1979hx}. Adding various curvature-dependent (non-minimal) terms does not improve the situation. The situation does not improve either if one restricts to Einstein backgrounds. The reason is simple: in checking gauge invariance, one has to commute covariant derivatives, and the commutator brings the four-index Riemann tensor, $[\nabla,\nabla] \xi\sim R\,\xi$. Its traceless component, the Weyl tensor, survives on Einstein backgrounds and destroys the gauge invariance. However, in view of the discussion above, one should phrase this result as the impossibility of putting Fronsdal fields on backgrounds more general than maximally symmetric space-times.

It has been known since \cite{Hughston:1979tq,Aragone:1979hx,Woodhouse:1985id}, see also \cite{Krasnov:2021nsq}, that higher-spin fields can propagate on self-dual backgrounds; see e.g.  \cite{Skvortsov:2025ohi}. However, to achieve this, one has to choose an appropriate Lorentz covariant field description, which originates from twistor theory, not the Fronsdal fields. It is also known that higher-spin fields can propagate on pp-wave backgrounds \cite{Metsaev:1997ut}, see also \cite{Tran:2025yzd}.

\section{Living on the Light-Front}\label{section3}

The light-front (or light-cone) approach to field dynamics dates back to Dirac \cite{Dirac:1949cp} and has been instrumental in making the first important steps in various directions. We refer to \cite{Ponomarev:2022vjb} for a pedagogical exposition of the subject. The present paper is based on \cite{Bengtsson:1986kh}, where the complete classification of cubic interactions of massless spinning fields was obtained, and on \cite{Metsaev:1991mt, Metsaev:1991nb, Ponomarev:2016lrm}, where the first quartic analysis of the higher spin problem was conducted. 

The main idea of the approach is to construct the charges of the Poincaré algebra, $P^A$ and $J^{AB}$, directly in terms of physical degrees of freedom (the light-cone gauge is unitary):\footnote{We choose the mostly plus convention for $\eta_{AB}$ and $A,B,...=0,...,d-1$. In fact, $d=4$. The light-front coordinates are $x^\pm=(x^3\pm x^0)/\sqrt{2}$, so that $\eta^{+-}=\eta^{-+}=1$. In $4d$, we replace $x^{1,2}$ with two complex conjugate variables $z$ and $\bar{z}$, so that $x^Ax^B \eta_{AB}=2x^+x^-+2z\bar{z}$. }
\begin{align}
[P^A,P^B]&=0\,,\\
[J^{AB},P^C]&=P^A\eta^{BC}-P^B\eta^{AC}\,,\\
[J^{AB},J^{CD}]&=J^{AD}\eta^{BC}-J^{BD}\eta^{AC}-J^{AC}\eta^{BD}+J^{BC}\eta^{AD}\,.
\end{align}
It is also convenient to choose $x^+$ as the light-cone time.\footnote{This choice may not be free of troubles, in principle, since $x^\pm$ lies along the characteristic surfaces. Various subtleties are discussed, e.g. in \cite{Neville:1971zk,Heinzl:2000ht,Heinzl:1993px,Barnich:2024aln}. As far as we can see, none of them matters for the present paper.} Most of the generators turn out to be kinematical, i.e. they do not receive any corrections due to interactions (or receive simple ones, e.g. $J^{+-}$, $J^{+a}$) and stay as in the free theory. The choice of the light-cone time minimises the number of generators that need to be deformed. It is well-known that the only relations to worry about are\footnote{The Latin indices $a,b,...=z,\bar{z}$ are reserved for the transverse directions.}
\begin{align}
    [J^{a-},H]&=0\,, &[J^{a-},J^{b-}]&=0\,,
\end{align}
where $H=P^-$ is the light-cone Hamiltonian. It can further be shown that, at least classically, the second relation is a consequence of the first one, see \cite{Ponomarev:2016lrm} for more details. 

One of the crucial simplifications of the light-cone gauge\footnote{It is worth mentioning that we do not have to impose any gauge. It is true that, given one or another covariant description, we can go into the light-cone gauge and get what is in the text. However, there is no general statement that allows one to uplift any light-cone result to a given covariant description, e.g. to express the result as a local interaction in terms of Fronsdal fields. The existence of the $S$-matrix relies on $J^{AB}$, $P^A$ rather than on having Lorentz-covariant fields with such indices.} in $4d$ is that a massless spin-$s$ field is represented by a complex scalar field $\phi(x)=\phi^{+ \lambda}(x)$ and its conjugate $\bar\phi(x)=\phi^{-\lambda}(x)$ that are helicity eigenstates. In Lorentzian signature, the one adopted here, they are complex conjugates of each other $\phi^{-\lambda}=(\phi^{+\lambda})^*$. The free action is $S=\tfrac{1}{2}\int d^4x\, \phi^{-\lambda}\Box\phi^{\lambda}$ and is real in any signature. 

It is convenient to work with Fourier-transformed fields
\begin{align}
    \phi(p,x^+)&=(2\pi)^{-\tfrac{d-1}2}  \int e^{-i(x^-p^++p\cdot x)} \phi(x,x^+)\, d^{d-1}x\,.
\end{align}
One more ingredient is the Poisson bracket:\footnote{When seen as coming from the covariant Fronsdal free theory as a result of imposing the light-cone gauge, this turns out to be a Dirac bracket.}
\begin{align}\label{equaltime}
    [\phi^{\mu}(p,x^+),\phi^{\lambda}(q,x^+)]&=\delta^{\mu,-\lambda}\frac{\delta^{3}(p+q)}{2p^+}\,.
\end{align}
Once generators $J^{AB}$ and $P^A$ satisfy the Poincaré algebra relations at $x^+=0$, they do so at any $x^+$. Therefore, another important simplification is to set $x^+=0$, which explains why we omit it from all formulas below. The free field realisation reads
\begin{align}\label{FreeFieldReal}
    h_2&=-\frac{p\pb}{\beta}\,, &&
    \begin{aligned}
        j^{z-}_2&= \pfrac{\pb} \frac{ p\pb}{\beta} +p \pfrac{\beta} +\lambda\frac{p}{\beta}\,,\\
         j^{\zb-}_2&= \pfrac{p} \frac{ p\pb}{\beta} +\pb \pfrac{\beta} -\lambda\frac{\pb}{\beta}  \,.     
    \end{aligned}
\end{align}
The corresponding Poincaré charges are
\begin{align}
Q_2&= \int p^+\, d^{3}p\,\phi^{-\mu}_{-p} O(p,\pl_p)\phi^{\mu}_p\,, && O=h_2\,, j^{z-}_2\,,j^{\zb-}_2\,.
\end{align}
To take into account interactions, $J^{a-}$ and $H$ are assumed to have local expansions in $\phi$. On general grounds, with some simple kinematical relations taken into account, we should have
{\allowdisplaybreaks\besubeqs\begin{align}
    H&=H_2+\sum_n\int d^{3n}q\,\deltas{\sum q_i} h_{\lambda_1,...,\lambda_n}^{q_1,...,q_n}\, \phi^{\lambda_1}_{q_1}...\,\phi^{\lambda_n}_{q_n}\,,\\
    J^{z-}&=J^{z-}_2+\sum_n\int d^{3n}q\, \deltas{\sum q_i}\left[ j_{\lambda_1,...,\lambda_n}^{q_1,...,q_n}-\frac{1}{n}h_{\lambda_1,...,\lambda_n}^{q_1,...,q_n}\left(\sum_j \pfrac{\bar{q}_j}\right)\right]\, \phi^{\lambda_1}_{q_1}...\,\phi^{\lambda_n}_{q_n}   \,,\\ 
    J^{\zb-}&=J^{\zb-}_2+\sum_n\int d^{3n}q\, \deltas{\sum q_i}\left[ \jb_{\lambda_1,...,\lambda_n}^{q_1,...,q_n}-\frac{1}{n} h_{\lambda_1,...,\lambda_n}^{q_1,...,q_n}\left(\sum_j \pfrac{q_j}\right)\right]\, \phi^{\lambda_1}_{q_1}...\,\phi^{\lambda_n}_{q_n}   \,,  
\end{align}\esubeqs}\noindent
where we have not made any assumptions regarding the spectrum of states, and all helicities are allowed. The kinematical generators constrain further the form of densities $h$, $j$ and $\jb$ to depend on 
\begin{align}
   \PP_{km}&=q_k\beta_m-q_m\beta_k\,, & \PPb_{km}&=\qb_k\beta_m-\qb_m\beta_k \,.
\end{align}
Here $k,m=1,...,N$ label fields $\phi^{\lambda_k}_{q_k}$ at order $N$. Due to momentum conservation, there are $N-2$ such independent variables among $\PP$ and the same for $\PPb$. Finally, the densities have to be
\begin{subequations}
\begin{align}
    h_{\lambda_1,...,\lambda_n}(q_1,...,q_n)&=h_{\lambda_1,...,\lambda_n}(\PP_{km},\PPb_{km},\beta_k)\,,\\
    j_{\lambda_1,...,\lambda_n}(q_1,...,q_n)&=j_{\lambda_1,...,\lambda_n}(\PP_{km},\PPb_{km},\beta_k)\,, \qquad \text{same for } \jb\,.
\end{align}
\end{subequations}
The remaining kinematical constraints (together with their origin indicated on the left) can be rewritten as
{\allowdisplaybreaks\besubeqs\label{kinematics}\begin{align}
    J^{z\zb}&: &&\left[\NPP -\NPPb+\sum_k \lambda_k\right]h_{\lambda_1,...,\lambda_n}^{q_1,...,q_n}\sim0\,,\\
    J^{-+}&: && \left[\NPP +\NPPb+\sum_k \beta_k \pfrac{\beta_k}\right] h_{\lambda_1,...,\lambda_n}^{q_1,...,q_n}\sim0\,,\\
    J^{-+}&: && \left[\NPP +\NPPb+\sum_k \beta_k \pfrac{\beta_k}\right]j_{\lambda_1,...,\lambda_n}^{q_1,...,q_n}\sim0\,,\\
    J^{-+}&: && \left[\NPP +\NPPb+\sum_k \beta_k \pfrac{\beta_k}\right]\jb_{\lambda_1,...,\lambda_n}^{q_1,...,q_n}\sim0\,,\\
    J^{z\zb}&: &&\left[\NPP -\NPPb+\sum_k \lambda_k-1\right]j_{\lambda_1,...,\lambda_n}^{q_1,...,q_n}\sim0\,,\\
    J^{z\zb}&: &&\left[\NPP -\NPPb+\sum_k \lambda_k+1\right]\jb_{\lambda_1,...,\lambda_n}^{q_1,...,q_n}\sim0\,,
\end{align}\esubeqs}\noindent
where $\NPP=\sum \PP \pfrac{\PP}$ (sum over $\PP$'s that the densities depend on), idem. for $\NPPb$. 
\paragraph{Cubic vertices.} The complete basis of cubic interactions is given by \cite{Bengtsson:1986kh,Metsaev:1991mt,Metsaev:1991nb} and has a very neat form of (there are no negative powers of $\PP$, $\PPb$, as explained below)
\besubeqs\label{famouscubic}\begin{align}
    h^{\lambda_i}_3\equiv h_{\lambda_1,\lambda_2,\lambda_3}&= C^{\lambda_1,\lambda_2,\lambda_3} \frac{\PPb^{\lambda_{123}}}{\beta_1^{\lambda_1}\beta_2^{\lambda_2}\beta_3^{\lambda_3}}+\bar{C}^{-\lambda_1,-\lambda_2,-\lambda_3} \frac{\PP^{-\lambda_{123}}}{\beta_1^{-\lambda_1}\beta_2^{-\lambda_2}\beta_3^{-\lambda_3}}\,,\\
    j^{\lambda_i}_3\equiv j_{\lambda_1,\lambda_2,\lambda_3}&=+\frac23 C^{\lambda_1,\lambda_2,\lambda_3} \frac{\PPb^{\lambda_{123}-1}}{\beta_1^{+\lambda_1}\beta_2^{+\lambda_2}\beta_3^{+\lambda_3}}\Lambda^{\lambda_1,\lambda_2,\lambda_3}\,,\\
    \jb^{\lambda_i}_3\equiv \jb_{\lambda_1,\lambda_2,\lambda_3}&=-\frac23\bar{C}^{-\lambda_1,-\lambda_2,-\lambda_3} \frac{\PP^{-\lambda_{123}-1}}{\beta_1^{-\lambda_1}\beta_2^{-\lambda_2}\beta_3^{-\lambda_3}}\Lambda^{\lambda_1,\lambda_2,\lambda_3}\,,
\end{align}\esubeqs
where
\begin{subequations}
\begin{align}
    &\lambda_{ijk\ell}=\lambda_i+\lambda_j+\lambda_k+\lambda_{\ell}\,,\qquad
    \lambda_{ijk}=\lambda_i+\lambda_j+\lambda_k\,,\qquad
    \lambda_{ij}=\lambda_i+\lambda_j\,,\\
    &\Lambda=\beta_1(\lambda_2-\lambda_3)+\beta_2(\lambda_3-\lambda_1)+\beta_3(\lambda_1-\lambda_2)\,.
\end{align}
\end{subequations}
There is just one independent $\PP$ at this order, $\PP_{12}=\PP_{23}=\PP_{31}$ (momentum conservation). It is convenient to define democratic variables, both for $\PP$ and $\PPb$,
\begin{align}
    \PP^a_{12}&=...=\PP^a=\frac13\left[ (\beta_1-\beta_2)q^a_3+(\beta_2-\beta_3)q^a_1+(\beta_3-\beta_1)q^a_2\right]\,.
\end{align}
The dynamical information about the theory is encoded in coupling constants $C^{\lambda_1,\lambda_2,\lambda_3}$ and $\bar{C}^{-\lambda_1,-\lambda_2,-\lambda_3}$. Note that $\lambda_1+\lambda_2+\lambda_3>0$ for $C^{\lambda_1,\lambda_2,\lambda_3}$ unless $\lambda_i=0$ and $\lambda_1+\lambda_2+\lambda_3<0$ for $\bar{C}^{-\lambda_1,-\lambda_2,-\lambda_3}$ unless $\lambda_i=0$. There is a unique $\lambda_i=0$ vertex, which corresponds to $(\phi_0)^3$. It is a matter of choice where to put this scalar cubic self-interaction, into $C$ or $\bar{C}$. To simplify notation, we assume that all coupling constants are packed in $\mathcal{C}^{\lambda_1,\lambda_2,\lambda_3}$ and it is equal to $C^{\lambda_1,\lambda_2,\lambda_3}$ for $\lambda_1+\lambda_2+\lambda_3>0$, to $\bar C^{\lambda_1,\lambda_2,\lambda_3}$ for $\lambda_1+\lambda_2+\lambda_3<0$ and $C^{0,0,0}$ is the coefficient of $(\phi_0)^3$. To have the right dimension, we should set 
\begin{align}
    C^{\lambda_1,\lambda_2,\lambda_3}&= (\ell_P)^{\lambda_{123}-1} c^{\lambda_1,\lambda_2,\lambda_3}\,,
\end{align}
where $c$ are dimensionless and $\ell_P$ is some constant of length dimension. Any choice of $c^{\lambda_1,\lambda_2,\lambda_3}$ leads to a theory that is consistent up to the cubic order. At this order, all interactions do not feel each other's presence since the main constraint
\begin{align}
    [H,J^{a-}]\Big|_3=[H_3,J^{a-}_2]-[J^{a-}_3,H_2]&=0\,,
\end{align}
is a linear equation with respect to $H_3$, $J_3$, i.e. any linear combination of solutions is a solution again. The basis is provided by \eqref{famouscubic}. 
An interesting feature of the cubic interactions, which is hard to see in any covariant description, is that the vertices split into holomorphic, $\PPb$-dependent, and anti-holomorphic, $\PP$-dependent. Indeed, $H_2$ at the three-particle level is proportional to $\PP \PPb$ and, hence, any non-holomorphic contribution can be redefined away (via an appropriate canonical transformation). Covariant descriptions usually lead to some $\PP\PPb$-terms by default.
\paragraph{Higher orders. } The first serious constraint comes at the quartic order. It is at the quartic order that the spectrum of a theory usually gets fixed, as well as most of the cubic couplings. In general, at order $n$, we have to solve
\begin{align}\label{LFNoether}
    [J^{a-}_2,H_n]-[H_2,J^{a-}_n]&=\sum_{\substack{i,j>2\\i+j-2=n}}[H_i^{\vphantom{a}},J^{a-}_j]\,.
\end{align}
Rewriting it in terms of the densities leads to 
\begin{align}\label{maineq}
  \mathbf{H}_2\,  j^{a-}_n&=\mathbf{J}^{a-}_2[ h_n]+\sum_{\substack{i,j>2\\i+j-2=n}}[H_i^{\vphantom{a}},J^{a-}_j]\,,
\end{align}
where the boldface operators are sums of the one-particle ones:
\begin{align}
\label{defHJ}
\mathbf{J}^{a-}_2&=\sum_i \tilde{j}^{a-}_2(q_i)\,, &
\mathbf{H}_2&=\sum_i h_2(q_i)\,.
\end{align}
Here $\tilde{j}^{a-}_2$ commutes with $\delta^3(\sum q_i)$ and reads
\begin{align}
\tilde{j}^{a-}_2(q_i)=\tilde{j}_2^{a-}(q_i,\pl_{q_i})= j^{a-}_2(q_i)^T-h_2(q_i)\frac{1}{n}\left(\sum_j \pfrac{q^a_j}\right)
\,,
\end{align}
where $h_2$ can be found in \eqref{FreeFieldReal} and 
\begin{align}\label{J2T}
    (j^{z-}_2)^T&= -\frac{ q\qb}{\beta}\pfrac{\qb} -q \pfrac{\beta} +\lambda\frac{q}{\beta}\,, &
    (j^{\zb-}_2)^T&= - \frac{ q\qb}{\beta}\pfrac{q} -\qb \pfrac{\beta} -\lambda\frac{\qb}{\beta}  \,.      
\end{align}
At first sight, it may seem puzzling how a single equation, \eqref{LFNoether}, involving two free functions, $j_n$ and $h_n$, could admit unique solutions. Indeed, as is evident from \eqref{maineq}, $\mathbf{H}_2$ represents merely a number (the total energy). Thus, one can always divide by $\mathbf{H}_2$ to determine $j_n$ for any given choice of $h_n$: 
\begin{align}\label{maineqB_full}
    j^{a-}_n&=\frac{1}{\mathbf{H}_2}\Big(\mathbf{J}^{a-}_2[ h_n]+\sum_{\substack{i,j>2\\i+j-2=n}}[H_i^{\vphantom{a}},J^{a-}_j]\Big)\,.
\end{align}
In doing so, we will find transverse momenta in the denominator, which makes $j_n$ non-local at this order and will induce a similar non-locality in $h_n$ at the next order. The proof of Poincaré invariance of $S$-matrix relies on $H$ and $J^{a-}$ being local, i.e. not having transverse momenta in denominators. It is this locality assumption that allows one to find unique (or almost unique) solutions for two functions from a single equation. If locality is abandoned, one can literally take any ``interaction'' $h_n$ at each order and prolong it to a ``formally consistent'' generator of Poincaré algebra.\footnote{As a side remark, let us note that the same effect also appears in the Noether procedure \cite{Barnich:1993vg}; i.e. one can always find a solution by abandoning locality. The same reasoning applies to theories in $(A)dS_d$, with the additional subtlety that the denominators take the form $p^2+...+\Lambda$, where $\Lambda$ is the cosmological constant. This allows one to expand the denominators and observe infinite derivative tails in powers of $p^2$, rather than an explicit non-locality in the denominator, which can make one erroneously feel that the situation in $(A)dS$ is somewhat better.}

\section{Quartic analysis}\label{section4}
A general procedure for bootstrapping theories goes as follows. One begins by making certain assumptions about the spectrum and cubic interactions, which translates into requiring some $C^{\lambda_1,\lambda_2,\lambda_3}$ and/or $\bar C^{\lambda_1,\lambda_2,\lambda_3}$ not to vanish.
The equation to be solved at the quartic order is
\begin{align}\label{maineqB}
  \mathbf{H}_2\,  j^{a-}_4&=\mathbf{J}^{a-}_2[ h_4]+[H_3^{\vphantom{a}},J^{a-}_3]\,.
\end{align}
This is understood as a quadratic equation for the couplings. Due to the (anti)-holomorphic factorization of the cubic vertices, the equation splits into $CC$, $\bar C\bar C$ (we call these chiral), and $C \bar C$ sectors. For each of these sectors, we find that $[H_3^{\vphantom{a}},J^{a-}_3]$ contributes as
\begin{align}
    \deltas{\sum q_i}C^{\lambda_1,\lambda_2,\omega}C^{-\omega,\lambda_3,\lambda_4} F_{\lambda_1,\lambda_2,\lambda_3,\lambda_4}(\PP_{12},\PP_{34},\PPb_{12},\PPb_{34},\beta_i)\, \phi^{\lambda_1}_{q_1}\phi^{\lambda_2}_{q_2}\phi^{\lambda_3}_{q_3}\phi^{\lambda_4}_{q_4}\,,
\end{align}
where the function $F$ depends on the sector and will be given below. An important technical aspect is whether some of $\phi$'s have the same helicity and, hence, the density should take this symmetry, i.e. bosonic symmetrisation (or anti-symmetry in the case of fermions), into account. For a given helicity $\lambda$, the field $\phi^\lambda$ may also come in several species, e.g. as in the case of Yang-Mills theory.  

\paragraph{Chiral theories.} A remarkable property of \eqref{maineqB} observed in \cite{Metsaev:1991mt,Metsaev:1991nb}, see also \cite{Ponomarev:2016lrm}, is that the chiral sectors receive no contribution from $H_4$ and $J_4$. In other words, there are two closed subsystems of equations, for $CC$ and for $\bar C\bar C$, that are independent of higher orders:
\begin{align}\label{HOLO}
     &[H_3(\PPb),J^{z-}_3(\PPb)]=0\,,&
    &[H_3(\PP),J^{\bar{z}-}_3(\PP)]=0\,.
\end{align}
These equations are highly constraining, particularly for higher-spin fields. Solving the holomorphic ($CC$) and anti-holomorphic ($\bar C\bar C$) quartic constraints allows one to determine the allowed spectrum of cubic vertices, by fixing the various couplings $C^{\lambda_1,\lambda_2,\lambda_3}$ and $\bar{C}^{\lambda_1,\lambda_2,\lambda_3}$. In particular, a general solution to the holomorphic quartic constraint was recently found in \cite{Serrani:2025owx}, where a complete classification of lower-derivative chiral theories was also achieved. 

It is worth noting that if one allows at least one higher-spin particle to self-interact via a coupling of the type $C^{s,s,-s}$, with $s>2$ for singlet fields and $s>1$ in the color case, the holomorphic constraint forces the inclusion of all possible vertices. This leads to the so-called Metsaev solution, which uniquely determines all cubic couplings to be
\begin{align}\label{Metsaev}
&C^{\lambda_1,\lambda_2,\lambda_3}=\frac{k(\ell_p)^{\lambda_{123}-1}}{\Gamma(\lambda_1+\lambda_2+\lambda_3)}\,,&
\bar C^{\lambda_1,\lambda_2,\lambda_3}=\frac{\bar{k}(\ell_p)^{\lambda_{123}-1}}{\Gamma(\lambda_1+\lambda_2+\lambda_3)}\,.&
\end{align}
Moreover, if we search for a unitary theory, the couplings above are related by\footnote{The reality condition for the fields is $(\phi^\lambda_q)^*=\phi^{-\lambda}_{-q}$. Complex conjugation swaps $\PP$ or $\PPb$. When the $(-q)$ is eliminated in the vertex, after the complex conjugation $\PP$ and $\PPb$ remain unchanged. In addition, the product of $\beta$'s yields $(-)^{\lambda_{123}}$. Therefore, to have a Hermitian action, every $\PP$ or $\PPb$ needs to be accompanied by $i$. This $i$ can be absorbed into $\ell_p$, and we do not write it anywhere below. After putting this $i$ under the rug, the reality conditions for the couplings are the ones written down.}
\begin{equation}
    \bar C^{-\lambda_1,-\lambda_2,-\lambda_3}=(C^{\lambda_1,\lambda_2,\lambda_3})^*\,.
\end{equation} 
In the analysis below, it is not necessary to assume Metsaev couplings; it suffices to follow the general solution to the holomorphic constraint presented in \cite{Serrani:2025owx}. Further details will be provided in the following paragraph.

\paragraph{$\mathbf{[H_3,J_3]}$ commutator.} We now analyse the quartic constraint \eqref{maineqB} in the general case,\footnote{We will focus on the quartic constraint \eqref{maineqB} for $a=z$. The complex conjugate constraint will be analysed in a few pages.} assuming that the starting cubic theory already satisfies both the holomorphic and anti-holomorphic constraints. Let us compute the commutator explicitly, both in the presence of an external gauge group $G$ and in its absence. Expressed in terms of the densities \eqref{famouscubic}, the commutator takes the following form:
\begin{equation}\label{constraint}
\begin{split}
[H_3,J_3^{z-}]= & \sum_{\lambda_i,\alpha_j}\int d^9p\,d^9q\, \delta\left(\sum_i q_i\right) \left[j_3^{\lambda_i}(q_i)-\frac{h_3^{\lambda_i}(q_i)}{3}\left(\sum_k\frac{\partial}{\partial \bar{q}_k}\right)\right] \times \\
 & \delta\left(\sum_j p_j\right)h_3^{\alpha_j}(p_j)\left[\mathrm{Tr}\prod_{i=1}^3\phi_{q_i}^{\lambda_i},\mathrm{Tr}\prod_{j=1}^3\phi_{p_j}^{\alpha_j}\right]\,,
 \end{split}
\end{equation}
where to keep the analysis general, we assume that fields take values in the matrix algebra (e.g. $\phi^\lambda=\phi^\lambda_a T^a$ with some generators $T_a$) --- analogous to what is done in Yang-Mills theory. As a result, the fields acquire a Lie algebra structure, and we apply the trace to ensure the appearance of an invariant tensor. The non-colored case corresponds to the trivial algebra of singlet fields, in which case the trace can be safely omitted.

We now choose to contract the fields $\phi_{q_3}^{\lambda_3}$ and $\phi_{p_3}^{\alpha_3}$, which correspond to the internal (exchanged) lines in the Feynman diagrams, and we get
\begin{equation}
    \sum_{\lambda_1,\lambda_2,\lambda_3}\sum_{\alpha_1,\alpha_2,\alpha_3}[\mathrm{Tr}(\phi^{\lambda_1}_{q_1}\phi^{\lambda_2}_{q_2}\phi^{\lambda_3}_{q_3}),\mathrm{Tr}(\phi^{\alpha_1}_{p_1}\phi^{\alpha_2}_{p_2}\phi^{\alpha_3}_{p_3})]=9\sum_{\lambda_1,\lambda_2}\sum_{\alpha_1,\alpha_2}\sum_{\omega}\mathrm{Tr}(\phi^{\lambda_1}_{q_1}\phi^{\lambda_2}_{q_2}\phi^{\alpha_1}_{p_1}\phi^{\alpha_2}_{p_2})[\phi^{\omega}_{q_3},\phi^{-\omega}_{p_3}]\,.
\end{equation}
We can now take a closer look at Eq.~\eqref{constraint} and observe that the first derivatives $\pl_{\bar{q}_1}$, $\pl_{\bar{q}_2}$ still act on the fields, while the derivative $\partial_{\bar{q}_3}$ acts on the delta function associated with momentum conservation. It is therefore natural to integrate by parts the derivatives acting on the fields, but not the one acting on the delta function.
Then, using the Poisson bracket and the symmetry under the exchange $p\leftrightarrow q$ within the integral, we arrive at the following expression:
\begin{equation}\label{constraint_rewritten}
\begin{split}
[H_3,J_3^{z-}]= & \sum_{\lambda_i,\alpha_j}\int d^9p\,d^9q\,\delta\left(\sum_i q_i\right) \delta\left(\sum_j p_j\right)9\,\delta^{\lambda_3,-\alpha_3}\frac{\delta(q_3+p_3)}{2q_3^+}\phi^{\lambda_1}_{q_1}\phi^{\lambda_2}_{q_2}\phi^{\alpha_1}_{p_1}\phi^{\alpha_2}_{p_2}\times\\
&\left(j_3^{\lambda_i}(q_i)+\sum_{k\neq 3}\frac{\partial}{\partial \bar{q}_{k}}\frac{h_3^{\lambda_i}(q_i)}{3}\right)h_3^{\alpha_j}(p_j)\,.
\end{split}
\end{equation}
As expected, momentum conservation factors out. This follows directly from the Poincaré algebra, which implies the relation
\begin{subequations}
\begin{align}
    [P^a,H]=[P^+,H]=0\,,\qquad
    [P^a,J^{b-}]=\delta^{ab}P^-&\,,\qquad
    [P^+,J^{a-}]=-P^a\,,\\
    [P^a,[J^{b-},H]]+[J^{b-},[H,P^a]]+[H,[P^a,J^{b-}]]&=0
    \;\implies\;
    [P^a,[J^{b-},H]]=0\,,\\
    [P^+,[J^{b-},H]]+[J^{b-},[H,P^+]]+[H,[P^+,J^{b-}]]&=0
    \;\implies\;
    [P^+,[J^{b-},H]]=0\,,
\end{align}
\end{subequations}
where we used the definition of the Poincaré algebra and the Jacobi identity. The last relation means that the commutator (at any order) must commute with $P^a$ and $P^+$, and therefore it must be proportional to a delta function ensuring momentum conservation.
\begin{figure}[H]
    \centering
    \begin{tikzpicture}
        \begin{feynman}
            \vertex (i1) at (-6, 1) {\(\lambda_2\)};
            \vertex (i2) at (-6,-1) {\(\lambda_1\)};
            \vertex (i3) at (-2, 1) {\(\lambda_3\)};
            \vertex (i4) at (-2,-1) {\(\lambda_4\)};

            \vertex (v1) at (-5, 0);
            \vertex (v3) at (-3, 0);

            \node at (-6.8, 0) {\(K^{\text{holo}}_{1234\omega}\)};
            \node at (-5.8, 0) {\(C\)};
            \node at (-2.2, 0) {\(C\)};
            \node at (-8.5,-0.2) {\Large $\sum\limits_{\lambda_i\in S_4}\sum\limits_{\omega}$};
            \node at (0,0) {\Large $=\;0$};
            
            \vertex at (-5, 0.5) {\(\omega\)};
            \vertex at (-4.1, 1.4) {\([H_3(\PPb),J^{z-}_3(\PPb)]\)};
            \vertex at (-3.2, 0.5) {\(-\omega\)};
            
            \diagram* {
                (i1) -- (v1),
                (i2) -- (v1),
                (v1) -- [plain] (v3),
                (v3) -- (i3),
                (v3) -- (i4),
            };

        \end{feynman}
    \end{tikzpicture}
    \caption{Holomorphic constraint.}
        \label{fig_holo}
\end{figure}
\begin{figure}[H]
    \centering
    \begin{tikzpicture}
        \begin{feynman}
            \vertex (i1) at (-6, 1) {\(\lambda_2\)};
            \vertex (i2) at (-6,-1) {\(\lambda_1\)};
            \vertex (i3) at (-2, 1) {\(\lambda_3\)};
            \vertex (i4) at (-2,-1) {\(\lambda_4\)};

            \vertex (ii1) at (1, 1) {\(\lambda_2\)};
            \vertex (ii2) at (1,-1) {\(\lambda_1\)};
            \vertex (ii3) at (3, 1) {\(\lambda_3\)};
            \vertex (ii4) at (3,-1) {\(\lambda_4\)};

            \vertex (v1) at (-5, 0);
            \vertex (v3) at (-3, 0);

            \node at (-6.8, 0) {\(K_{1234\omega}\)};
            \node at (-5.8, 0) {\(C\)};
            \node at (-2.2, 0) {\(\bar{C}\)};
            \node at (-8.5,-0.2) {\Large $\sum\limits_{\lambda_i\in S_4}\sum\limits_{\omega}$};
            \node at (-0.5,0) {\Large $+$};
            \node at (4,0) {\Large $=\;0$};
            
            \vertex at (-5, 0.5) {\(\omega\)};
            \vertex at (-4.1, 1.4) {\([H_3,J^{z-}_3]\)};
            \vertex at (-3.2, 0.5) {\(-\omega\)};
            
            \vertex at (2, 1.4) {\(\mathbf{J}^{z-}_2[ h_4]\)};
            
            \diagram* {
                (i1) -- (v1),
                (i2) -- (v1),
                (v1) -- [plain] (v3),
                (v3) -- (i3),
                (v3) -- (i4),
            };
            \diagram* {
                (ii1) -- (ii4),
                (ii2) -- (ii3),
            };

        \end{feynman}
    \end{tikzpicture}
    \caption{Non-holomorphic constraint.}
    \label{fig_nonholo}
\end{figure}
\noindent
Starting from Eq.~\eqref{constraint_rewritten} and using the explicit form of the densities \eqref{famouscubic}, we obtain the following expression:
\begin{align}\label{commutator_singlet}
     \begin{split}
       [H_3,J^{z-}_3]=&\sum_{\lambda_i,\omega}\int d^{12}q\,\delta\left(\sum_i q_i\right)\frac{9}{2}\Big[(-)^{\omega}\frac{\beta_1(\lambda_1+\omega-\lambda_2)-\beta_2(\lambda_2+\omega-\lambda_1)}{(\beta_1+\beta_2)^{2\omega+1}}\frac{\beta_3^{\lambda_3}\beta_4^{\lambda_4}}{\beta_1^{\lambda_1}\beta_2^{\lambda_2}}\times\\
       &\mathcal{C}^{1234\omega}\PPb_{12}^{\lambda_{12}+\omega-1}\PP_{34}^{-\lambda_{34}+\omega}\Big]\phi^{\lambda_1}_{q_1}\phi^{\lambda_2}_{q_2}\phi^{\lambda_3}_{q_3}\phi^{\lambda_4}_{q_4}\,,
       \end{split}
\end{align}
where we denote $\mathcal{C}^{1234\omega}\equiv C^{\lambda_1,\lambda_2,\omega}\bar{C}^{-\omega,\lambda_3,\lambda_4}$ the product of the couplings, in analogy with \cite{Serrani:2025owx}, and we have used the fact that the holomorphic part of the commutator already vanishes. A diagrammatic representation of the holomorphic and anti-holomorphic quartic constraints is shown in Figures \ref{fig_holo} and \ref{fig_nonholo}, respectively. In the case of matrix-valued fields, we have
\begin{align}\label{commutator_color}
     \begin{split}
       [H_3,J^{z-}_3]=&\sum_{\lambda_i,\omega}\int d^{12}q\,\delta\left(\sum_i q_i\right)\frac{9}{2}\Big[(-)^{\omega}\frac{\beta_1(\lambda_1+\omega-\lambda_2)-\beta_2(\lambda_2+\omega-\lambda_1)}{(\beta_1+\beta_2)^{2\omega+1}}\frac{\beta_3^{\lambda_3}\beta_4^{\lambda_4}}{\beta_1^{\lambda_1}\beta_2^{\lambda_2}}\times\\
       &\mathcal{C}^{1234\omega}\PPb_{12}^{\lambda_{12}+\omega-1}\PP_{34}^{-\lambda_{34}+\omega}\Big]\mathrm{Tr}(\phi^{\lambda_1}_{q_1}\phi^{\lambda_2}_{q_2}\phi^{\lambda_3}_{q_3}\phi^{\lambda_4}_{q_4})\,.
       \end{split}
\end{align}
In order to be as general as possible, and following \cite{Serrani:2025oaw}, let us compute the commutator for the most general class of cubic vertices. We extend the field content by assigning to each field an additional index, indicating that it belongs to a representation of a gauge Lie algebra $\mathfrak{g}$ with structure constants $f_{abc}$. A field of helicity $\lambda$ is then assumed to take values in a representation $V_{\lambda}$ of $\mathfrak{g}$. We will denote by $\fA^{\lambda_1,\lambda_2,\lambda_3}_{abc}$ the tensor specifying the cubic coupling, using the same Latin indices for all modules $V_{\lambda}$. The most general form of the cubic Hamiltonian and boost generators is then given by
\begin{align}
    H_3&=\sum_{\lambda_1,\lambda_2,\lambda_3}\fA_{abc}^{\lambda_1,\lambda_2,\lambda_3}\int d^9 q\;\delta\Big(\sum_i q_i\Big)h_3^{\lambda_i}(q_i)\,(\phi^{\lambda_1}_{q_1})^{a}(\phi^{\lambda_2}_{q_2})^{b}(\phi^{\lambda_3}_{q_3})^{c}\,,\\
    J^{z-}_3&=\sum_{\lambda_1,\lambda_2,\lambda_3}\fA_{abc}^{\lambda_1,\lambda_2,\lambda_3}\int d^9q\,\delta\Big(\sum_i q_i\Big)\Big[j_3^{\lambda_i}(q_i)\,-\frac{1}{3}\,h_3^{\lambda_i}(q_i)\,\Big(\sum_j\frac{\partial}{\partial \bar{q}_j}\Big)\Big](\phi^{\lambda_1}_{q_1})^{a}(\phi^{\lambda_2}_{q_2})^{b}(\phi^{\lambda_3}_{q_3})^{c}\,,
\end{align}
where we have summed over all $\lambda_{1,2,3}$. This choice is convenient, especially for computations. However, to remain consistent, we need to take into account the appropriate symmetrisation\footnote{For bosons, and then integer helicities, this is a consequence of Bose symmetry. Fermions would get an extra minus sign.} \cite{Serrani:2025oaw} of the coupling constants:
\begin{equation}\label{f_sym}
\fA_{a_{\sigma_1}a_{\sigma_2}a_{\sigma_3}}^{\lambda_{\sigma_1},\lambda_{\sigma_2},\lambda_{\sigma_3}}=(-)^{\lambda_{123}}\fA_{a_1 a_2 a_3}^{\lambda_1,\lambda_2,\lambda_3}\,.
\end{equation}
Note that this imposes a symmetry property on the cubic couplings, though only in the case of identical fields, such as 
\begin{align}\label{sym_ff}
    &\fA_{a_2a_1a_3}^{\lambda,\lambda,s}=(-)^{2\lambda+s}\fA_{a_1 a_2 a_3}^{\lambda,\lambda,s}&
    &\implies&
    &\fA_{a_2a_1a_3}=(-)^{2\lambda+s}\fA_{a_1 a_2 a_3}\,.
\end{align}
Considering the most general class of cubic vertices, the commutator takes the form 
\begin{align}\label{commutator_general}
    \begin{split}
    [H_3,J_3^{z-}]=&\sum_{\lambda_i,\omega}\int d^{12}q\;\delta \left(\sum_i q_i\right)\frac{1}{2}\Big[(-)^{\omega}\frac{\beta_1(\lambda_1+\omega-\lambda_2)-\beta_2(\lambda_2+\omega-\lambda_1)}{(\beta_1+\beta_2)^{2\omega+1}}\frac{\beta^{\lambda_3}_3\beta_4^{\lambda_4}}{\beta_1^{\lambda_1}\beta_2^{\lambda_2}}\,\times\\
    &\mathcal{F}^{1234\omega} \PPb_{12}^{\lambda_{12}+\omega-1}\PPb_{34}^{\lambda_{34}-\omega}\,(\phi^{\lambda_1}_{q_1})^{a_1}(\phi^{\lambda_2}_{q_2})^{a_2}(\phi^{\lambda_3}_{q_3})^{a_3}(\phi^{\lambda_4}_{q_4})^{a_4}\Big]\,,
    \end{split}
\end{align}
where, for convenience, we rewrite the tensor couplings as $\fA^{\lambda_1,\lambda_2,\lambda_3}_{abc}\equiv \fA_{abc}C^{\lambda_1,\lambda_2,\lambda_3}$ and define $\mathcal{F}^{1234\omega}=\fA_{a_1a_2c}\fA^c_{\phantom{c}a_3a_4}C^{\lambda_1,\lambda_2,\omega}\bar{C}^{-\omega,\lambda_3,\lambda_4}$. Therefore, in the singlet case, we have $\mathcal{F}^{1234\omega}\equiv \mathcal{C}^{1234\omega}$.\footnote{ Here, $\fA\fdu{a_1a_2}{c}\fA_{c\,a_3a_4}$ indicates the natural pairing between the positive and negative helicity fields $(\phi^{+\lambda})^a$ and $(\phi^{-\lambda})_b$, where the negative helicity fields take values in the dual vector space (note that the Poisson bracket pairs the fields of opposite helicities). For example, if $f^a_{\phantom{a}bc}$ are structure constants of some Lie algebra $\mathfrak{g}$, and $(\phi^{+\lambda})^a$ transforms in the adjoint representation, then $(\phi^{-\lambda})_b$ transforms in the canonical dual to the adjoint, i.e. in the coadjoint one. Since we consider only the case of (semi)-simple Lie algebras, one can use the invariant metric tensor $k_{ab}$ instead of $\delta^a_b$. For the case where generators $T_a$ are given, this is the Killing form $k_{ab}=\mathrm{Tr}(T_aT_b)$.} The contraction is performed with the standard Poisson bracket for fields decorated with vector space indices and reads 
\begin{equation}\label{commutator_f}
    [(\phi^{\lambda}_q)^a,(\phi^{s}_p)_b]= \delta^{\lambda,-s}\delta^a_b\frac{\delta^3(q+p) }{2q^+}\,.
\end{equation}
Although this point is not directly relevant for our purposes, let us emphasise that, following \cite{Serrani:2025owx,Serrani:2025oaw,Skvortsov:2020wtf}, the factor $(-)^{\omega}$ appearing in $[H_3,J_3]$ can be removed by choosing even-helicity fields to be Hermitian and odd-helicity fields to be anti-Hermitian matrices (with singlets treated as $1\times1$ matrices).

At this stage, we have all the necessary tools to begin our analysis of quartic interactions. In particular, to find $h_4$, as suggested in \cite{Ponomarev:2016lrm}, we can use the following trick:
\begin{equation}\label{quartic}
    \textbf{H}_2 j_4^{z-}=\textbf{J}_2^{z-}[h_4]+[H_3,J_3^{z-}]\implies (\textbf{J}_2^{z-}[h_4]+[H_3,J_3^{z-}])|_{\textbf{H}_2=0}=0\,.
\end{equation}
In this way, we first determine $h_4$, and by substituting it back into the first equation above, we then obtain the corresponding $j_4$. Note that this is not as strong as being on-shell, where, in addition to the full four-momentum conservation, one also has $p_i^2=0$. 

Let us highlight some features of the commutator \eqref{commutator_general} in comparison with the holomorphic one studied in \cite{Ponomarev:2016lrm,Serrani:2025owx}. The symmetry properties of the cubic couplings remain unchanged. Moreover, the kinematical part of the commutator --- disregarding the couplings and the fields --- is, as for the holomorphic one, (anti-)symmetric under the exchange  $1\leftrightarrow 2$ or $3\leftrightarrow 4$ for (odd-)even-derivative cubic vertices. However, two important differences emerge due to the presence of both the holomorphic $\PPb$ and the anti-holomorphic $\PP$ terms:
\begin{itemize}
    \item First, once the ordering of the external helicities $\lambda_i$ is fixed, each $\omega$ in the sum contributes independently (there are more independent monomials that can be constructed from $\PP$ and $\PPb$ than from just $\PP$ as was the case for the holomorphic constraints). As a result, only the various permutations of $\lambda_i$ can contribute to the same quartic constraint, while for the (anti)holomorphic constraints, different $\omega$ could 'talk' to each other. This is precisely what makes the constraint stronger and more difficult to solve than the holomorphic one, where the sum over $\omega$ played a crucial role in solving it in the presence of higher-derivative vertices.
    \item Second, the sum over pairs such as $(1234)$ and $(3412)$ no longer leads to the clean simplifications we obtained in terms of only three independent variables $A,B,C$, as in the holomorphic constraint \cite{Ponomarev:2016lrm,Serrani:2025owx}. As we will see later, it is now possible for one of these permutations to be present while the other is absent --- a situation that did not occur before. In the holomorphic case, the presence of $\mathcal{F}^{1234\omega}$ automatically implies the presence of $\mathcal{F}^{3412\omega}$. Here, however, the two products of couplings are generally different, so summing over $(1234)$ and $(3412)$ would implicitly assume that both products are nonzero.
\end{itemize}
This is, in a sense, expected: in the present case, the commutators are not required to vanish by themselves but rather to cancel out through the inclusion of a suitable quartic interaction, as shown in \eqref{quartic}. Let us now rewrite the quartic commutator for singlet fields \eqref{commutator_singlet} as\footnote{As already noted, although the full commutator involves a sum over $\omega$, it decomposes into several independent constraints. Each constraint receives contributions only from terms with identical external helicities, and with the helicity $\omega$ of the exchanged field fixed by requiring not to change the powers of $\PPb$ and $\PP$, ensuring they belong to the same constraint.}
\begin{equation}\label{sym_constraint}
    [H_3,J_3^{z-}]\sim\text{Sym}\Big[(-)^{\omega}\mathcal{C}^{1234\omega}K_{1234\omega}\Big]\,,
\end{equation}
where we have defined the following kinematical factor:
\begin{equation}\label{kinematical_factor}
    K_{1234\omega}=\frac{\beta_1(\lambda_1+\omega-\lambda_2)-\beta_2(\lambda_2+\omega-\lambda_1)}{(\beta_1+\beta_2)^{2\omega+1}}\frac{\beta_3^{\lambda_3}\beta_4^{\lambda_4}}{\beta_1^{\lambda_1}\beta_2^{\lambda_2}}
       \PPb_{12}^{\lambda_{12}+\omega-1}\PP_{34}^{-\lambda_{34}+\omega}\,,
\end{equation}
and where Sym denotes the sum over the following $6$ distinct permutations of the external helicities:
\begin{equation}\label{sym_terms}
    (1234)+(1324)+(1423)+(3412)+(2413)+(2314)\,.
\end{equation}
Notice that the presence of these $6$  contributions is fully analogous to the holomorphic constraint. The key difference, as already emphasised, is that all of them must be treated as independent. Summing over terms such as $(1234)$ and $(3412)$ would not only fail to simplify the expressions but would also be inconsistent if we wish to consider the most general case.

In the case where the fields take values in the matrix algebra \eqref{commutator_color}, we have to consider color-ordered constraints. For the color ordering $[1234]$ corresponding to
\begin{equation}
    [1234]=(1234)+(2341)+(3412)+(4123)\,,
\end{equation}
we get
\begin{equation}\label{cycl_constraint}
    [H_3,J_3^{z-}]\sim\text{Cycl}\Big[(-)^{\omega}\theta_{\omega}\mathcal{C}^{1234\omega}K_{1234\omega}\Big]\,,
\end{equation}
where Cycl denotes the sum over the cyclic permutations, and $\theta_{\omega}$ arises from the freedom to choose a phase factor in the definition of the Poisson bracket; see \cite{Skvortsov:2020wtf,Serrani:2025owx,Serrani:2025oaw}.

Assuming the most general class of cubic vertices as in \eqref{commutator_general}, the commutator could include any of the possible $4!=24$ permutations of $(1234)$ depending on the symmetries of the generic cubic couplings $\mathfrak{f}_{a_1a_2a_3}^{\lambda_1,\lambda_2,\lambda_3}$.

In the following, our goal is to solve the quartic constraint \eqref{quartic}, where the commutator $[H_3,J^{z-}_3]$ takes either the form given in \eqref{sym_constraint}, the cyclic form in \eqref{cycl_constraint}, or a more general one if we consider \eqref{commutator_general}. To be as general as possible, we will not assume any specific form for the couplings. However, as already pointed out above, the kinematical factor \eqref{kinematical_factor} is (anti-)symmetric under the exchange  $1\leftrightarrow 2$ or $3\leftrightarrow 4$ for (odd-)even-derivative cubic vertices. 
Therefore, when searching for the most general solutions of the quartic constraint \eqref{maineqB}, choosing the exchange $(1234)$ or any of the other permutations $(2134)$, $(1243)$, or $(2143)$ does not make any difference. We can thus focus on the six independent kinematical factors \eqref{sym_terms}; once a solution is found, we can further analyse the possible types of structure constants that may arise.

\paragraph{$\mathbf{[H,J^{z-}]}$ and $\mathbf{[H,J^{\zb-}]}$ constraints.} Let us recall that we need to solve the two independent constraints~\eqref{maineqB}, for both $a=z,\zb$. Together, at the quartic order, we need to solve the following system
\begin{equation}\label{quartic_system}
    \begin{cases}
    &\textbf{H}_2 j_4^{z-}=\textbf{J}_2^{z-}[h_4]+[H_3,J_3^{z-}]\implies (\textbf{J}_2^{z-}[h_4]+[H_3,J_3^{z-}])|_{\textbf{H}_2=0}=0\\
    &\textbf{H}_2 j_4^{\bar{z}-}=\textbf{J}_2^{\bar{z}-}[h_4]+[H_3,J_3^{\bar{z}-}]\implies (\textbf{J}_2^{\bar{z}-}[h_4]+[H_3,J_3^{\bar{z}-}])|_{\textbf{H}_2=0}=0\,.\\
    \end{cases}
\end{equation}
The operators $\textbf{J}_2^{z-}$ and $\textbf{J}_2^{\bar{z}-}$ are described in \eqref{defHJ} and \eqref{J2T}, while the commutator $[H_3,J_3^{\zb-}]$ is written as
\begin{equation}
\begin{split}
[H_3,J_3^{\zb-}]= & \sum_{\lambda_i,\alpha_j}\int d^9p\,d^9q\,\delta\left(\sum_i q_i\right) \delta\left(\sum_j p_j\right)9\,\delta^{\lambda_3,-\alpha_3}\frac{\delta(q_3+p_3)}{2q_3^+}\phi^{\lambda_1}_{q_1}\phi^{\lambda_2}_{q_2}\phi^{\alpha_1}_{p_1}\phi^{\alpha_2}_{p_2}\times\\
&\left(\jb_3^{\lambda_i}(q_i)+\sum_{k\neq 3}\frac{\partial}{\partial q_{k}}\frac{h_3^{\lambda_i}(q_i)}{3}\right)h_3^{\alpha_j}(p_j)\,.
\end{split}
\end{equation}
Using the explicit form of the densities \eqref{famouscubic}, we obtain
\begin{align}
     \begin{split}
       [H_3,J^{\zb-}_3]=&\sum_{\lambda_i,\omega}\int d^{12}q\,\delta\left(\sum_i q_i\right)\frac{9}{2}\Big[(-)^{\omega}\frac{\beta_2(\lambda_2+\omega-\lambda_1)-\beta_1(\lambda_1+\omega-\lambda_2)}{(\beta_1+\beta_2)^{-2\omega+1}}\frac{\beta_1^{\lambda_1}\beta_2^{\lambda_2}}{\beta_3^{\lambda_3}\beta_4^{\lambda_4}}\times\\
       &\bar{C}^{\lambda_1,\lambda_2,\omega}C^{-\omega,\lambda_3,\lambda_4}\PP_{12}^{-\lambda_{12}-\omega-1}\PPb_{34}^{\lambda_{34}-\omega}\Big]\phi^{\lambda_1}_{q_1}\phi^{\lambda_2}_{q_2}\phi^{\lambda_3}_{q_3}\phi^{\lambda_4}_{q_4}\,.
       \end{split}
\end{align}
This corresponds to $[H_3,J_3^{z-}]$ in \eqref{commutator_singlet} after 
\begin{align}
    &q_i\longleftrightarrow \bar{q}_i\,,&
    &\lambda_i\longleftrightarrow -\lambda_i\,,&
    &C^{\lambda_1,\lambda_2,\lambda_3}\longleftrightarrow \bar{C}^{-\lambda_1,-\lambda_2,-\lambda_3}\,.&
\end{align}
In the case of matrix-valued fields, we need to take the trace, i.e. use $\mathrm{Tr}(\cdot)$ and, for the most general class of cubic vertices, introduce the tensor couplings $\fA^{\lambda_1,\lambda_2,\lambda_3}_{abc}\equiv \fA_{abc}C^{\lambda_1,\lambda_2,\lambda_3}$ as in \eqref{commutator_general}. Notice that these are the same transformations obtained after applying parity (see Appendix \ref{AppendixA}). This observation will be important later when we discuss quartic vertices for specific, highly symmetric cases.

For our purposes, we would like both constraints in the system \eqref{quartic_system} to act on the same Hamiltonian density $h_4$. It is therefore convenient to rewrite the commutator in the following equivalent form:
\begin{align}
     \begin{split}
       [H_3,J^{\zb-}_3]=&\sum_{\lambda_i,\omega}\int d^{12}q\,\delta\left(\sum_i q_i\right)\frac{9}{2}\Big[(-)^{\omega}\frac{\beta_4(\lambda_4-\omega-\lambda_3)-\beta_3(\lambda_3-\omega-\lambda_4)}{(\beta_3+\beta_4)^{2\omega+1}}\frac{\beta_3^{\lambda_3}\beta_4^{\lambda_4}}{\beta_1^{\lambda_1}\beta_2^{\lambda_2}}\times\\
       &C^{\lambda_1,\lambda_2,\omega}\bar{C}^{-\omega,\lambda_3,\lambda_4}\PPb_{12}^{\lambda_{12}+\omega}\PP_{34}^{-\lambda_{34}+\omega-1}\Big]\phi^{\lambda_1}_{q_1}\phi^{\lambda_2}_{q_2}\phi^{\lambda_3}_{q_3}\phi^{\lambda_4}_{q_4}\,,
       \end{split}
\end{align} 
where the following relabelling has been performed:
\begin{align}
    &(q_1,\qb_1,\beta_1,\lambda_1)\leftrightarrow (q_3,\qb_3,\beta_3,\lambda_3)\,,&
    &(q_2,\qb_2,\beta_2,\lambda_2)\leftrightarrow (q_4,\qb_4,\beta_4,\lambda_4)\,,&
    & C\leftrightarrow \bar{C}\,,&
    & \omega\leftrightarrow -\omega\,.
\end{align} 

\subsection{Construction of the ansätze}
Here, we construct the most general ansatz for the quartic local Hamiltonian density $h_4$ and local boost density $j_4$. To do so, we examine the quartic constraint \eqref{quartic}, from which we can determine the required powers of $\PP$, $\PPb$, and $\beta$ to build the ansatz. We denote a quartic vertex by the pair $(n,m)$, where $n$ and $m$ correspond to the number of derivatives in the holomorphic and anti-holomorphic cubic vertices appearing in the exchange term $[H_3,J^{z-}_3]$ and $[H_3,J^{\zb-}_3]$ that participate in the same quartic constraint. Specifically:
\begin{itemize}
    \item The $n$-derivative holomorphic vertex with helicity $(\lambda_1,\lambda_2,\omega)$ and cubic coupling $C^{\lambda_1,\lambda_2,\omega}$, with $n=\lambda_{12}+\omega>0$.

    \item The $m$-derivative anti-holomorphic vertex with helicity $(\lambda_3,\lambda_4,-\omega)$ and cubic coupling $\bar{C}^{-\omega,\lambda_3,\lambda_4}$, with $m=-\lambda_{34}+\omega>0$.

\end{itemize}
In the following, we refer to lower-derivative quartic vertices as those with $n,m\leq 2$, which participate in the same constraint as the lower-derivative cubic vertices.
Conversely, higher-derivative quartic vertices are those with $n,m>2$. Note also that the total number of derivatives in a quartic vertex is $n+m-2$. From the structure of the constraint \eqref{quartic}, we can find 
\begin{align}
        &[H_3,J^{z-}_3]\sim \beta^{\lambda_{34}-\lambda_{12}-2\omega}\bar{\PP}^{\lambda_{12}+\omega-1}\PP^{-\lambda_{34}+\omega}\sim \beta^{-n-m}\PPb^{n-1}\PP^m\,,\\\label{h4_dependence}
        &\textbf{J}_2^{z-}\cdot h_4\sim \frac{\PP}{\beta^2}h_4
        \implies
        h_4\sim \beta^{\lambda_{34}-\lambda_{12}-2\omega+2}\bar{\PP}^{\lambda_{12}+\omega-1}\PP^{-\lambda_{34}+\omega-1}\sim \beta^{2-n-m}\PPb^{n-1}\PP^{m-1}\,,\\
        &\textbf{H}_2\cdot j_4^{z-}\sim \frac{\PP\,\PPb}{\beta^3}j_4^{z-}
        \implies
        j_4^{z-}\sim \beta^{\lambda_{34}-\lambda_{12}-2\omega+3}\bar{\PP}^{\lambda_{12}+\omega-2}\PP^{-\lambda_{34}+\omega-1}\sim \beta^{3-n-m}\PPb^{n-2}\PP^{m-1}\,.
\end{align}
We immediately observe a particular feature of the $(1,1)$ quartic vertices: they admit no contribution from $j_4$, but only $h_4$. As we will see, this is precisely the case for Yang–Mills theory. Moreover, Yang–Mills theory is complete at the quartic order, meaning $h_n=0$ and $j_n=0$ for $n\geq 5$.

We now need to identify the independent variables. As mentioned above, for a four-point scattering, we have two independent $\PP$ and two independent $\PPb$. We choose $\PP_{12}$, $\PP_{34}$, $\PPb_{12}$, and $\PPb_{34}$ as our independent variables, while the remaining ones can be expressed in terms of these using the relations collected in Appendix \ref{AppendixB}. Additionally, there are three independent $\beta$ variables, since one can be eliminated using momentum conservation. Altogether, this gives us a total of seven independent variables. We will show how to further eliminate two of them, reducing the number of independent variables to five.

First, due to the fixed homogeneity in $\beta$ of $h_4$ and $j_4$, we can remove one further variable by performing a change of variables. Of the new variables, only one must carry homogeneity different from zero. For example, assuming $\beta_1$, $\beta_2$, and $\beta_3$ as independent variables, we can make the following change of variables:
\begin{align}
    &\tilde{x}=\frac{\beta_2}{\beta_1}\quad
    \implies\quad
    \beta_2=\tilde{x}\beta_1\,,&
    &\tilde{y}=\frac{\beta_3}{\beta_1}\quad
    \implies\quad
    \beta_3=\tilde{y}\beta_1\,.
\end{align}
Using the variables $\tilde{x}$, $\tilde{y}$ and $\beta_1$, the only one that carries non-zero homogeneity in $\beta$ is $\beta_1$. Therefore, we can drop it in the construction of $h_4$ and $j_4$. Once a solution is found, we can reconstruct the $\beta_1$ dependence from the required homogeneity.

Due to them having better symmetry properties, in the following, we use the following slightly different variables:
\begin{align}\label{xy_variables}
    &x=\frac{\beta_1-\beta_2}{\beta_1+\beta_2}\quad
    \implies\quad
    \beta_2=\frac{1-x}{x+1}\beta_1\,,&
    &y=\frac{\beta_3-\beta_4}{\beta_3+\beta_4}\quad
    \implies\quad
    \beta_3=\frac{y+1}{1-y}\beta_4\,,
\end{align}
and upon using momentum conservation, the same comments as before apply. Second, we can also remove one $\PP$ or $\PPb$ by using, as we have already pointed out in \eqref{quartic}, the energy conservation (i.e. with $\textbf{H}_2=0$) for the external fields. This gives us a further relation of the form
\begin{align}\label{on_shell}
    &\textbf{H}_2=\frac{1}{\beta_1+\beta_2}\left(\frac{\PPb_{12}\PP_{12}}{\beta_1\beta_2}-\frac{\PPb_{34}\PP_{34}}{\beta_3\beta_4}\right)=0&
    &\implies&
    &\PPb_{12}=\frac{\beta_1\beta_2}{\beta_3\beta_4}\frac{\PPb_{34}\PP_{34}}{\PP_{12}}=\frac{x^2-1}{y^2-1}\frac{\PPb_{34}\PP_{34}}{\PP_{12}}\,,
\end{align}
where the variables $x$ and $y$ are those introduced in \eqref{xy_variables}.
The most natural way to take this constraint into account is to choose as independent variables $\PP_{12}\PPb_{34}$, $\PPb_{12}\PP_{34}$ and $\frac{\PPb_{12}\PP_{12}}{\beta_1\beta_2}+\frac{\PPb_{34}\PP_{34}}{\beta_3\beta_4}$. This choice works when $n=m$, whereas for $n\ge m$ we complete the set by including the variables $\PPb_{12}$ and $\PPb_{34}$, as we see in the ansatz below. 

We now present the most general ansatz used in our analysis.
We do not write the ansatz explicitly for $j^{z-}_4$, as it coincides with that of $h_4$ in the $(n-1,m)$ case. In what follows, we denote $(\beta_1+\beta_2)^2\Big(\frac{\PPb_{12}\PP_{12}}{\beta_1\beta_2}+\frac{\PPb_{34}\PP_{34}}{\beta_3\beta_4}\Big)\equiv \mathbf{s_{12}}$ and take each function $f_i$ to depend on the variables $x$ and $y$, i.e. $f_i=f_i(x,y)$. The ansätze are
{\allowdisplaybreaks\besubeqs\begin{align}
    h_4^{(1,1)}&=f(x,y)\,,\\
    h_4^{(2,1)}&=\frac{1}{\beta_1+\beta_2}\left(\PPb_{12}f_1+\PPb_{34}f_2\right)\,,\\
    h_4^{(3,1)}&=\frac{1}{(\beta_1+\beta_2)^2}\left(\PPb_{12}^2f_1+\PPb_{12}\PPb_{34}f_2+\PPb_{34}^2f_3\right)\,,\\\
    h_4^{(n,1)}&=\frac{1}{(\beta_1+\beta_2)^{n-1}}\left(\PPb_{12}^{n-1}f_1+\PPb_{12}^{n-2}\PPb_{34}f_2+\cdots+ \PPb_{12}\PPb_{34}^{n-2}f_{n-1}+\PPb_{34}^{n-1}f_n\right)\,,\\
    h_4^{(2,2)}&=\frac{1}{(\beta_1+\beta_2)^2}\left(\PP_{12}\PPb_{34}f_1+\mathbf{s_{12}}f_2+\PPb_{12}\PP_{34}f_3\right)\,,\\
    \nonumber
     h_4^{(n,n)}&=\frac{1}{(\beta_1+\beta_2)^{2n-2}}\Big(\PP_{12}^{n-1}\PPb_{34}^{n-1}f_1+\PP_{12}^{n-2}\PPb_{34}^{n-2}\mathbf{s_{12}}f_2+\cdots+\mathbf{s_{12}}^{n-1}f_n+\cdots\\
     &\qquad\qquad\qquad\qquad+\PPb_{12}^{n-2}\PP_{34}^{n-2}\mathbf{s_{12}}f_{2n-2}+\PPb_{12}^{n-1}\PP_{34}^{n-1}f_{2n-1}\Big)\,,\\
     h_4^{(3,2)}&=\frac{1}{(\beta_1+\beta_2)^3}\left(\PP_{12}\PPb^2_{34}f_1+\PPb_{34}\mathbf{s_{12}}f_2+\PPb_{12}\mathbf{s_{12}}f_3+\PP_{34}\PPb^2_{12}f_4\right)\,,\\
     \nonumber
    h_4^{(n\geq m)}&=\frac{1}{(\beta_1+\beta_2)^{n+m-2}}\Big(\PP_{12}^{m-1}\PPb_{34}^{n-1}f_1+\PP_{12}^{m-2}\PPb_{34}^{n-2}\mathbf{s_{12}}f_2+\cdots+\PPb_{34}^{n-m}\mathbf{s_{12}}^{m-1}f_m+\cdots\\
    &\qquad\PPb_{12}^{n-m}\mathbf{s_{12}}^{m-1}f_n+\cdots+\PPb_{12}^{n-2}\PP_{34}^{m-2}\mathbf{s_{12}}f_{n+m-2}+\PPb_{12}^{n-1}\PP_{34}^{m-1}f_{n+m-1}\Big)\,.
\end{align}\esubeqs}\noindent
To improve symmetry, we consistently use $(\beta_1+\beta_2)$ as the homogeneity factor. Furthermore, the number of free functions to be determined at each order is given by:
\begin{equation}
    \#f=n+m-1\,.
\end{equation}
We have focused on the case $n\geq m$; the complementary case $n\leq m$ is easily obtained by taking the parity transformed system of constraints \eqref{quartic_system}. This gives the same set of equations, with the products of couplings mapped to the $n\leq m$ case. This simply reflects the fact that there should be no distinction between positive and negative helicities: the two are interchangeable, and their roles can always be exchanged:
\begin{align}
    &C^{\lambda_1,\lambda_2,\omega}\bar{C}^{-\omega,\lambda_3,\lambda_4}&
    &\longleftrightarrow&
    \bar{C}^{-\lambda_1,-\lambda_2,-\omega}C^{\omega,-\lambda_3,-\lambda_4}\,.
\end{align}
The ansätze for the case $n\leq m$ are then the same after exchanging $\PP$ and $\PPb$. The same ansätze were already used by Metsaev in \cite{Metsaev:1991nb}. We will comment more on \cite{Metsaev:1991nb} in the last section of the paper when discussing non-localities.

We do not write any ansatz for $(n,0)$ quartic vertices because no local quartic vertex of this type can exist. Indeed, from \eqref{h4_dependence} we see that it would require negative powers of $\PP$ or $\PPb$, leading to non-local quartic vertices. 

The scalar cubic coupling $C^{0,0,0}$ appears in the holomorphic quartic constraint. As shown in \cite{Serrani:2025owx}, generic solutions to the holomorphic constraint do not accommodate a scalar self-coupling. Nevertheless, in special cases --- and in particular for certain lower-spin couplings --- the scalar cubic vertex can remain consistent. We discuss these possibilities in Appendix \ref{AppendixC}.
\subsection{Solving the quartic constraint}
The analysis of the system of quartic constraints \eqref{quartic_system} based on the ansatz described above leads to a system of coupled first-order PDEs that can be approached using various methods. We describe two methods: the first is for finding explicit solutions; however, it becomes increasingly intractable as the number of derivatives grows.\footnote{If the system of PDEs is consistent and a solution does exist, it is fast, but if the system is inconsistent, it becomes rapidly intractable.} A second method allows us to determine whether a solution may exist by exploiting the Frobenius condition (integrability condition) on flat space $d^2=0$.\footnote{The Frobenius theorem states that a distribution (locally defined by a first-order system of PDEs) is integrable if and only if it is involutive. In flat space, involutivity of such a system reduces to the statement that successive derivatives commute, i.e. to the nilpotency $d^2=0$; in particular, in Cartesian coordinates, one has $[\partial_x,\partial_y]=0$.} We also note that, in principle, each solution can be complex and written as
\begin{align}
    &h_4(x,y)=\Re[h_4(x,y)]+ i \Im[h_4(x,y)]\,,
\end{align}
where $h_4(x,y)$ is a complex function of two real variables. However, the commutator $[H_3,J^{z-}_3]$ is real (up to the possible complex value of the product of coupling), and the only potentially non-trivial part of the vertex --- i.e. the part that contributes to solving the quartic constraint (and that may not exist) --- is its real component. The imaginary part corresponds to solutions of the homogeneous PDEs. In the following, we will therefore concentrate on the real part of $h_4(x,y)$ while separately pointing out the corresponding homogeneous solutions. 

\paragraph{Structure of PDEs.} Once we use the appropriate ansatz described before and go on the energy shell via \eqref{on_shell}, the generic form of the PDEs arising from \eqref{quartic_system} is
\begin{align}\label{constraint_generic}
&G(\PP_{12},\PP_{34},\PPb_{34},x,y,\partial_x,\partial_y)=0\,,&
&x=\frac{\beta_1-\beta_2}{\beta_1+\beta_2}\,,&
&y=\frac{\beta_3-\beta_4}{\beta_3+\beta_4}\,.
\end{align}
The system of PDEs arises from isolating independent terms, corresponding to terms with different powers of $\PP_{12}$, $\PP_{34}$, and $\PPb_{34}$. For the $[H_3,J^{\bar{z}-}_3]$ constraint and a $(n,m)$ quartic vertex, this always leads to a coupled system of PDEs of the form
\begin{equation}\label{system_of_PDEs}
    \begin{cases}
    P^{(1)}_1f_1+P^{(1)}_2\partial_xf_1+P^{(1)}_3=0\\
    P^{(2)}_1f_2+P^{(2)}_2\partial_xf_2+P^{(2)}_3f_1+P^{(2)}_4\partial_yf_1+P^{(2)}_5=0\\
    P^{(3)}_1f_3+P^{(3)}_2\partial_xf_3+P^{(3)}_3f_2+P^{(3)}_4\partial_yf_2+P^{(3)}_5=0\\
    ...\\
    P^{(k-2)}_1f_{k-2}+P^{(k-2)}_2\partial_xf_{k-2}+P^{(k-2)}_3f_{k-3}+P^{(k-2)}_4\partial_yf_{k-3}+P^{(k-2)}_5=0\\
    P^{(k-1)}_1f_{k-1}+P^{(k-1)}_2\partial_xf_{k-1}+P^{(k-1)}_3f_{k-2}+P^{(k-1)}_4\partial_yf_{k-2}+P^{(k-1)}_5=0\\
    P^{(k)}_1f_{k-1}+P^{(k)}_2\partial_yf_{k-1}+P^{(k)}_3=0\,,
    \end{cases}
\end{equation}
and another similar system of PDEs for the $[H_3,J^{z-}_3]$ constraint:
\begin{equation}\label{system_of_PDEs_2}
    \eqref{system_of_PDEs}\quad
    \text{with}\quad 
    P_{\bullet}^{(\bullet)}\leftrightarrow \bar{P}_{\bullet}^{(\bullet)}\quad \partial_x\leftrightarrow\partial_y\,.
\end{equation}
where $2k=2(n+m)$ is the number of PDEs, $P^{(i)}_j=P^{(i)}_j(x,y)$ and $\bar{P}^{(i)}_j=\bar{P}^{(i)}_j(x,y)$ are known rational functions of $x$, $y$, and $f_i=f_i(x,y)$, with $i=1,...,k-1$ being the free functions in the ansatz to be determined. This system of PDEs is a consequence of the structure of the linear differential operators $J^{\bar{z}-}_2$ and $J^{z-}_2$ acting on $h_4$.

The PDEs in \eqref{system_of_PDEs} and \eqref{system_of_PDEs_2} are linear first-order partial differential equations. The Cauchy–Kovalevskaya theorem ensures that, under mild assumptions --- such as working over the field of real $\mathbb{R}$ or complex numbers $\mathbb{C}$ and having coefficients $P^{(i)}_j(x,y)$ and $\bar{P}^{(i)}_j(x,y)$ that are analytic functions of $x$ and $y$  (in our case, they are actually rational functions) --- the system always admits a local solution with real or complex analytic functions. These can be obtained, for instance, using the method of characteristics or by direct (brute-force) integration. The system does not usually have a unique solution since it admits homogeneous solutions, which parameterize quartic vertices that are consistent on their own.

Importantly, the system consists of $2k$ PDEs for only $k-1$ unknown functions. As a result, although each equation can be solved, the existence of non-trivial solutions requires satisfying the integrability conditions. In flat space, these conditions are equivalent to the commutativity of partial derivatives, $d^2=0$, or $\partial_x\partial_yf=\partial_y\partial_xf$.

\paragraph{Solving the PDEs.} The structure of the systems \eqref{system_of_PDEs} and \eqref{system_of_PDEs_2} allows them to be solved straightforwardly. Starting from the first equation (and analogously from the last), we solve for $f_1(x,y)$, substitute the result into the subsequent PDEs, and then solve for $f_2(x,y)$. This procedure can be iterated to determine all functions $f_i(x,y)$.

At each step, the integration introduces a free constant. Once the final function $f_{k-1}(x,y)$ has been determined and substituted back into the last two PDEs, we are left with a pair of algebraic equations
\begin{align}\label{algebraic_system}
&\sum_{i=1}^{k-1} A_i(x,y) a_i = 0\,,&
&\sum_{i=1}^{k-1} B_i(x,y) a_i = 0\,.
\end{align}
where $A_i$ and $B_i$ are known rational functions of $x$ and $y$, while the $a_i$ are the integration constants arising from the solution of the various PDEs. Solving the algebraic system \eqref{algebraic_system} fixes some of these free coefficients. In particular, as we will see, all free coefficients are fixed except in the case where one considers the homogeneous system alone, i.e. in the absence of exchange contributions.

Notice that, once a solution for the density $h_4$ is found, we can use the full form of the constraint in \eqref{quartic} to extract the boost density $j_4$. This is always possible: indeed, once the Hamiltonian of the system is constructed, we know, in principle, everything about the system. Determining the explicit expressions for the boost densities then becomes merely an exercise.

Let us emphasise that, for this procedure to work, a solution must exist. Otherwise, not only would we fail to find one, but additional complications could arise, for instance, overly cumbersome integrations, the appearance of irrational functions or square roots, or situations that render the process overly involved. For this reason, having an alternative method to check integrability without performing the explicit integrations is extremely valuable. We now proceed to describe such a method.

\paragraph{Existence of a solution.}

The method we present here allows us to check the Frobenius integrability condition and, consequently, the existence of solutions without the need to solve any of the PDEs explicitly. Starting from the initial system of PDEs, we differentiate the equations and impose the condition $\partial_x\partial_y=\partial_y\partial_x$. The idea is to take successive higher-order derivatives until the number of equations exceeds the number of variables.\footnote{Here, the variables refer to the unknown functions and their derivatives. The system effectively becomes an algebraic system of equations to be solved.} In practice, it is often necessary to go further and generate a substantially larger set of equations due to the possible presence of non-trivial relations among them. This behaviour is characteristic of overdetermined systems of differential equations subject to integrability conditions.

Starting from \eqref{system_of_PDEs}, which we rewrite schematically, we apply both derivatives $\partial_x$ and $\partial_y$ as follows:
\begin{equation}
    \begin{cases}
    (\partial_x)^{h-1}(\partial_y)^{h-1}(f_1+\partial_xf_1)=0\\
    (\partial_x)^{h-1}(\partial_y)^{h-1}(f_2+\partial_xf_2+f_1+\partial_yf_1)=0\\
   (\partial_x)^{h-1}(\partial_y)^{h-1}(f_3+\partial_xf_3+f_2+\partial_yf_2)=0\\
    ...\\
    (\partial_x)^{h-1}(\partial_y)^{h-1}(f_{k-2}+\partial_xf_{k-2}+f_{k-3}+\partial_yf_{k-3})=0\\
    (\partial_x)^{h-1}(\partial_y)^{h-1}(f_{k-1}+\partial_xf_{k-1}+f_{k-2}+\partial_yf_{k-2})=0\\
    (\partial_x)^{h-1}(\partial_y)^{h-1}(f_{k-1}+\partial_yf_{k-1})=0\,.
    \end{cases}
\end{equation}
The same is done for the system \eqref{system_of_PDEs_2}.
To maximise efficiency, we take the same number of $\partial_x$ and $\partial_y$ derivatives.

Once a sufficient number of steps have been performed, to concretely solve the algebraic system, we fix $x$ and $y$ to specific values (while generic rational values are sufficient, we specifically choose $(x,y)=(2,0)$, as we have observed that in practice this choice significantly speeds up the integrability method) and search for a solution to this system. If a solution is found, the original system of first-order PDEs is integrable, meaning that a solution exists; otherwise, the system of PDEs is inconsistent.\footnote{More precisely, integrability would be guaranteed if one were to consider derivatives of arbitrarily high order and check the existence of a solution. In practice, however, it is sufficient to go to a sufficiently high derivative order: if a solution exists at that stage, this ensures the actual existence of a solution to the full system. Conversely, if no solution exists already for a finite subset of equations, then the full system admits no solution.}

We will use this method to verify and formulate general conjectures about consistent and inconsistent quartic vertices, both for lower-derivative higher-spin interactions and, more interestingly, for higher-derivative ones, since we will find some yes-go results. This method is computationally far more efficient for proving no-go's than the previous approach based on explicitly solving the PDEs. Once this method establishes the existence of a solution, we can then construct it explicitly using the approach outlined above.

\section{Quartic vertices for lower-derivative theories}\label{section5}
We begin by considering the system of quartic constraint \eqref{quartic_system}, involving the exchange terms $[H_3,J^{z-}_3]$ and $[H_3,J^{\zb-}_3]$ with one- and two-derivative cubic vertices. We first examine the integrability of the resulting PDEs, and when a solution exists, we proceed to solve them explicitly to find the corresponding densities. 

As expected, we recover the no-go theorems for local minimal vertices of higher-spin fields, alongside the well-known positive results for lower-spin quartic vertices, such as those in Yang–Mills theory and gravity. We consider all possible lower-derivative quartic vertices, then three cases: $(1,1)$, $(2,1)$, and $(2,2)$ quartic vertices.

To cover all possible cases, we organise the analysis as follows.
As shown in \eqref{sym_terms}, the commutator, due to its symmetries, can be organised in terms of the following relevant exchanges:
\begin{equation}\label{various_ordering}
    \overset{s}{(1234)}+\overset{t}{(2341)}+\overset{s}{(3412)}+\overset{t}{(4123)}+\overset{u}{(1324)}+\overset{u}{(2413)}\,.
\end{equation}
We have slightly reorganised these terms so that the first four naturally correspond to the exchanges relevant to the color case, while the remaining two are additional.
This arrangement also facilitates the reading of the tables presented below, as it makes it straightforward to recognise patterns that can be traced back to the color case.

\paragraph{Quartic vertices.} Before presenting the results, it is important to understand the logic behind them. Let us briefly recall what we are doing and what we are looking for. We proceed to solve the constraint\footnote{The other constraint is analogous; thus, we restrict attention to this one.} \eqref{quartic}. Quartic vertices are uniquely specified by the pair $(n,m)$ and the helicities of the external fields $(\lambda_1,\lambda_2,\lambda_3,\lambda_4)$, where $\lambda_4=n-m-\lambda_{123}$. Let us write down \eqref{quartic} explicitly, as 
\begin{align}\label{general_structure}
    \begin{split}
    &\Big(\sum_{i=1}^4\left(-\frac{ q_i\qb_i}{\beta_i}\pfrac{\qb_i} -q_i \pfrac{\beta_i} +\lambda\frac{q_i}{\beta_i}\right)h^{(n,m)}_{\lambda_1,\lambda_2,\lambda_3,\lambda_4} g^{\lambda_1,\lambda_2,\lambda_3,\lambda_4}_{a_1a_2a_3a_4}+(-)^{\omega}\Big(K_{1234\omega}\mathcal{F}_{1234\omega}\\
    &+K_{2341\omega}\mathcal{F}_{2341\omega}+K_{3412\omega}\mathcal{F}_{3412\omega}+K_{4123\omega}\mathcal{F}_{4123\omega}+K_{1324\omega}\mathcal{F}_{1324\omega}\\
    &+K_{2413\omega}\mathcal{F}_{2413\omega}\Big)\Big)(\phi^{\lambda_1}_{q_1})^{a_1}(\phi^{\lambda_2}_{q_2})^{a_2}(\phi^{\lambda_3}_{q_3})^{a_3}(\phi^{\lambda_4}_{q_4})^{a_4}=0\,,
    \end{split}
\end{align}
where $g^{\lambda_1,\lambda_2,\lambda_3,\lambda_4}_{a_1a_2a_3a_4}$ represents the coupling constant of the quartic vertex.
The system of PDEs above admits solutions whose form depends on the presence or absence of the various coefficients $\mathcal{F}_{\bullet\bullet\bullet\bullet\bullet}$. 

If we set all $\mathcal{F}_{\bullet\bullet\bullet\bullet\bullet}$ to zero (i.e. we set the cubic vertices to zero), the problem reduces to solving the homogeneous PDEs. These determine the quartic vertices, which are well-defined even in the absence of any cubic interactions.\footnote{When all cubic vertices are set to zero, the quartic constraint admits the trivial solution $h_4=0$ as well.}
Alternatively, one may look for solutions when at least one exchange term $\mathcal{F}_{\bullet\bullet\bullet\bullet\bullet}$ is non-zero. In this case, the quartic vertices become necessary to ensure the Lorentz invariance of the theory. 
We emphasise that, as already mentioned, the kinematical factors are (anti-)symmetric under the relevant permutations.\footnote{We recall that $K_{1234\omega}=(-)^{\lambda_{12}+\omega}K_{2134\omega}=(-)^{\lambda_{34}+\omega}K_{1243\omega}=(-)^{\lambda_{1234}}K_{2143\omega}$.} Consequently, once a solution exists, one can, in principle, choose which of the four possible kinematical structures $K_{\bullet\bullet\bullet\bullet\bullet}$ is present. In highly symmetric situations, this distinction disappears.\footnote{In general $K_{1234\omega}\mathcal{F}_{1234\omega}$ and $K_{2134\omega}\mathcal{F}_{2134\omega}$, are unrelated. For instance, one can have $f\fdu{a_1a_2}{c}\neq 0$ while $f\fdu{a_2a_1}{c}=0$. However, in special cases --- e.g. when $\lambda_1=\lambda_2$ --- one finds $K_{1234\omega}\mathcal{F}_{1234\omega}=K_{2134\omega}\mathcal{F}_{2134\omega}$, so the two contributions coincide and the distinction disappears.}

The main problem then reduces to determining whether \eqref{general_structure} --- together with the $J^{\bar{z}-}$ constraint --- admits a solution with six free coefficients, each activating one of the six independent kinematical structures.

\paragraph{Integrability.} Using the integrability method, we can construct tables listing all quartic vertices that admit a solution to the quartic constraint \eqref{quartic_system} for lower-derivative theories. The tables below were obtained as follows:
\begin{itemize}
    \item Assuming we are searching for an $(n,m)$ quartic vertex, we check the Frobenius integrability of the system of quartic constraint \eqref{quartic_system} by evaluating the commutators $[H_3,J^{z-}_3]$ and $[H_3,J^{\zb-}_3]$ with all possible exchanges \eqref{various_ordering} as explained above. There are six orderings that contribute to the same quartic constraint, each weighted by an independent free coefficient $k_i=(-)^{\omega}\mathcal{F}_{\bullet\bullet\bullet\bullet\bullet}$. For instance, for $(1234)$ we have $k_1=(-1)^{\omega}\mathcal{F}_{1234\omega}$ and analogously for the other orderings.

    \item The helicities $\omega$ appearing in the exchanges are automatically fixed by the requirement of having an $(n,m)$ quartic vertex. In addition, we can fix the value of $\lambda_4$ via the relation $n-m=\lambda_{1234}$. As a result, the problem reduces to searching for exchanges of the form:
    \begin{align}
        (1234)\equiv(\lambda_1,\lambda_2,\omega,\lambda_3,\lambda_4)=(\lambda_1,\lambda_2,n-\lambda_{12},\lambda_3,n-m-\lambda_{123})\,.
    \end{align}

    \item We set the maximum helicity $\lambda_{\text{max}}>|\lambda_1|,|\lambda_2|,|\lambda_3|$ and apply the integrability condition to all such cases. Whenever a solution to the Frobenius integrability condition is found, we record in the tables whether it also fixes any of the coefficients $k_i$. In particular, in all three tables presented below, we have checked the existence of a solution to the quartic constraint up to $\lambda_{\text{max}}=10$.
    
    \item The tables list only inequivalent solutions. Whenever a solution exists, permutations of the external helicities generally generate additional valid solutions. However, these are already taken into account, as they can always be deduced from the combined symmetries of the kinematical structures and the coupling constants.
    
\end{itemize}
Therefore, the tables collect the quartic couplings (``yes-go'') that admit a local solution to the quartic constraint \eqref{quartic_system}, while at the same time indicating that all quartic vertices not listed do not satisfy it and thus fall into the category of ``no-go''.

The first entry $(\lambda_1,\lambda_2,\omega,\lambda_3,\lambda_4)$ specifies the helicities associated with the contribution from the $(1234)$ ordering in the exchange $[H_3,J^{z-}_3]$ and $[H_3,J^{\zb-}_3]$. The remaining entries correspond to the other orderings listed in \eqref{various_ordering}. Due to the structure of the $\PP^\bullet\PPb^\bullet$ monomials, different orderings present in the quartic constraint can ``talk'' to each other, which may restrict and relate the products of the cubic couplings ($k_i$ in our shorthand notation). These constraints/relations are summarised in the tables below. If the entry is zero, it means that the corresponding product of couplings must vanish. If the entry contains $k_i$, it means that two (or more) $k_i$ turns out to be related or unconstrained. We summarise the results in Tables~\ref{tab(1,1)}, \ref{tab(2,1)}, and \ref{tab(2,2)}. For example, in Table~\ref{tab(1,1)} the second column has $k_2$ under $k_2$, which means that $k_2$ is unconstrained. The third column has $k_1$ under $k_3$, which means that $k_3=k_1$. The last column has $k_1-k_2$ under $k_6$, which means that $k_6=k_1-k_2$. We also keep $k_1$ as the initial free coupling; hence, there is no reason to include it in the tables. 

A final remark concerning the tables is the following. For exchanges of the form $(\lambda_1,\lambda_1,0,\lambda_2,\lambda_2)$, both commutators $[H_3,J^{z-}_3]$ and $[H_3,J^{\zb-}_3]$ vanish. In this case, the corresponding exchange does not enter the system of constraint \eqref{quartic_system}, and the associated coefficient $k_i$ is therefore left unconstrained. In the tables, we therefore list the coefficient as free. However, it should be kept in mind that, for example, this coefficient will not appear in the solution for the quartic vertex. 

\begin{table}[H]
    \centering
    \begin{tabular}{|c|c|c|c|c|c|c|}
\hline
$(\lambda_1,\lambda_2,\omega,\lambda_3,\lambda_4)$ & $k_2=(2341)$ & $k_3=(3412)$ & $k_4=(4123)$ & $k_5=(1324)$ & $k_6=(2413)$ \\ \hline
$ (-1,1,1,-1,1)$  & $k_2$ & $k_1$ & $k_2$ & $0$ & $k_1-k_2$ \\ \hline
$ (-1,1,1,0,0)$ & $k_2$ & $k_1$ & $0$ & $0$ & $k_1-k_2$ \\ \hline
$(0,0,1,0,0)$  & $k_2$ & $k_1$ & $k_2$ & $k_5$ & $k_5$ \\ \hline
    \end{tabular}
    \caption{$(1,1)$ quartic vertices satisfying the quartic constraints.}
    \label{tab(1,1)}
\end{table}

\begin{table}[H]
    \centering
    \begin{tabular}{|c|c|c|c|c|c|c|}
\hline
$(\lambda_1,\lambda_2,\omega,\lambda_3,\lambda_4)$ & $k_2=(2341)$ & $k_3=(3412)$ & $k_4=(4123)$ & $k_5=(1324)$ & $k_6=(2413)$ \\ \hline
$(-1,2,1,-1,1)$  & $k_1$ & $0$ & $0$ & $0$ & $k_1$ \\ \hline
$(-1,2,1,0,0)$  & $k_1$ & $0$ & $0$ & $0$ & $k_1$ \\ \hline
$(0,1,1,-1,1)$ & $0$ & $0$ & $k_4$ & $0$ & $k_1-k_4$ \\ \hline
$(0,1,1,0,0)$  & $k_2$ & $0$ & $0$ & $0$ & $k_6$ \\ \hline
    \end{tabular}
    \caption{$(2,1)$ quartic vertices satisfying the quartic constraints.}
    \label{tab(2,1)}
\end{table}

\begin{table}[H]
    \centering
    \begin{tabular}{|c|c|c|c|c|c|c|}
\hline
$(\lambda_1,\lambda_2,\omega,\lambda_3,\lambda_4)$ & $k_2=(2341)$ & $k_3=(3412)$ & $k_4=(4123)$ & $k_5=(1324)$ & $k_6=(2413)$ \\ \hline
$(-2,2,2,-2,2)$  & $k_1$ & $k_1$ & $k_1$ & $0$ & $k_1$ \\ \hline
$(-2,2,2,-1,1)$  & $k_1$ & $k_1$ & $0$ & $0$ & $k_1$ \\ \hline
$(-2,2,2,0,0)$  & $k_1$ & $k_1$ & $0$ & $0$ & $k_1$ \\ \hline
$(-1,1,2,-1,1)$ & $k_2$ & $k_1$ & $k_2$ & $0$ & $k_6$ \\ \hline
$(-1,1,2,0,0)$ & $k_2$ & $k_1$ & $0$ & $0$ & $k_6$ \\ \hline
$(-1,2,1,-1,0)$  & $k_1$ & $0$ & $0$ & $0$ & $k_1$ \\ \hline
$(0,0,2,0,0)$ & $k_2$ & $k_1$ & $k_2$ & $k_5$ & $k_5$ \\ \hline
$(0,1,1,-2,1)$  & $0$ & $0$ & $k_1$ & $0$ & $k_1$ \\ \hline
    \end{tabular}
    \caption{$(2,2)$ quartic vertices satisfying the quartic constraints.}
    \label{tab(2,2)}
\end{table}
\noindent
Several free coefficients remain, corresponding to solutions that are consistent (at this order!). For instance, in the Yang–Mills case $ (-1,1,1,-1,1)$, the presence of two free coefficients, $k_1$ and $k_2$, is related to the color structure of the theory. We have two different quartic contributions for the [++$--$] and [+$-$+$-$] sectors. Instead, for gravity $(-2,2,2,-2,2)$, there is only one free coefficient, reflecting the no-go for a multi-graviton theory (a color graviton would have two different color-ordered amplitudes). 

Let us stress once again that these tables represent consistency at the quartic order. The quintic light-cone constraint may impose further conditions. For instance, in Yang–Mills theory, the quintic Hamiltonian should vanish since the theory has no quintic vertex; therefore, the quintic constraint should be satisfied automatically without requiring a nonzero quintic vertex.

In the following, we prove the existence (at the quartic order) and uniqueness of Yang–Mills theory and gravity using the results collected in the tables above. Our strategy proceeds as follows:
\begin{itemize}
\item Firstly, we list the unique cubic vertices using \eqref{famouscubic}.

\item Secondly, we solve the holomorphic constraints \eqref{HOLO} that give relations among products of (anti-)holomorphic $CC$ ($\bar{C}\bar{C}$) couplings.

\item Thirdly, we use the information provided by the tables above, which establish the existence of a quartic vertex solving the quartic constraint \eqref{quartic_system} and determine the relations among the products of $C\bar{C}$ couplings.

\item Fourthly, we use the CPT symmetry.
\end{itemize}
As we will see below, the holomorphic constraint gives us the Jacobi identity for the self-dual sectors, while the solution of the non-holomorphic quartic constraint \eqref{quartic_system} gives us, together with the CPT symmetry, among others, a relation that in \cite{Fonseca:2025mzj} was called non-ambiguity condition (NAC). This allows us to relate the two self-dual sectors and find a unique Lie algebra.

\paragraph{Existence and uniqueness of Yang-Mills theory.}  We analyse the case of a spin-$1$ self-interacting theory. We start with the one-derivative holomorphic and anti-holomorphic spin-$1$ cubic vertices:
\begin{equation}\label{YM_cubic}
    h^{\text{YM}}_3=\fA^{a}_{[bc]}\frac{\PPb \beta_1}{\beta_2\beta_3}(\phi^-_{q_1})_a(\phi^+_{q_2})^b(\phi^+_{q_3})^c+\bar{\fA}^{[ab]}_{c}\frac{\PP \beta_3}{\beta_1\beta_2}(\phi^-_{p_1})_a(\phi^-_{p_2})_b(\phi^+_{p_3})^c\,,
\end{equation}
where $\fA^{a}_{[bc]}=-\fA^{a}_{[cb]}$ and $\bar{\fA}^{[ab]}_{c}=-\bar{\fA}^{[ba]}_{c}$ are antisymmetric as a consequence of \eqref{sym_ff}. In what follows, we omit the brackets $[\;]$. The two sets of indices (upper indices for $\phi^{+}$ and lower indices for $\phi^-$) belong, in general, to different vector spaces. The only requirement needed to define the kinetic term ($\sim \partial_{\mu}(\phi^{-})^a\partial^{\mu}(\phi^{+})_a$) and to allow contractions in the commutator \eqref{commutator_f} is the existence of a pairing between these spaces. Accordingly, we assume that the positive- and negative-helicity fields $(\phi^+)^a$ and $(\phi^-)_b$ take values in vector spaces that are dual to each other.

Both couplings $\fA$ and $\bar{\fA}$ satisfy their own Jacobi identity as a direct consequence of the holomorphic quartic constraint \cite{Serrani:2025owx}. Let us briefly recall how this arises. For the cubic vertices \eqref{YM_cubic}, only a single commutator contributes to the holomorphic constraint \eqref{HOLO} (see \cite{Serrani:2025owx}):
\begin{equation}
    \fA^{a_1}_{a_2a_3}\fA^{b_1}_{b_2b_3}[(\phi^-_{q_1})_{a_1}(\phi^+_{q_2})^{a_2}(\phi^+_{q_3})^{a_3},(\phi^-_{p_1})_{b_1}(\phi^+_{p_2})^{b_2}(\phi^+_{p_3})^{b_3}]\,.
\end{equation}
The holomorphic constraint gives
\begin{align}
    \begin{split}
    [H_3(\PPb),J_3^{z-}]=\int d^{12}q\;\delta \left(\sum_i q_i\right)&\left(\frac{\beta_1-\beta_2}{\beta_1+\beta_2}\PPb_{34}+\frac{\beta_3+3\beta_4}{\beta_3+\beta_4}\PPb_{12}\right)\\
    &\fA^c_{a_1a_2}\fA^{a_3}_{a_4c}(\phi^+_{q_1})^{a_1}(\phi^+_{q_2})^{a_2}(\phi^-_{q_3})_{a_3}(\phi^+_{q_4})^{a_4}=0\,.
    \end{split}
\end{align}
We can use the three independent variables $A,B,C$ defined as 
\begin{align}\label{ABC_variables}
\begin{split}
    2A&=\PPb_{12}+\PPb_{34}=\PPb_{23}-\PPb_{14}\,,\qquad
    2B=\PPb_{13}-\PPb_{24}=\PPb_{34}-\PPb_{12}\,,\\
    2C&=\PPb_{14}+\PPb_{23}=-\PPb_{13}-\PPb_{24}\,,
    \end{split}
\end{align}
and notice that Bose symmetrisation imposes that the equation above must be symmetric under the exchange of fields $1,2,4$. This leads to the form\footnote{Alternatively, we could apply the general formula in \eqref{holo_constraint} directly, summing over all possible contributions.} 
    \begin{align}
    \begin{split}
    [H_3(\PPb),J_3^{z-}]=&\int d^{12}q\;\delta \left(\sum_i q_i\right)(A-B-C)\Big(\fA^c_{a_1a_2}\fA^{a_3}_{a_4c}+\fA^c_{a_2a_4}\fA^{a_3}_{a_1c}\\
    &+\fA^c_{a_4a_1}\fA^{a_3}_{a_2c}\Big)(\phi^+_{q_1})^{a_1}(\phi^+_{q_2})^{a_2}(\phi^-_{q_3})_{a_3}(\phi^+_{q_4})^{a_4}=0\,.
    \end{split}
\end{align}
The unique nontrivial solution is given by the Jacobi identity
\begin{equation}\label{SDJI_YM}
    \fA^c_{a_1a_2}\fA^{a_3}_{a_4c}+\fA^c_{a_2a_4}\fA^{a_3}_{a_1c}+\fA^c_{a_4a_1}\fA^{a_3}_{a_2c}=0\,.
\end{equation}
The same steps can be repeated for $\bar{\fA}$, leading to the Jacobi identity in the dual space. Therefore, within the light-cone formalism, the Jacobi identities for both $\fA$ and $\bar{\fA}$ follow directly from the holomorphic quartic constraint. As a result, already the self-dual sectors are governed by a Lie algebra; a complex Lie algebra in the generic case.

Let us now turn to the analysis of the non-holomorphic quartic constraint. As we will see, the conditions that arise are closely related to those obtained by imposing consistent factorisation or via constructibility using BCFW. An analysis along these lines was recently carried out in \cite{Fonseca:2025mzj}. We will adopt the same terminology here for the various conditions we will find.

We describe in detail the information that can be extracted from the first line of Table \ref{tab(1,1)}. We have $(-1,1,1,-1,1)$, corresponding to the exchange $C^{-1,1,1}\bar{C}^{-1,-1,1}$, with two free coefficients, $k_1$ and $k_2$. We find two independent quadratic conditions on the coupling constants:
\begin{align}
\nonumber
    &k_1=k_3\implies (-)^{\omega}\mathcal{F}_{1234}=(-)^{\omega}\mathcal{F}_{3412}\\\label{NAC_YM}
    &\implies C^{-1,1,1}\bar{C}^{-1,-1,1}\fA^{a_1}_{a_2c}\bar{\fA}^{c\,a_3}_{a_4}=C^{-1,1,1}\bar{C}^{-1,-1,1}\fA^{a_3}_{a_4c}\bar{\fA}^{c\,a_1}_{a_2}\implies \boxed{\fA^{a_1}_{a_2c}\bar{\fA}^{c\,a_3}_{a_4}=\fA^{a_3}_{a_4c}\bar{\fA}^{c\,a_1}_{a_2}}\,,\\
    \nonumber
    &k_1-k_2=k_6\implies (-)^{\omega}\mathcal{F}_{1234}-(-)^{\omega}\mathcal{F}_{2341}=(-)^{\omega}\mathcal{F}_{2413}\\
    \nonumber
    &\implies C^{-1,1,1}\bar{C}^{-1,-1,1}\fA^{a_1}_{a_2c}\bar{\fA}^{c\,a_3}_{a_4}-C^{1,-1,1}\bar{C}^{-1,1,-1}\fA^{a_3}_{a_2c}\bar{\fA}^{c\,a_1}_{a_4}=C^{1,1,-1}\bar{C}^{1,-1,-1}\fA^{c}_{a_2a_4}\bar{\fA}^{a_1a_3}_{c}\\
    \nonumber
    &\implies C^{-1,1,1}\bar{C}^{-1,-1,1}\fA^{a_1}_{a_2c}\bar{\fA}^{c\,a_3}_{a_4}-C^{-1,1,1}\bar{C}^{-1,-1,1}\fA^{a_3}_{a_2c}\bar{\fA}^{c\,a_1}_{a_4}=C^{-1,1,1}\bar{C}^{-1,-1,1}\fA^{c}_{a_2a_4}\bar{\fA}^{a_1a_3}_{c}\\\label{JI_YM}
    &\implies \boxed{\fA^{a_1}_{a_2c}\bar{\fA}^{c\,a_3}_{a_4}-\fA^{a_3}_{a_2c}\bar{\fA}^{c\,a_1}_{a_4}=\fA^{c}_{a_2a_4}\bar{\fA}^{a_1a_3}_{c}}\,,
\end{align}
where we have used the symmetry property \eqref{f_sym} of $C^{\lambda_1,\lambda_2,\lambda_3}\fA^{a_1a_2a_3}$. We also notice that the third condition $k_2=k_4$ would lead to the same constraint as \eqref{NAC_YM}. In analogy with \cite{Fonseca:2025mzj}, we call the condition \eqref{JI_YM} the modified Jacobi identity and \eqref{NAC_YM} the non-ambiguity condition (NAC).

These relations are, in principle, enough for the existence of a ``Yang-Mills'' theory, up to the quartic order. However, if we are willing to relate the holomorphic and anti-holomorphic couplings, we need a further assumption, CPT symmetry. CPT symmetry implies the following relation between $\fA$ and $\bar{\fA}$ (see \cite{Fonseca:2025mzj}):
\begin{equation}\label{CPT_YM}
    \bar{\fA}^{ab}_{c}=(\fA^{c}_{ab})^*\,.
\end{equation}
Notice that this is also a consequence of imposing unitarity.

In \cite{Fonseca:2025mzj} it was proven that \eqref{NAC_YM} together with \eqref{CPT_YM} implies the existence of a basis where the $F_a=(\fA_a)^c_b$ are Hermitian $F_a=F^{\dag}_a$; then $\fA^c_{ab}=(\fA^b_{ac})^*$. We can now further use \eqref{NAC_YM} to show that $\fA$ is indeed imaginary and fully antisymmetric $\fA_{abc}=\fA_{[abc]}$.
The Jacobi identity for $\fA$ is now a consequence either of the Jacobi identity for the self-dual sector \eqref{SDJI_YM} or of the modified Jacobi identity \eqref{JI_YM}. Therefore, the self-coupling of a spin $1$ massless particle has the structure of a Lie algebra, with the couplings $\fA^c_{ab}$ that play the role of the structure constants. 

The results just obtained can also be derived using the Noether procedure, starting from the free Maxwell Lagrangian in the covariant formulation. In particular, this problem can be addressed in a more systematic way within the framework of BV-BRST cohomology. For Yang–Mills theory, this analysis was carried out in \cite{Wald:1986bj, Barnich:2000zw}, and we find perfect agreement.

\paragraph{Existence and uniqueness of gravity.} We can perform the same analysis we have done for Yang-Mills with gravity. We start with the two cubic spin-$2$ vertices:
\begin{equation}\label{GR_cubic}
    h^{\text{GR}}_3=g^{a}_{(bc)}\frac{\PPb^2 \beta^2_1}{\beta^2_2\beta^2_3}(\phi^{-2}_{q_1})_a(\phi^{+2}_{q_2})^b(\phi^{+2}_{q_3})^c+\bar{g}^{(ab)}_{c}\frac{\PP^2 \beta^2_3}{\beta^2_1\beta^2_2}(\phi^{-2}_{q_1})_a(\phi^{-2}_{q_2})_b(\phi^{+2}_{q_3})^c\,,
\end{equation}
where $g^{a}_{(bc)}=g^{a}_{(cb)}$ and $\bar{g}^{(ab)}_{c}=\bar{g}^{(ba)}_{c}$ are symmetric as a consequence of \eqref{sym_ff}. In what follows, we omit the brackets $(\;)$.

Both couplings $g$ and $\bar{g}$ form an associative and commutative algebra, as a direct consequence of the holomorphic quartic constraint \cite{Serrani:2025owx}. Let us briefly recall how this arises. For the cubic vertices \eqref{GR_cubic}, only a single commutator contributes to the holomorphic constraint \eqref{HOLO} (see \cite{Serrani:2025owx}):
\begin{equation}
    g^{a_1}_{a_2a_3}g^{b_1}_{b_2b_3}[(\phi^{-2}_{q_1})_{a_1}(\phi^{+2}_{q_2})^{a_2}(\phi^{+2}_{q_3})^{a_3},(\phi^{-2}_{p_1})_{b_1}(\phi^{+2}_{p_2})^{b_2}(\phi^{+2}_{p_3})^{b_3}]\,,
\end{equation}
The holomorphic constraint gives
\begin{align}
    \begin{split}
    [H_3(\PPb),J_3^{z-}]=\int d^{12}q\;\delta \left(\sum_i q_i\right)&\left(\frac{2(\beta_2-\beta_1)}{\beta_1+\beta_2}\PPb_{12}\PPb_{34}^2-\frac{2(\beta_3+3\beta_4)}{\beta_3+\beta_4}\PPb_{34}\PPb_{12}^2\right)\\
    &g^c_{a_1a_2}g^{a_3}_{a_4c}(\phi^{+2}_{q_1})^{a_1}(\phi^{+2}_{q_2})^{a_2}(\phi^{-2}_{q_3})_{a_3}(\phi^{+2}_{q_4})^{a_4}=0\,.
    \end{split}
\end{align}
We can use the variables $A,B,C$ defined in \eqref{ABC_variables} and notice that Bose symmetrisation imposes that the equation above must be symmetric under the exchange of fields $1,2,4$. This leads to the form
\begin{align}
    \begin{split}
    [&H_3(\PPb),J_3^{z-}]=\int d^{12}q\;\delta \left(\sum_i q_i\right)(-A+B+C)\Big((A^2-B^2)g^c_{a_1a_2}g^{a_3}_{a_4c}\\
    &+(B^2-C^2)g^c_{a_2a_4}g^{a_3}_{a_1c}+(C^2-A^2)g^c_{a_4a_1}g^{a_3}_{a_2c})\Big)(\phi^{+2}_{q_1})^{a_1}(\phi^{+2}_{q_2})^{a_2}(\phi^{-2}_{q_3})_{a_3}(\phi^{+2}_{q_4})^{a_4}=0\,.
    \end{split}
\end{align}
The unique nontrivial solution is given by
\begin{equation}\label{holoGR_solution}
    g^c_{a_1a_2}g^{a_3}_{a_4c}=g^c_{a_2a_4}g^{a_3}_{a_1c}=g^c_{a_4a_1}g^{a_3}_{a_2c}\,.
\end{equation}
If we identify $g^a_{bc}$ with the structure constants of a finite $N$-dimensional algebra, with $V$ as internal space and define the product as
\begin{equation}
    (x\cdot y)^a= g^a_{bc} \,x^b y^c\,,\qquad
    \forall\;x,y\in V\,.
\end{equation}
Effectively, the condition \eqref{holoGR_solution} coincides with the associativity of the algebra:
\begin{align}
    &(x\cdot y)\cdot z= x\cdot (y\cdot z)&
    &\implies&
    &g^c_{a_1[a_2}g^{a_3}_{a_4]c}=g^c_{a_1a_2}g^{a_3}_{a_4c}-g^c_{a_1a_4}g^{a_3}_{a_2c}=0\,.
\end{align}
Moreover, the condition $g^{a}_{bc}=g^{a}_{cb}$ implies that the algebra is commutative.
The same steps can be repeated for $\bar{g}$, leading to the commutativity and associativity of the dual algebra. Therefore, within the light-cone formalism, the associativity for both $g$ and $\bar{g}$ follow directly from the holomorphic quartic constraint. As a result, already the self-dual sectors are governed by a commutative and associative algebra.

We now turn to the analysis of the non-holomorphic quartic constraint. We describe in detail the information that can be extracted from the first line of Table \ref{tab(2,2)}. We have $(-2,2,2,-2,2)$ corresponding to the exchange $C^{-2,2,2}\bar{C}^{-2,-2,2}$, with only one free coefficient, $k_1$. We find the following independent quadratic conditions on the structure constants: 
\begin{align}
\nonumber
    &k_1=k_3\implies (-)^{\omega}\mathcal{F}_{1234}=(-)^{\omega}\mathcal{F}_{3412}\\
    \nonumber
    &\implies C^{-2,2,2}\bar{C}^{-2,-2,2}g^{a_1}_{a_2c}\bar{g}^{c\,a_3}_{a_4}=C^{-2,2,2}\bar{C}^{-2,-2,2}g^{a_3}_{a_4c}\bar{g}^{c\,a_1}_{a_2}\\\label{ASS_GR}
    &\implies \boxed{g^{a_1}_{a_2c}\bar{g}^{c\,a_3}_{a_4}=g^{a_3}_{a_4c}\bar{g}^{c\,a_1}_{a_2}}\,,\\
    \nonumber
    &k_1=k_2=k_6\implies (-)^{\omega}\mathcal{F}_{1234}=(-)^{\omega}\mathcal{F}_{2341}=(-)^{\omega}\mathcal{F}_{2413}\\
    \nonumber
    &\implies C^{-2,2,2}\bar{C}^{-2,-2,2}g^{a_1}_{a_2c}\bar{g}^{c\,a_3}_{a_4}=C^{2,-2,2}\bar{C}^{-2,2,-2}g^{a_3}_{a_2c}\bar{g}^{c\,a_1}_{a_4}=C^{2,2,-2}\bar{C}^{2,-2,-2}g^{c}_{a_2a_4}\bar{g}^{a_1a_3}_{c}\\\label{NAC_GR}
    &\implies \boxed{g^{a_1}_{a_2c}\bar{g}^{c\,a_3}_{a_4}=g^{a_3}_{a_2c}\bar{g}^{c\,a_1}_{a_4}=g^{c}_{a_2a_4}\bar{g}^{a_1a_3}_{c}}\,.
\end{align}
The other condition, $k_1=k_4$, is a consequence of the one above. We call the condition \eqref{ASS_GR} the modified associativity and \eqref{NAC_GR} the non-ambiguity condition.

These relations are, in principle, sufficient to allow the existence of a theory of ``gravity'', up to the quartic order. However, if we are willing to relate the holomorphic and anti-holomorphic couplings, we need to assume CPT symmetry.  CPT symmetry implies the following relation between $g$ and $\bar{g}$, see \cite{Fonseca:2025mzj}:
\begin{equation}\label{CPT_GR}
    \bar{g}^{ab}_{c}=(g^{c}_{ab})^*\,.
\end{equation}
Notice that this is also a consequence of unitarity. 

In \cite{Fonseca:2025mzj} was proven that  \eqref{NAC_GR} together with \eqref{CPT_GR} implies the existence of a basis where the $G_a=(g_a)^c_b$ are Hermitian $G_a=G^{\dag}_a$, then $g^c_{ab}=(g^b_{ac})^*$. We can now further use \eqref{NAC_GR} to show that $g$ is indeed real and fully symmetric $g_{abc}=g_{(abc)}$.
The associativity for $g$ is now a consequence either of the associativity for the self-dual sector or of the modified associativity \eqref{ASS_GR}. Therefore, the self-coupling of a spin $2$ massless particle has the structure of a commutative, symmetric, and associative algebra, with structure constant $g^c_{ab}$.

Once again, the same results can be derived using the Noether procedure within the framework of BV-BRST cohomology. For gravity, this analysis was carried out in \cite{Wald:1986bj, Boulanger:2000rq}, and we find perfect agreement. In \cite{Boulanger:2000rq}, it was also shown that finite-dimensional real algebras endowed with a positive-definite scalar product, which are commutative, symmetric, and associative, necessarily have a trivial structure: they decompose into a direct sum of one-dimensional ideals. Consequently, for the graviton, only self-interactions are possible.

Following the same line of reasoning applied to the cubic couplings of Yang–Mills theory and gravity, a similar analysis can be carried out for any other set of cubic couplings. For example, looking at the second line of Table \ref{tab(1,1)}, where we have $(-1,1,1,0,0)$. In this case, we find the following condition:
\begin{align}
\nonumber
    &k_1=k_3\implies (-)^{\omega}\mathcal{F}_{1234}=(-)^{\omega}\mathcal{F}_{3412}\\\label{YMScalar_cond1}
    &\implies \boxed{C^{-1,1,1}\bar{C}^{-1,0,0}\fA^{a_1}_{a_2c}\bar{\ell}^{c}_{a_3a_4}=C^{0,0,1}\bar{C}^{-1,-1,1}\ell^{a_3a_4}_{c}\bar{\fA}^{c\,a_1}_{a_2}}\,,\\
    \nonumber
    &k_1-k_2=k_6\implies (-)^{\omega}\mathcal{F}_{1234}-(-)^{\omega}\mathcal{F}_{2341}=(-)^{\omega}\mathcal{F}_{2413}\\
    \nonumber
    &\implies -C^{-1,1,1}\bar{C}^{-1,0,0}\fA^{a_1}_{a_2c}\bar{\ell}^{c}_{a_3a_4}-C^{1,0,0}\bar{C}^{0,0,-1}\ell^{a_3c}_{a_2}\bar{\ell}^{a_1}_{c\,a_4}=C^{1,0,0}\bar{C}^{0,-1,0}\ell^{a_4c}_{a_2}\bar{\ell}^{a_1}_{c\,a_3}\\\label{YMScalar_cond2}
    &\implies \boxed{C^{-1,1,1}\bar{C}^{-1,0,0}\fA^{a_1}_{a_2c}\bar{\ell}^{c}_{a_3a_4}+C^{1,0,0}\bar{C}^{0,0,-1}\ell^{a_3c}_{a_2}\bar{\ell}^{a_1}_{c\,a_4}=C^{1,0,0}\bar{C}^{-1,0,0}\ell^{a_4c}_{a_2}\bar{\ell}^{a_1}_{c\,a_3}}\,.
\end{align}
While analysing the third line of Table \ref{tab(1,1)}, where we have $(0,0,1,0,0)$, gives
\begin{align}
\nonumber
    &k_1=k_3\implies (-)^{\omega}\mathcal{F}_{1234}=(-)^{\omega}\mathcal{F}_{3412}\\\label{Scalar_cond1}
    &\implies C^{0,0,1}\bar{C}^{-1,0,0}\ell^{a_1a_2}_{c}\bar{\ell}^{c}_{a_3a_4}=C^{0,0,1}\bar{C}^{-1,0,0}\ell^{a_3a_4}_{c}\bar{\ell}^{c}_{a_1a_2}\implies \boxed{\ell^{a_1a_2}_{c}\bar{\ell}^{c}_{a_3a_4}=\ell^{a_3a_4}_{c}\bar{\ell}^{c}_{a_1a_2}}\,.
\end{align}
Of particular interest would be to perform this analysis explicitly for the most general lower-spin cubic couplings, in order to classify all allowed lower-spin theories up to quartic order and to identify whether any new possibilities emerge. The advantage of the light-cone approach is threefold:
\begin{itemize}
\item It provides the most general framework for studying perturbation theory, allowing for a complete search in which nothing is missed.

\item It gives natural access to the self-dual sectors.

\item By treating all helicities on the same footing, as complex scalar fields, it allows us to find solutions for all higher-spin fields, as we will see in the next section.

\end{itemize}
We also emphasise that, in spirit, light-cone perturbation theory is closer to on-shell methods \cite{Benincasa:2007xk}. To study the self-dual sector within the BV-BRST cohomology framework, one would likely need to start from a chiral version of the free action and use chiral variables, along the lines of \cite{Krasnov:2021nsq}. In particular, it would be interesting to understand how the holomorphic constraint emerges in that approach.

\subsection{Lower-spin theories}

As a warm-up exercise and a consistency check of our method, we compute the quartic vertices $h_4$ for Yang-Mills theory and the quartic vertex $h_4$ along with its related boost generators $j^{z-}_4$ and $j^{\bar{z}-}_4$ for gravity. 
These are already known in light-cone gauge \cite{Brink:1982pd,Bengtsson:1983vn,Siegel:1999ew,Chakrabarti:2005ny,Chakrabarti:2006mb}, and we reproduce them in our formalism. The idea is to start from the most general ansatz described above and solve the system of quartic constraints \eqref{quartic_system}.

We employ two different definitions for $x$ and $y$. The choice of definition is dictated by the symmetry properties of the specific example. In particular, we may use either of the following:
\begin{align}\label{xy_1234}
    &x=\frac{\beta_1-\beta_2}{\beta_1+\beta_2}\,,&
    &y=\frac{\beta_3-\beta_4}{\beta_3+\beta_4}\,.
\end{align}
We then denote the quartic vertex using round brackets as $h^{(n,m)}_{(\lambda_1,\lambda_2,\lambda_3,\lambda_4)}$. Alternatively, we may use the following definition:
\begin{align}\label{xy_1324}
    &x=\frac{\beta_1-\beta_3}{\beta_1+\beta_3}\,,&
    &y=\frac{\beta_2-\beta_4}{\beta_2+\beta_4}\,.
\end{align}
We then denote the quartic vertex using square brackets as $h^{(n,m)}_{[\lambda_1,\lambda_2,\lambda_3,\lambda_4]}$. This second definition is chosen because it leads to simpler transformation properties under cyclic permutations. We will adopt it in situations where color-like structures are present, as for Yang–Mills theory.

\paragraph{(1,1) quartic vertices.}
These vertices solve the system of quartic constraints \eqref{quartic_system} when the commutators $[H_3,J_3^{z-}]$ and $[H_3,J_3^{\zb-}]$ involve solely cubic one-derivative interactions. The ansatz is
\begin{align}\label{ansatz_YM}
    &h^{(1,1)}_{[\lambda_1,\lambda_2,\lambda_3,\lambda_4]}=f(x,y)\,,
\end{align}
where $f(x,y)$ is a real function of two real variables. 

\paragraph{(2,2) quartic vertices.}
These vertices solve the system of quartic constraints \eqref{quartic_system} when the commutators $[H_3,J_3^{z-}]$ and $[H_3,J_3^{\zb-}]$ involve solely cubic two-derivative interactions. The ansatz is
\begin{align}\label{ansatz_GR}
     \begin{split}
    &h^{(2,2)}_{(\lambda_1,\lambda_2,\lambda_3,\lambda_4)}=\frac{\PP_{12}\PPb_{34}}{(\beta_1+\beta_2)^2}f_1+\left(\frac{\PPb_{12}\PP_{12}}{\beta_1\beta_2}+\frac{\PPb_{34}\PP_{34}}{\beta_3\beta_4}\right)f_2+\frac{\PPb_{12}\PP_{34}}{(\beta_1+\beta_2)^2}f_3\,,\\
     &j^{z-\,(1,2)}_{(\lambda_1,\lambda_2,\lambda_3,\lambda_4)}=\frac{\PP_{12}g_1+\PP_{34}g_2}{(\beta_1+\beta_2)}\,,\qquad
     j^{\zb-\,(2,1)}_{(\lambda_1,\lambda_2,\lambda_3,\lambda_4)}=\frac{\PPb_{12}\bar{g}_1+\PPb_{34}\bar{g}_2}{(\beta_1+\beta_2)}\,,
     \end{split}
\end{align}
where $f_1(x,y)$, $f_2(x,y)$, $f_3(x,y)$, $g_1(x,y)$, $g_2(x,y)$, $\bar{g}_1(x,y)$, and $\bar{g}_2(x,y)$ are real functions of two real variables. 
\paragraph{Yang-Mills theory.} Searching for Yang-Mills theory, we take our fields to transform in the adjoint representation of a Lie algebra. In practice, we just take them to be matrix-valued. Due to the cyclicity of the trace, the two color-ordered contributions, namely [++$--$] and [+$-$+$-$], remain independent and do not mix. The quartic Hamiltonian takes the form:
\begin{equation}\label{YM_quartic_Hamiltonian}
H_4=\int d^{12}q\,\delta\left(\sum q_i\right) \left(h_{[+,+,-,-]}^{(1,1)}\mathrm{Tr}[\phi^+_{q_1}\phi^+_{q_2}\phi^-_{q_3}\phi^-_{q_4}]+h_{[+,-,+,-]}^{(1,1)}\mathrm{Tr}[\phi^+_{q_1}\phi^-_{q_2}\phi^+_{q_3}\phi^-_{q_4}]\right)\,,
\end{equation}
It is well known that Yang-Mills theory provides the unique unitary and parity-invariant non-abelian quartic completion of the cubic vertices $C^{1,1,-1}$ and $\bar{C}^{-1,-1,1}$. Therefore, as a consistency check of our method, we expect to find a unique solution for the quartic vertex. 

Since $j^{z-}_4$ and $j^{\bar{z}-}_4$ are absent in the $(1,1)$ case, the quartic Hamiltonian densities must obey the following system of quartic constraints:
\begin{equation}
    \begin{cases}
    &\textbf{J}_2^{z-}[h_4]+[H_3,J_3^{z-}]=0\\
    &\textbf{J}_2^{\bar{z}-}[h_4]+[H_3,J_3^{\bar{z}-}]=0\,.
    \end{cases}
\end{equation}
Using the independence of $\PP_{12}$ and $\PP_{34}$, and of $\PPb_{12}$ and $\PPb_{34}$, respectively, we end up solving a system of inhomogeneous first-order PDEs. Solving the PDEs gives us the following quartic Hamiltonian densities:
\begin{align}\label{h4_1_YM}
h^{(1,1)}_{[+,+,-,-]}(x,y)&=k_1\frac{x^2+y^2-2}{2(x+y)^2}\,,\\\label{h4_2_YM}
h^{(1,1)}_{[+,-,+,-]}(x,y)&=k_1\frac{2 \left(2 x^2 y^2-x^2-y^2\right)}{\left(x^2-y^2\right)^2}\,,
\end{align}
where $k_1=-\,C^{1,1,-1}C^{1,-1,-1}=-\,C^{1,-1,1}C^{-1,1,-1}$. Note that $h^{(1,1)}_{[+,-,+,-]}(x,y)=h^{(1,1)}_{[+,-,+,-]}(y,x)$ follows from the symmetry $(1234)\leftrightarrow (3412)$, implied by the cyclicity of the trace
\begin{equation}
    h_{[+,-,+,-]}^{(1,1)}(x,y)\mathrm{Tr}[\phi^+_{q_1}\phi^-_{q_2}\phi^+_{q_3}\phi^-_{q_4}]=h_{[+,-,+,-]}^{(1,1)}(x,y)\mathrm{Tr}[\phi^+_{q_3}\phi^-_{q_4}\phi^+_{q_1}\phi^-_{q_2}].
\end{equation}
Moreover, we can check the parity and unitarity invariance of the quartic vertices. For the definitions of parity and unitarity, as well as their action on the densities $h_4$, we refer to Appendix \ref{AppendixA}. In the present case, parity acts by exchanging positive and negative helicities. Thanks to the use of the smart variables \eqref{xy_1324}, this action effectively translates into simple symmetry of the quartic vertices. Both solutions are therefore parity invariant, as we have 
\begin{align}\label{YM1sym}
    &h^{(1,1)}_{[+,+,-,-]}(x,y)\overset{P}{=}h^{(1,1)}_{[-,-,+,+]}(x,y)&
    &\overset{(1234)\leftrightarrow (3412)}{\Longrightarrow}&
    &h^{(1,1)}_{[+,+,-,-]}(x,y)=h^{(1,1)}_{[+,+,-,-]}(-x,-y)\,,\\\label{YM2sym1}
    &h^{(1,1)}_{[+,-,+,-]}(x,y)\overset{P}{=}h^{(1,1)}_{[-,+,-,+]}(x,y)&
    &\overset{(1234)\leftrightarrow (2341)}{\Longrightarrow}&
    &h^{(1,1)}_{[+,-,+,-]}(x,y)=h^{(1,1)}_{[+,-,+,-]}(y,-x)\,,\\\label{YM2sym2}
    &h^{(1,1)}_{[+,-,+,-]}(x,y)\overset{P}{=}h^{(1,1)}_{[-,+,-,+]}(x,y)&
    &\overset{(1234)\leftrightarrow (4123)}{\Longrightarrow}&
    &h^{(1,1)}_{[+,-,+,-]}(x,y)=h^{(1,1)}_{[+,-,+,-]}(-y,x)\,.
\end{align}
As we can see, $h^{(1,1)}_{[+,-,+,-]}(x,y)=h^{(1,1)}_{[+,-,+,-]}(y,x)$, together with the symmetries \eqref{YM2sym1} and \eqref{YM2sym2} explains why $h^{(1,1)}_{[+,-,+,-]}=h^{(1,1)}_{[+,-,+,-]}(x^2,y^2)$.

We can compare \eqref{h4_1_YM} and \eqref{h4_2_YM} with an explicit computation of the quartic vertex of Yang–Mills theory using light-cone coordinates. The calculation is straightforward and begins with the Lagrangian of Yang–Mills theory:
\begin{equation}
    \mathcal{L}^{\text{YM}}(A_{\mu})=-\frac{1}{4}\mathrm{Tr}(F_{\mu\nu}F^{\mu\nu})\,.
\end{equation}
By imposing the light-cone gauge $A^+=0$ and rewriting the Yang–Mills Lagrangian in terms of light-cone coordinates, we obtain the following quartic interaction terms:
\begin{align}
    \begin{split}
H^{\text{YM}}_4&=\int d^{12}q\,\delta\left(\sum q_i\right) \Big(\frac{g^2}{4}\left(\frac{(\beta_2-\beta_3)(\beta_1-\beta_4)}{(\beta_2+\beta_3)(\beta_1+\beta_4)}-1\right)\mathrm{Tr}(A^+_{q_1}A^+_{q_2}A^-_{q_3}A^-_{q_4})\\
        &+\frac{g^2}{4}\left(\frac{(\beta_2-\beta_3)(\beta_1-\beta_4)}{(\beta_2+\beta_3)(\beta_1+\beta_4)}-\frac{(\beta_1-\beta_2)(\beta_3-\beta_4)}{(\beta_1+\beta_2)(\beta_3+\beta_4)}+2\right)\mathrm{Tr}(A^+_{q_1}A^-_{q_2}A^+_{q_3}A^-_{q_4})\Big)\,.
        \end{split}
\end{align}
where we denote by $g$ the coupling constant, which, upon using momentum conservation, corresponds precisely to the quartic vertices given in \eqref{h4_1_YM} and \eqref{h4_2_YM} upon identifying $k_1=-g^2$. This also shows that by employing suitable variables, the form of the vertices can be simplified. We have also verified our result by comparison with those obtained in \cite{Chakrabarti:2005ny, Chakrabarti:2006mb}, finding complete agreement; see also \cite{Ananth:2017pio}.

\paragraph{Gravity.} Repeating the same procedure but for gravity, using for $h^{(2,2)}_{(+2,+2,-2,-2)}$ the ansatz \eqref{ansatz_GR}, we find
\begin{subequations}\label{GR}
\begin{align}
    f_1(x,y)&=-\frac{1}{\left(x^2-y^2\right)^4}\big(32 x y \left(x^2 \left(\left(x^2-4\right) y^2+y^4+1\right)+y^2\right)\big)\,,\\
    \nonumber
    f_2(x,y)&=\frac{1}{\left(x^2-y^2\right)^4}\big(2 \big(x^6 \left(-4 y^4-7 y^2+5\right)+x^4 \left(-4 y^6+38 y^4-17 y^2+1\right)\\
    &\qquad\qquad\qquad +x^2 y^2 \big(-7 y^4-17 y^2+6\big)+5 y^6+y^4\big)\big)\,,\\
    f_3(x,y)&=-\frac{1}{\left(x^2-y^2\right)^4}\big(32 x y \left(x^4 \left(y^2+8\right)+x^2 \left(y^4-20 y^2+1\right)+8 y^4+y^2\right)\big)\,,\\
    \nonumber
    g_1(x,y)&=\frac{1}{\left(x^2-1\right)
   \left(x^2-y^2\right)^4}\big(16 x \big(-5 x^6+\left(3 x^4+5 x^2-3\right) y^6+\left(8 x^4+10 x^2-3\right) x^2 y^2\\
   &\qquad\qquad\qquad+\left(3 x^6-31 x^4+16 x^2-3\right) y^4\big)\big)\,,\\
   \nonumber
   g_2(x,y)&=-\frac{1}{\left(y^2-1\right)
   \left(x^2-y^2\right)^4}\big(16 y \big(x^6 \left(3 y^4+27 y^2-29\right)+x^4 \left(3 y^6-67 y^4+58 y^2+3\right)\\
   &\qquad\qquad\qquad+x^2 y^2 \left(22 y^4-12 y^2-7\right)+y^4 \left(y^2-2\right)\big)\big)\,.
\end{align}
\end{subequations}
The symmetry under the exchange $(1234)\leftrightarrow (2134)$ implies
\begin{align}
    &f_1(x,y)=-f_1\left(-x,y\right)\,,&
    &f_2(x,y)=f_2\left(-x,y\right)\,,&
    &f_3(x,y)=-f_3\left(-x,y\right)\,.
\end{align}
Similarly, under the exchange $(1234)\leftrightarrow (1243)$, it implies
\begin{align}
    &f_1(x,y)=-f_1\left(x,-y\right)\,,&
    &f_2(x,y)=f_2\left(x,-y\right)\,,&
    &f_3(x,y)=-f_3\left(x,-y\right)\,.
\end{align}
Moreover, some symmetries follow from parity invariance. In particular, parity relates
\begin{align}
    &h^{(2,2)}_{(+2,+2,-2,-2)}(x,y)\overset{P}{=}\bar{h}^{(2,2)}_{(-2,-2,+2,+2)}(x,y)\,.
\end{align}
This leads to additional relations among the $f_i$. For instance, the symmetry under the exchange $(1234)\leftrightarrow (3412)$ implies
\begin{align}
    &f_1(x,y)=f_1(y,x)\,,&
    &f_2(x,y)=f_2(y,x)\,,&
    &f_3(x,y)=f_3(y,x)\,.
\end{align}
The symmetry under the exchange $(1234)\leftrightarrow (2134)$ gives
\begin{align}
    &f_1(x,y)=-f_3\left(-x,y\right)\,,&
    &f_2(x,y)=f_2\left(-x,y\right)\,,
\end{align}
and under the exchange $(1234)\leftrightarrow (1243)$ gives
\begin{align}
    &f_1(x,y)=-f_3\left(x,-y\right)\,,&
    &f_2(x,y)=f_2\left(x,-y\right)\,.
\end{align}
Finally, note that the two constraints appearing in \eqref{quartic_system} are related by a parity transformation. Concretely, under parity, one has
\begin{align}\label{parity_related_constraint}
    &\mathbf{H}_2 j_4^{z-}=\mathbf{J}_2^{z-}[h_4]+[H_3,J_3^{z-}] &
    &\overset{P}{\longrightarrow}&
    &\mathbf{H}_2 j_4^{\zb-}=\mathbf{J}_2^{\zb-}[h_4]+[H_3,J_3^{\zb-}]\,,
\end{align}
where the Hamiltonians are parity invariant. Therefore, $j^{\bar{z}-\,(2,1)}_{(-2,-2,+2,+2)}=P\,j^{z-\,(1,2)}_{(+2,+2,-2,-2)}$.

\paragraph{Others lower-spin quartic vertices.}
We follow the same procedure used for Yang–Mills theory and gravity to construct quartic vertices for lower-spin theories with one- and two-derivative interactions. For clarity and readability of the main text, all results are collected in Appendix \ref{AppendixD}.

\subsection{Four-point amplitudes}\label{subsection5.2}

All massless cubic vertices found in the light-cone formalism can also be recovered in the spinor-helicity one (as amplitudes). This indicates the close relationship between the two approaches. When restricted to on-shell external fields, the two coincide; see \cite{Ponomarev:2016cwi}. For all known massless lower-spin theories, four-point amplitudes are well-known and can be expressed using the spinor-helicity formalism \cite{Elvang:2013cua}. 

Here, we employ the explicit on-shell map between the spinor-helicity and light-cone formalisms to compare our results and verify their agreement. This provides a strong consistency check for the quartic Hamiltonian densities we have derived. We illustrate this explicitly for Yang–Mills theory and gravity. The same analysis extends straightforwardly to the other cases and to higher spins, and additional amplitudes will be discussed in a following section.

It is essential to emphasise that our results extend beyond simply reproducing known on-shell amplitudes; they are genuinely new, as they hold off-shell. The quartic Hamiltonian densities we have constructed are off-shell quantities. For instance, our off-shell formulation allows one to write a Lagrangian and compute off-shell quantities (e.g. form factors). This is the trade-off for the more involved procedure compared to the spinor-helicity method, which builds amplitudes using little-group scaling for the cubic ones and applies on-shell recursion relations (like BCFW) for higher-point ones. While efficient, the latter, in its simpler form, assumes ``good'' high-energy behaviour in order to efficiently drop the contour integral at infinity, making it less suited for constructing vertices for higher-spin fields \cite{Benincasa:2007xk}. We review the on-shell relation between the light-cone and spinor-helicity formalisms in Appendix \ref{AppendixE}. 

The total four-point amplitude in light-cone gauge is computed by summing all nonvanishing exchange contributions together with the quartic vertex:
\begin{align}\label{total_A}
        \begin{split}
        \mA&(1_{\lambda_1}2_{\lambda_2}3_{\lambda_3}4_{\lambda_4})=k_1\mA^{\text{e}}(1_{\lambda_1}2_{\lambda_2}3_{\lambda_3}4_{\lambda_4})+k_2\mA^{\text{e}}(2_{\lambda_2}3_{\lambda_3}4_{\lambda_4}1_{\lambda_1})+k_3\mA^{\text{e}}(3_{\lambda_3}4_{\lambda_4}1_{\lambda_1}2_{\lambda_2})\\
        &+k_4\mA^{\text{e}}(4_{\lambda_4}1_{\lambda_1}2_{\lambda_2}3_{\lambda_3})+k_5\mA^{\text{e}}(1_{\lambda_1}3_{\lambda_3}2_{\lambda_2}4_{\lambda_4})+k_6\mA^{\text{e}}(2_{\lambda_2}4_{\lambda_4}1_{\lambda_1}3_{\lambda_3})+h^{(n,m)}_{(\lambda_1,\lambda_2,\lambda_3,\lambda_4)}\,,
        \end{split}
\end{align}
where $\mA^{\text{e}}$ denotes the exchange diagram that, in the light-cone gauge, takes the form
\begin{align}\label{exchange_diagram}
    \begin{split}
    \mA^{\text{e}}(1_{\lambda_1}2_{\lambda_2}3_{\lambda_3}4_{\lambda_4})=&
    \frac{\PPb_{12}^{\lambda_1+\lambda_2+\omega}}{\beta^{\lambda_1}_1\beta^{\lambda_2}_2(-\beta_1-\beta_2)^{\omega}}\frac{1}{(q_1+q_2)^2}\frac{\PP_{34}^{-\lambda_3-\lambda_4+\omega}}{\beta^{-\lambda_3}_3\beta^{-\lambda_4}_4(\beta_1+\beta_2)^{\omega}}\\ 
    =&\frac{(-)^{\omega}}{2}\frac{\PPb_{12}^{\lambda_1+\lambda_2+\omega-1}\PP_{34}^{-\lambda_3-\lambda_4+\omega}}{\PP_{12}(\beta_1+\beta_2)^{2\omega}}\frac{\beta_3^{\lambda_3}\beta_4^{\lambda_4}}{\beta_1^{\lambda_1-1}\beta_2^{\lambda_2-1}}\,.
    \end{split}
\end{align}
The sum over six distinct contributions may seem unusual when compared to the standard $s$-, $t$-, and $u$-channel exchange contributions in the covariant formulation. The reason for more diagrams is that the same diagram can have $-+$ or $+-$ on the internal line, and these are different diagrams in the light-cone gauge. In the case of singlet fields (as for gravity), \eqref{exchange_diagram} coincides with the sum over the $s$-, $t$-, and $u$-channel contributions in the covariant language:
\begin{equation}
    \mA(1_{\lambda_1}2_{\lambda_2}3_{\lambda_3}4_{\lambda_4})=\mA_s(1_{\lambda_1}2_{\lambda_2}3_{\lambda_3}4_{\lambda_4})+\mA_t(1_{\lambda_1}2_{\lambda_2}3_{\lambda_3}4_{\lambda_4})+\mA_u(1_{\lambda_1}2_{\lambda_2}3_{\lambda_3}4_{\lambda_4})\,.
\end{equation}
While for color amplitudes, it coincides with the expression in terms of color-ordered amplitudes as
\begin{equation}\label{color_ordered}
    \mA(12\cdots n)=\sum_{\sigma\in S_n/Z_n}\mathrm{Tr} (T^{a_{\sigma_1}}T^{a_{\sigma_2}}\cdots T^{a_{\sigma_n}})\mtA(\sigma_1\sigma_2\cdots\sigma_n)\,.
\end{equation}
We now proceed to compute the amplitudes in the light-cone formulation for Yang–Mills theory and gravity.

\paragraph{Yang-Mills theory.} 
For the color ordering $[1^+2^+3^-4^-]$, the four-point amplitude obtained from \eqref{total_A} and \eqref{exchange_diagram}, with the coefficients $k_i$ fixed as in Table~\ref{tab(1,1)}, reads
\begin{equation}
    \mA(1^+2^+3^-4^-)=\mA^{\text{e}}(1^+2^+3^-4^-)+\mA^{\text{e}}(2^+3^-4^-1^+)+\mA^{\text{e}}(4^-1^+2^+3^-)+h^{(1,1)}_{[+,+,-,-]}\,,
\end{equation}
where we set $k_1=1$. By substituting the values of the exchanges and the quartic vertex \eqref{h4_1_YM}, we obtain
\begin{align}
    \mtA(1^+2^+3^-4^-)&=\frac{\langle 34\rangle^4}{\langle 12\rangle\langle 23\rangle\langle 34\rangle\langle 41\rangle}\,.
\end{align}
Similarly, for the color ordering $[1^+2^-3^+4^-]$, we have 
\begin{align}
    \begin{split}
    \mA(1^+2^-3^+4^-)&=\mA^{\text{e}}(1^+2^-3^+4^-)+\mA^{\text{e}}(2^-3^+4^-1^+)\\
    &\;\;\;+\mA^{\text{e}}(3^+4^-1^+2^-)+\mA^{\text{e}}(4^-1^+2^-3^+)+h^{(1,1)}_{[+,-,+,-]}\,.
    \end{split}
\end{align}
By substituting the values of the exchanges and the quartic vertex \eqref{h4_2_YM}, we obtain
\begin{align}
    \mtA(1^+2^-3^+4^-)&=\frac{\langle 24\rangle^4}{\langle 12\rangle\langle 23\rangle\langle 34\rangle\langle 41\rangle}\,.
\end{align}
\paragraph{Gravity.} For gravity, the total four-point amplitude corresponds to
\begin{align}
    \begin{split}
    \mA(1^+2^+3^-4^-)&=\mA^{\text{e}}(1^+2^+3^-4^-)+\mA^{\text{e}}(1^+4^-3^-2^+)+\mA^{\text{e}}(3^-2^+1^+4^-)\\
    &\;\;\;+\mA^{\text{e}}(1^+3^-2^+4^-)+\mA^{\text{e}}(2^+4^-1^+3^-)+h^{(2,2)}_{(+2,+2,-2,-2)}\,,
    \end{split}
\end{align}
where again we set $k_1=1$. By substituting the values of the exchanges and the quartic vertex \eqref{GR}, we obtain
\begin{align}
    \mA(1^+2^+3^-4^-)&=-\frac{[12]^4\langle 34\rangle^4}{s t u}\,.
\end{align}

\section{Quartic vertices for higher-derivative theories}\label{section6}

We now turn to the case of higher-derivative interactions. As we will see, the solution space grows with derivatives, highlighting the deep relation between higher-spin fields and the necessity of higher derivatives.

As in the lower-derivative cases, we begin with an integrability-based analysis. Local higher-derivative quartic vertices are considerably more cumbersome, and for this reason, we avoid presenting them explicitly. Tables for $(3,1)$ \ref{tab(3,1)}, $(3,2)$ \ref{tab(3,2)}, and $(3,3)$ \ref{tab(3,3)} quartic vertices satisfying the quartic constraints are given below.

\begin{table}[H]
    \centering
    \begin{tabular}{|c|c|c|c|c|c|c|}
\hline
$(\lambda_1,\lambda_2,\omega,\lambda_3,\lambda_4)$ & $k_2=(2341)$ & $k_3=(3412)$ & $k_4=(4123)$ & $k_5=(1324)$ & $k_6=(2413)$ \\ \hline
$(-1,2,2,-1,2)$ & $k_2$ & $k_1$ & $k_2$ & $0$ & $k_1-k_2$ \\ \hline
$(-1,2,2,0,1)$ & $k_2$ & $k_1$ & $0$ & $0$ & $k_1-k_2$ \\ \hline
$(0,1,2,0,1)$ & $k_2$ & $k_1$ & $k_2$ & $0$ & $k_6$ \\ \hline
$(0,2,1,0,0)$  & $k_2$ & $0$ & $0$ & $0$ & $k_1-k_2$ \\ \hline
$(1,1,1,-1,1)$  & $0$ & $0$ & $k_4$ & $0$ & $k_1-k_4$ \\ \hline
    \end{tabular}
    \caption{$(3,1)$ quartic vertices satisfying the quartic constraints.}
    \label{tab(3,1)}
\end{table}
\begin{table}[H]
    \centering
    \begin{tabular}{|c|c|c|c|c|c|c|}
\hline
$(\lambda_1,\lambda_2,\omega,\lambda_3,\lambda_4)$ & $k_2=(2341)$ & $k_3=(3412)$ & $k_4=(4123)$ & $k_5=(1324)$ & $k_6=(2413)$ \\ \hline
$(-1,2,2,-2,2)$ & $0$ & $0$ & $k_1$ & $0$ & $k_1$ \\ \hline
$(-1,2,2,-1,1)$  & $k_2$ & $0$ & $0$ & $0$ & $k_1-k_2$ \\ \hline
$(-1,2,2,0,0)$  & $k_2$ & $0$ & $0$ & $0$ & $k_1-k_2$ \\ \hline
$(0,1,2,-2,2)$  & $0$ & $0$ & $k_1$ & $0$ & $k_1$ \\ \hline
$(0,1,2,-1,1)$  & $0$ & $0$ & $k_4$ & $0$ & $k_6$ \\ \hline
$(0,1,2,0,0)$  & $k_2$ & $0$ & $0$ & $0$ & $k_6$ \\ \hline
$(1,1,1,-2,1)$  & $0$ & $0$ & $k_1$ & $0$ & $k_1$ \\ \hline
    \end{tabular}
    \caption{$(3,2)$ quartic vertices satisfying the quartic constraints.}
    \label{tab(3,2)}
\end{table}
\begin{table}[H]
    \centering
    \begin{tabular}{|c|c|c|c|c|c|c|}
\hline
$(\lambda_1,\lambda_2,\omega,\lambda_3,\lambda_4)$ & $k_2=(2341)$ & $k_3=(3412)$ & $k_4=(4123)$ & $k_5=(1324)$ & $k_6=(2413)$ \\ \hline
$(-2,2,3,-2,2)$ & $k_2$ & $k_1$ & $k_2$ & $0$ & $k_1-k_2$ \\ \hline
$(-2,2,3,-1,1)$ & $k_2$ & $k_1$ & $0$ & $0$ & $k_1-k_2$ \\ \hline
$(-2,2,3,0,0)$ & $k_2$ & $k_1$ & $0$ & $0$ & $k_1-k_2$ \\ \hline
$(-1,1,3,-1,1)$ & $k_2$ & $k_1$ & $k_2$ & $0$ & $k_6$ \\ \hline
$(-1,1,3,0,0)$ & $k_2$ & $k_1$ & $0$ & $0$ & $k_6$ \\ \hline
$(-1,2,2,-1,0)$ & $k_2$ & $0$ & $0$ & $0$ & $k_1-k_2$ \\ \hline
$(0,0,3,0,0)$ & $k_2$ & $k_1$ & $k_2$ & $k_5$ & $k_5$ \\ \hline
$(0,1,2,-2,1)$ & $0$ & $0$ & $k_4$ & $0$ & $k_1-k_4$ \\ \hline
    \end{tabular}
    \caption{$(3,3)$ quartic vertices satisfying the quartic constraints.}\label{tab(3,3)}
\end{table}
\noindent
Additional tables are presented in Appendix~\ref{AppendixF}, where we display results up to $(4,4)$. Although our explicit analysis extends to quartic vertices up to $(8,8)$ (with additional checks beyond this range), we do not include those tables here, as they are too lengthy. Nevertheless, we used them as well to identify patterns across the various cases, allowing us to infer the general behaviour of a generic $(n,m)$ quartic vertex, which we summarise below.

One possible way to present the results would be to write tables for a generic $(n,m)$ quartic vertex. Although this could be done in principle, such an approach neither provides intuition about the underlying structure nor offers a convenient formulation for further analysis. Therefore, we proceed differently. First, we describe the pattern that emerges from inspecting all tables up to $(8,8)$. Once this is established, we present another formulation of the result, which will prove useful for addressing various questions.

\begin{itemize}
    \item From the complete list of $(n,m)$ quartic vertices satisfying the quartic constraint \eqref{quartic_system}, we focus on those containing the maximal and minimal external helicities. They correspond to the first line in the various tables we present.\footnote{This is the condition which, for $(1,1)$, gives the consistency of the Yang–Mills cubic vertices, and for $(2,2)$, ensures the consistency of the gravitational cubic vertices.} This depends on the specific type of quartic vertex, whether it is of the form $(even,even)$, $(odd,odd)$, $(even,odd)$, or $(odd,even)$. Let us write them here:
    \begin{subequations}
    \begin{align}
        &(even,even):&
        &(1234)=\left(-\frac{m+2}{2},\frac{n+2}{2},\frac{n+m}{2},-\frac{m+2}{2},\frac{n+2}{2}\right)\,,\\
        &\phantom{(even,even):}&
        &(2413)=\left(\frac{n+2}{2},\frac{n+2}{2},-2,-\frac{m+2}{2},-\frac{m+2}{2}\right)\,,\\
        &(odd,odd):&
        &(1234)=\left(-\frac{m+1}{2},\frac{n+1}{2},\frac{n+m}{2},-\frac{m+1}{2},\frac{n+1}{2}\right)\,,\\
        &\phantom{(odd,odd)}&
        &(2413)=\left(\frac{n+1}{2},\frac{n+1}{2},-1,-\frac{m+1}{2},-\frac{m+1}{2}\right)\,,\\
        &(even,odd):&
        &(1234)=\left(-\frac{m+1}{2},\frac{n+2}{2},\frac{n+m-1}{2},-\frac{m+1}{2},\frac{n}{2}\right)\,,\\
        &\phantom{(even,odd)}&
        &(2413)=\left(\frac{n+2}{2},\frac{n}{2},-1,-\frac{m+1}{2},-\frac{m+1}{2}\right)\,,\\
        &(odd,even):&
        &(1234)=\left(-\frac{m}{2},\frac{n+1}{2},\frac{n+m-1}{2},-\frac{m+2}{2},\frac{n+1}{2}\right)\,,\\
        &\phantom{(odd,even)}&
        &(2413)=\left(\frac{n+1}{2},\frac{n+1}{2},-1,-\frac{m}{2},-\frac{m+2}{2}\right)\,,
    \end{align}
    \end{subequations}
    where we omit the remaining exchanges, as they are straightforward to obtain. We display only $(1234)$ and $(2413)$, which also highlight the special role of spin-$1$ and spin-$2$.
    We now select from them the maximal and minimal external helicities, which we call respectively $M_{\text{ext}}$ and $m_{\text{ext}}$, and the maximal and minimal internal (exchanged) helicities, which we call respectively $M_{\text{int}}$ and $m_{\text{int}}$.
    \begin{align}
        &(even,even)&
        &M_{\text{ext}}=\frac{n+2}{2}\,,&
        &m_{\text{ext}}=-\frac{m+2}{2}\,,&
        &M_{\text{int}}=\frac{n+m}{2}\,,&
        &m_{\text{int}}=-2\,,\\
        &(odd,odd)&
        &M_{\text{ext}}=\frac{n+1}{2}\,,&
        &m_{\text{ext}}=-\frac{m+1}{2}\,,&
        &M_{\text{int}}=\frac{n+m}{2}\,,&
        &m_{\text{int}}=-1\,,\\
        &(even,odd)&
        &M_{\text{ext}}=\frac{n+2}{2}\,,&
        &m_{\text{ext}}=-\frac{m+2}{2}\,,&
        &M_{\text{int}}=\frac{n+m-1}{2}\,,&
        &m_{\text{int}}=-1\,,\\
        &(odd,even)&
        &M_{\text{ext}}=\frac{n+2}{2}\,,&
        &m_{\text{ext}}=-\frac{m+2}{2}\,,&
        &M_{\text{int}}=\frac{n+m-1}{2}\,,&
        &m_{\text{int}}=-1\,.
    \end{align}
    \item We now outline a set of rules that determine the specific elements appearing in the table.
    \begin{enumerate}
        \item All the exchanges $(\lambda_1,\lambda_2,\omega,\lambda_3,\lambda_4)$ involving either external or exchanged helicities that exceed the maximal value or fall below the minimal value identified in the previous point are set to zero.\footnote{This implies that no solution to the quartic constraints with a non-zero exchange contribution exists, in the sense that no local quartic vertex can satisfy the system.} Therefore, exchange diagrams must satisfy
        \begin{align}\label{rule1}
        &m_{\text{ext}}\leq\lambda_1,\lambda_2,\lambda_3,\lambda_4\leq M_{\text{ext}}\,,&
        &m_{\text{int}}\leq \omega\leq M_{\text{int}}
        \end{align}

        \item If at least one of the external fields has either maximal or minimal helicity, the existence of a solution requires turning on all exchange diagrams allowed by the first rule. 

        \item For exchange diagrams allowed by the first rule, where none of the external helicities are extremal, the following applies. If an exchange diagram $(1234)$ is allowed and the corresponding diagram $(3412)$ is also allowed, then both must be included with the same value of $k_i$. Otherwise, $(1234)$ appears alone.

    \end{enumerate}
\end{itemize}
We now make a few interesting comments using the results above:
\begin{itemize}
    \item As the number of derivatives of the cubic vertices involved and the derivatives of the quartic vertices (i.e. $n+m$ and $n+m-2$, respectively) increases, the number of consistent quartic vertices does the same. This is a characteristic feature of higher-spin theories.

    \item The quartic consistency of the cubic non-abelian self-interaction for a field of helicity $\lambda$ of the type $C^{\lambda,\lambda,-\lambda}$ is satisfied only for helicities $\lambda=1,2$. Indeed, for $\lambda$ even, a $(\lambda,\lambda)$ quartic vertex has, as external helicities, at most $\frac{\lambda+2}{2}$, while $(\lambda,-\lambda,\lambda,\lambda-\lambda)$ has $\lambda$ as external helicity. Therefore, consistency requires $\frac{\lambda+2}{2}=\lambda\implies \lambda=2$. The same happens for $\lambda$ odd, replacing $2$ with $1$.
    
    \item For $(n,n)$ quartic vertices, there exists a two parameter family of solutions where the exchange diagrams are of the form $(-a,a,n,-b,b)$, with $a,b\leq\frac{n+2}{2}$ for $n$ even and $a,b\leq\frac{n+1}{2}$ for $n$ odd. However, even though the local quartic vertex exists to solve the quartic constraint, the cubic vertices do not form a consistent local theory. Indeed, if $a=b$ is immediate to see that we would need to satisfy a new constraint because of the exchange $(n,-a,a,a,-n)$. However, the associated quartic constraint does not admit a solution because $n$ exceeds the maximal helicity (with the exceptions of $\lambda=1,2$).
    
    If instead $a\neq b$, one must consider the holomorphic ($CC$) quartic constraint generated by the exchange $(-a,a,n,n,n)$. Following \cite{Serrani:2025owx}, solving this holomorphic constraint requires, among others, the presence of the cubic vertex $(-n,n,n)$, which would imply the spectrum of the full chiral higher-spin theory. Consequently, one could then construct arbitrary ($C\bar{C}$) exchanges using the $\bar{C}$ cubic vertex $(-n,-b,b)$, many of which fail to satisfy the quartic constraint.

    Therefore, for instance, even though Tables \ref{tab(3,1)}--\ref{tab(3,3)} might suggest that color gravity could exist, a full analysis of all newly generated quartic constraints shows that no local solutions are possible for a multi-graviton theory.
\end{itemize}
The last comment was made to draw attention to the meaning of the tables. 
Knowing the solutions to all quartic constraints provides information about the consistency of cubic vertices with Lorentz invariance and also tells us which quartic higher-spin amplitudes turn out to be local. However, this is not yet enough to construct a consistent local higher-spin theory at the quartic order. Given a set of cubic couplings, one must examine all the quartic constraints they generate, and every such constraint must admit at least one local solution.

This situation is reminiscent of the holomorphic constraint discussed in \cite{Serrani:2025owx}. The main difference is that, in that case, by successively introducing additional cubic vertices --- typically, but not always, leading to the full chiral higher-spin spectrum with all holomorphic cubic vertices turned on --- a solution always exists. Here, by contrast, identifying a set of cubic couplings that solves all quartic constraints is not always possible, and determining all admissible local higher-spin theories is considerably more challenging.

Moreover, recall that we must always verify that the cubic vertices that solve the non-holomorphic quartic constraint also solve the holomorphic one. A complete solution to the quartic light-cone constraint is obtained only when the cubic couplings satisfy both constraints simultaneously. We will illustrate how this works through specific examples in the next part. To approach this more intricate problem, we will reformulate the solutions to the quartic constraint in a more natural way.

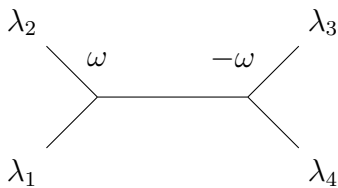
\begin{figure}[H]
    \centering
    \begin{tikzpicture}
        \begin{feynman}
            \vertex (i1) at (-6, 1) {\(\lambda_2\)};
            \vertex (i2) at (-6,-1) {\(\lambda_1\)};
            \vertex (i3) at (-2, 1) {\(\lambda_3\)};
            \vertex (i4) at (-2,-1) {\(\lambda_4\)};
            \vertex (v1) at (-5, 0);
            \vertex (v3) at (-3, 0);
            \vertex at (-5, 0.5) {\(\omega\)};
            \vertex at (-3.2, 0.5) {\(-\omega\)};
            \diagram* {
                (i1) -- (v1),
                (i2) -- (v1),
                (v1) -- [plain] (v3),
                (v3) -- (i3),
                (v3) -- (i4),
            };
        \end{feynman}
    \end{tikzpicture}
        \caption{Exchange diagrams for a generic pair $C\bar{C}$.}
    \label{figure_generic}
\end{figure}
\noindent
\paragraph{Reformulation.} A reformulation of the solutions is as follows. Starting from an $(n,m)$ exchange diagram $(\lambda_1,\lambda_2,\omega,\lambda_3,\lambda_4)$ as in Figure \ref{figure_generic}, where $n=\lambda_{12}+\omega$ and $m=-\lambda_{34}+\omega$. 
We start from the condition given in \eqref{rule1}. In particular, we focus on the stronger requirement obtained by excluding both maximal and minimal external helicities. Under this assumption, we find
\begin{subequations}
\begin{align}\label{raw_1}
    &-\frac{m}{2}\leq\lambda_1\leq \frac{n}{2}\,,&
    &-\frac{m}{2}\leq\lambda_2\leq \frac{n}{2}\,,&
    &-\frac{m}{2}\leq\lambda_3\leq \frac{n}{2}\,,&
    &-\frac{m}{2}\leq\lambda_4\leq \frac{n}{2}\,,\\\label{raw2}
    &0\leq\omega\leq \frac{n+m}{2}\,,&
    & n,m>0\,,
\end{align}
\end{subequations}
where $\omega>0$ is a consequence of \eqref{raw_1}. These conditions are equivalent to the following minimal set of conditions 
\begin{subequations}\label{kinda_trinagular}
\begin{align}\label{trinagular}
    &\lambda_1\leq \lambda_2+\omega\,,&
    &\lambda_2\leq \lambda_1+\omega\,,&
    &\lambda_3\leq \lambda_4+\omega\,,&
    &\lambda_4\leq \lambda_3+\omega\,,\\\label{extra_condition}
    &\lambda_{12}+\omega>0\,,&
    &-\lambda_{34}+\omega> 0\,,&
    &\lambda_{12}\geq \lambda_{34}\,.
\end{align}
\end{subequations}
In this case, a solution to the system of quartic constraints \eqref{quartic_system} does exist without the need for other exchange diagrams.\footnote{In the tables, this exchange would correspond to a free coefficient $k_i$.} In the case $\lambda_{12}=\lambda_{34}$, both $(1234)$ and $(3412)$ satisfy the condition. Indeed, equation \eqref{kinda_trinagular} becomes symmetric under the exchange $(\lambda_1,\lambda_2)\leftrightarrow (\lambda_3,\lambda_4)$. In this situation, both exchange diagrams must therefore be included, with the same value of $k_i$. This is an analogue of the third rule discussed above. The conditions \eqref{kinda_trinagular} show a clear difference between how the helicity in the exchange and in the external states is constrained.  
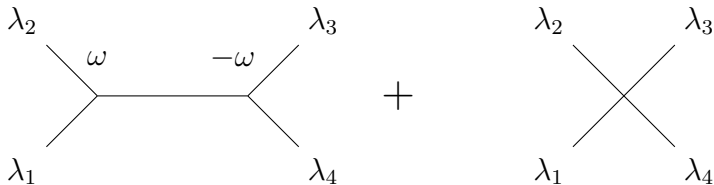
\begin{figure}[H]
    \centering
    \begin{tikzpicture}
        \begin{feynman}
            \vertex (i1) at (-6, 1) {\(\lambda_2\)};
            \vertex (i2) at (-6,-1) {\(\lambda_1\)};
            \vertex (i3) at (-2, 1) {\(\lambda_3\)};
            \vertex (i4) at (-2,-1) {\(\lambda_4\)};

            \vertex (ii1) at (1, 1) {\(\lambda_2\)};
            \vertex (ii2) at (1,-1) {\(\lambda_1\)};
            \vertex (ii3) at (3, 1) {\(\lambda_3\)};
            \vertex (ii4) at (3,-1) {\(\lambda_4\)};

            \vertex (v1) at (-5, 0);
            \vertex (v3) at (-3, 0);

            \node at (-1,0) {\Large $+$};
            
            \vertex at (-5, 0.5) {\(\omega\)};
            \vertex at (-3.2, 0.5) {\(-\omega\)};
            
            \diagram* {
                (i1) -- (v1),
                (i2) -- (v1),
                (v1) -- [plain] (v3),
                (v3) -- (i3),
                (v3) -- (i4),
            };
            \diagram* {
                (ii1) -- (ii4),
                (ii2) -- (ii3),
            };

        \end{feynman}
    \end{tikzpicture}
        \caption{A quartic vertex with a single exchange exists when the conditions \eqref{kinda_trinagular} are satisfied.}
\end{figure}
\noindent
The remaining possibilities, which follow from the first and second rules stated above, arise when some of the external helicities are allowed to be either maximal or minimal, thereby violating the conditions in \eqref{kinda_trinagular}. This corresponds to configurations in which at least one of the external helicities (any of the four) surpasses \eqref{kinda_trinagular}. Then, at least one of these conditions is true
\begin{equation}\label{surpass_trinagular}
    \lambda_1=\lambda_2+\omega+n_1\,,\quad
    \lambda_2=\lambda_1+\omega+n_2\,,\quad
    \lambda_3=\lambda_4+\omega+n_3\,,\quad
    \lambda_4=\lambda_3+\omega+n_4\,,
\end{equation}
where $n_{1,2,3,4}=1,2$. In this case, the quartic vertex exists if and only if we turn on all other exchanges that satisfy \eqref{kinda_trinagular}, with the possibility for some of the helicities to satisfy \eqref{surpass_trinagular}.

\begin{figure}[H]
    \centering
    \begin{tikzpicture}
        \begin{feynman}
            \vertex (i1) at (-6, 1) {\(5\)};
            \vertex (i2) at (-6,-1) {\(2\)};
            \vertex (i3) at (-2, 1) {\(-2\)};
            \vertex (i4) at (-2,-1) {\(-5\)};
            \vertex (v1) at (-5, 0);
            \vertex (v3) at (-3, 0);
            \vertex at (-5, 0.5) {\(1\)};
            \vertex at (-3.2, 0.5) {\(-1\)};
            \diagram* {
                (i1) -- (v1),
                (i2) -- (v1),
                (v1) -- [plain] (v3),
                (v3) -- (i3),
                (v3) -- (i4),
            };
        \end{feynman}
    \end{tikzpicture}
        \caption{Exchange diagrams between $C^{2,5,1}$ and $\bar{C}^{-2,-5,-1}$.}
        \label{figExample}
\end{figure}
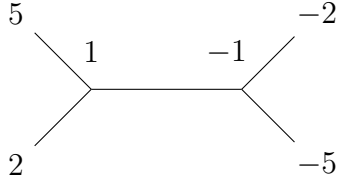

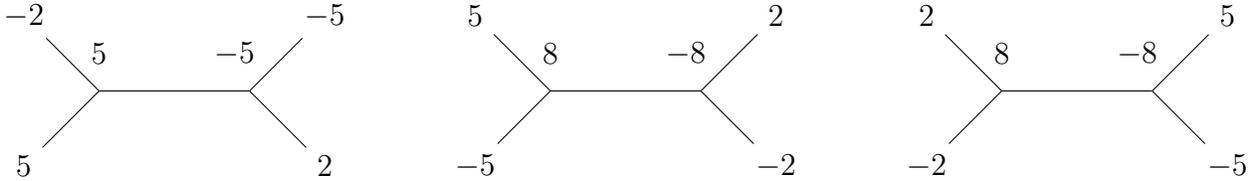
\begin{figure}[H]
    \centering
    \begin{tikzpicture}
        \begin{feynman}
            \vertex (i1) at (-6, 1) {\(-2\)};
            \vertex (i2) at (-6,-1) {\(5\)};
            \vertex (i3) at (-2, 1) {\(-5\)};
            \vertex (i4) at (-2,-1) {\(2\)};

            \vertex (ii1) at (0, 1) {\(5\)};
            \vertex (ii2) at (0,-1) {\(-5\)};
            \vertex (ii3) at (4, 1) {\(2\)};
            \vertex (ii4) at (4,-1) {\(-2\)};

            \vertex (iii1) at (6, 1) {\(2\)};
            \vertex (iii2) at (6,-1) {\(-2\)};
            \vertex (iii3) at (10, 1) {\(5\)};
            \vertex (iii4) at (10,-1) {\(-5\)};
            
            \vertex (v1) at (-5, 0);
            \vertex (v3) at (-3, 0);

            \vertex (vv1) at (1, 0);
            \vertex (vv3) at (3, 0);
            
            \vertex (vvv1) at (7, 0);
            \vertex (vvv3) at (9, 0);
            
            \vertex at (-5, 0.5) {\(5\)};
            \vertex at (-3.2, 0.5) {\(-5\)};

            \vertex at (1, 0.5) {\(8\)};
            \vertex at (2.8, 0.5) {\(-8\)};

            \vertex at (7, 0.5) {\(8\)};
            \vertex at (8.8, 0.5) {\(-8\)};

            \diagram* {
                (i1) -- (v1),
                (i2) -- (v1),
                (v1) -- [plain] (v3),
                (v3) -- (i3),
                (v3) -- (i4),
            };
            \diagram* {
                (ii1) -- (vv1),
                (ii2) -- (vv1),
                (vv1) -- [plain] (vv3),
                (vv3) -- (ii3),
                (vv3) -- (ii4),
            };
            \diagram* {
                (iii1) -- (vvv1),
                (iii2) -- (vvv1),
                (vvv1) -- [plain] (vvv3),
                (vvv3) -- (iii3),
                (vvv3) -- (iii4),
            };
        \end{feynman}
    \end{tikzpicture}
        \caption{Relevant exchange diagrams for parity invariant pairs of cubic abelian vertices.}
        \label{three_diagrams}
\end{figure}

Let us provide an explicit example for the $(8,8)$ quartic vertex in Figure \ref{figExample}.
\noindent
This exchange diagram respects the conditions \eqref{kinda_trinagular}, with some of the fields having maximal helicity \eqref{surpass_trinagular}. In particular, we have 
\begin{subequations}
\begin{align}
    &\lambda_1\leq\lambda_2+\omega\,,\quad
    \lambda_2=\lambda_1+\omega+2\,,\quad
    \lambda_3=\lambda_4+\omega+2\,,\quad
    \lambda_4\leq \lambda_3+\omega\,,\\
    &\lambda_{12}+\omega>0\,,\quad
    -\lambda_{34}+\omega>0\,,\quad
    \lambda_{12}=\lambda_{34}\,.
\end{align}
\end{subequations}
We then need to consider all other possible exchange diagrams, obtained by permuting the various external helicities, that still satisfy the conditions \eqref{kinda_trinagular} with some maximal helicity \eqref{surpass_trinagular}. These diagrams are represented in Figure \ref{three_diagrams}.
\noindent
If we activate all the cubic vertices in these diagrams, we can solve the system of quartic constraints \eqref{quartic_system}. Three possible structures arise:
\begin{enumerate}
    \item If the exchange diagram satisfies \eqref{kinda_trinagular}, it is sufficient to consider only the $s$-channel diagram $(1234)$, and, if allowed, $(3412)$, with the same $k_i$. As we will see, the amplitudes obtained by summing the exchange contributions and the quartic vertex exhibit a $\frac{1}{s}$ pole. We refer to these as single-channel amplitudes.

    \item If the exchange diagram has at least one external helicity that surpasses \eqref{kinda_trinagular} with at most $n_i=1$, the quartic vertices exhibit a structure similar to that of Yang–Mills theory. There are three possible quartic vertices associated with the three channel pairs $st$, $us$, and $tu$. Only two of them are independent, corresponding to the two cyclic orderings $[1234]$ and $[1324]$. In Yang–Mills theory, these are the standard color-ordered amplitudes with helicity configurations $[++--]$ and $[+-+-]$. The amplitudes obtained by summing the exchange contributions and the quartic vertex then exhibit poles of the form $\frac{1}{st}$, $\frac{1}{us}$, and $\frac{1}{tu}$, respectively. We refer to these as Yang–Mills-like (YM-like) amplitudes.

    \item If the exchange diagram has at least one external helicity that surpasses \eqref{kinda_trinagular} with $n_i=2$, the quartic vertices exhibit a structure similar to that of gravity. In this case, there is a single quartic vertex that requires the presence of the $s$-, $t$-, and $u$-channel exchange diagrams. The amplitudes obtained by summing the exchange contributions and the quartic vertex then display a $\frac{1}{stu}$ pole. We refer to these as gravity-like (GR-like) amplitudes.
\end{enumerate}

\paragraph{Homogeneous solutions for the quartic vertices.} One aspect that has not yet been discussed is the study of the homogeneous solutions of the quartic constraint, i.e. solutions to \eqref{quartic_system} without any exchange contributions. Once again, we can employ the integrability method. We show tables up to $(4,4)$ in Appendix \ref{AppendixF}. Performing this analysis, we find that a homogeneous solution exists if the following conditions are satisfied:
\begin{equation}\label{Homo_constraints}
    \lambda_1\leq \lambda_2+\omega-1\,,\quad
    \lambda_2\leq \lambda_1+\omega-1\,,\quad
    \lambda_3\leq \lambda_4+\omega-1\,,\quad
    \lambda_4\leq \lambda_3+\omega-1\,.
\end{equation}
Here, $\omega$ no longer represents an exchange field; instead, it parametrizes the number of derivatives carried by the quartic vertex, denoted by $D$. Indeed $D=n+m-2=\lambda_{12}-\lambda_{34}+2\omega-2$. Equivalently, one may eliminate $\omega$ in favour of $D$, and we get
\begin{equation}
    D-3\lambda_1+\lambda_{234}\geq 0\,,\quad
    D-3\lambda_2+\lambda_{134}\geq 0\,,\quad
    D+3\lambda_3-\lambda_{124}\geq 0\,,\quad
    D+3\lambda_4-\lambda_{123}\geq 0\,.
\end{equation}
This pattern will become clearer when we study four-point amplitudes in detail in Section \ref{section7}.

\subsection{Initial results}
We start by collecting a series of initial results obtained using the reformulation of the full set of local solutions to the quartic constraint.

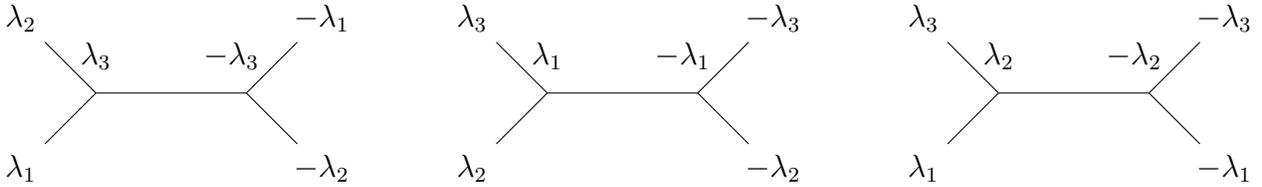
\begin{figure}[H]
    \centering
    \begin{tikzpicture}
        \begin{feynman}
            \vertex (i1) at (-6, 1) {\(\lambda_2\)};
            \vertex (i2) at (-6,-1) {\(\lambda_1\)};
            \vertex (i3) at (-2, 1) {\(-\lambda_1\)};
            \vertex (i4) at (-2,-1) {\(-\lambda_2\)};

            \vertex (ii1) at (0, 1) {\(\lambda_3\)};
            \vertex (ii2) at (0,-1) {\(\lambda_2\)};
            \vertex (ii3) at (4, 1) {\(-\lambda_3\)};
            \vertex (ii4) at (4,-1) {\(-\lambda_2\)};

            \vertex (iii1) at (6, 1) {\(\lambda_3\)};
            \vertex (iii2) at (6,-1) {\(\lambda_1\)};
            \vertex (iii3) at (10, 1) {\(-\lambda_3\)};
            \vertex (iii4) at (10,-1) {\(-\lambda_1\)};
            
            \vertex (v1) at (-5, 0);
            \vertex (v3) at (-3, 0);

            \vertex (vv1) at (1, 0);
            \vertex (vv3) at (3, 0);
            
            \vertex (vvv1) at (7, 0);
            \vertex (vvv3) at (9, 0);
            
            \vertex at (-5, 0.5) {\(\lambda_3\)};
            \vertex at (-3.2, 0.5) {\(-\lambda_3\)};

            \vertex at (1, 0.5) {\(\lambda_1\)};
            \vertex at (2.8, 0.5) {\(-\lambda_1\)};

            \vertex at (7, 0.5) {\(\lambda_2\)};
            \vertex at (8.8, 0.5) {\(-\lambda_2\)};

            \diagram* {
                (i1) -- (v1),
                (i2) -- (v1),
                (v1) -- [plain] (v3),
                (v3) -- (i3),
                (v3) -- (i4),
            };
            \diagram* {
                (ii1) -- (vv1),
                (ii2) -- (vv1),
                (vv1) -- [plain] (vv3),
                (vv3) -- (ii3),
                (vv3) -- (ii4),
            };
            \diagram* {
                (iii1) -- (vvv1),
                (iii2) -- (vvv1),
                (vvv1) -- [plain] (vvv3),
                (vvv3) -- (iii3),
                (vvv3) -- (iii4),
            };
        \end{feynman}
    \end{tikzpicture}
        \caption{$C\bar{C}$ exchange diagrams for parity invariant pairs of cubic abelian vertices.}
    \label{figure_abelian}

    \end{figure}

\paragraph{Abelian cubic vertices.}
Here, we study the consistency of abelian cubic vertices. As already discussed in \cite{Serrani:2025owx}, where the holomorphic quartic constraint was studied in detail, self-dual abelian cubic vertices are consistent on their own because they never generate any $CC$ exchange diagrams. In contrast, in the presence of a scalar field, $CC$ exchanges can occur, making the holomorphic constraint nontrivial. The only exception is the special case of the pair $C^{\lambda_1,\lambda_1,0}C^{0,\lambda_2,\lambda_2}$, which is unconstrained: in this case, the holomorphic constraint vanishes identically. However, once we consider the presence of both holomorphic and anti-holomorphic abelian vertices, they start to form $C\bar{C}$ diagrams, and we must then consider the non-holomorphic constraint. Let us first examine the case of a generic parity-related pair of abelian vertices, $C^{\lambda_1,\lambda_2,\lambda_3}$ and $\bar{C}^{-\lambda_1,-\lambda_2,-\lambda_3}$, with $\lambda_{1,2,3}>0$. 
The relevant exchange diagrams for the system of quartic constraints \eqref{quartic_system} are presented in Figure \ref{figure_abelian}.
\noindent
Following \eqref{kinda_trinagular}, we find the conditions
\begin{align}
&\lambda_1 \leq \lambda_2+\lambda_3\,,&
&\lambda_2 \leq \lambda_1+\lambda_3\,,&
&\lambda_3 \leq \lambda_1+\lambda_2\,.
\end{align}
These are the usual triangular inequalities. Therefore, each triangle with integer unit length defines an allowed pair of abelian cubic vertices. Notice the interesting analogy with the holomorphic constraint: here as well, the only allowed scalar cubic couplings are $C^{\lambda,\lambda,0}$.

\begin{figure}[H]
    \centering
    \begin{tikzpicture}
        \begin{feynman}
            \vertex (i1) at (-6, 1) {\(s\)};
            \vertex (i2) at (-6,-1) {\(s\)};
            \vertex (i3) at (-2, 1) {\(-s\)};
            \vertex (i4) at (-2,-1) {\(-s\)};

            \vertex (ii1) at (1, 1) {\(s\)};
            \vertex (ii2) at (1,-1) {\(s\)};
            \vertex (ii3) at (3, 1) {\(-s\)};
            \vertex (ii4) at (3,-1) {\(-s\)};

            \vertex (v1) at (-5, 0);
            \vertex (v3) at (-3, 0);

            \node at (-0.5,0) {\Large $+$};
            
            \vertex at (-5, 0.5) {\(s\)};
            \vertex at (-3.2, 0.5) {\(-s\)};
            
            \diagram* {
                (i1) -- (v1),
                (i2) -- (v1),
                (v1) -- [plain] (v3),
                (v3) -- (i3),
                (v3) -- (i4),
            };
            \diagram* {
                (ii1) -- (ii4),
                (ii2) -- (ii3),
            };

        \end{feynman}
    \end{tikzpicture}
        \caption{Linearised curvature terms $(R^s)^3$ for higher-spins are consistent to the quartic level.}
    \label{figure_Linearisedcurvature}
\end{figure}
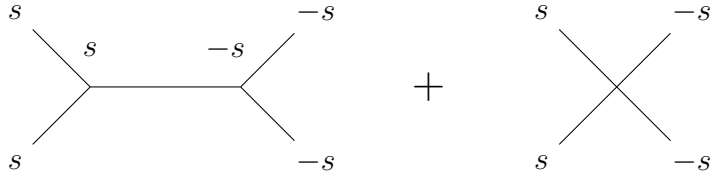

We have just described allowed unitary higher-spin theories with abelian vertices. Particular examples of the one described above are the generalisations of linearised curvatures such as $F^3$ and $R^3$, but for higher-spin fields, see Figure \ref{figure_Linearisedcurvature}.
\noindent

The same triangular inequalities were already found in \cite{Damour:1987vm,Damour:1987fp}. The underlying idea is to construct linearised higher-spin curvatures $R^s_{\mu_1\nu_1,...,\mu_s\nu_s}$ (de Wit–Freedman curvatures) and use them to build invariant tensors, such as $R^{s_1}R^{s_2}R^{s_3}$. By counting the number of free indices, one finds that such constructions are possible only if the spins satisfy the triangular inequalities. Let us note that imposing the light-cone gauge in a theory with only abelian cubic interactions can lead to higher order vertices in $H$, as in the process, one has to solve non-linear equations for auxiliary fields. 

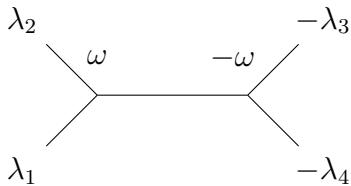
\begin{figure}[H]
    \centering
    \begin{tikzpicture}
        \begin{feynman}
            \vertex (i1) at (-6, 1) {\(\lambda_2\)};
            \vertex (i2) at (-6,-1) {\(\lambda_1\)};
            \vertex (i3) at (-2, 1) {\(-\lambda_3\)};
            \vertex (i4) at (-2,-1) {\(-\lambda_4\)};
            \vertex (v1) at (-5, 0);
            \vertex (v3) at (-3, 0);
            \vertex at (-5, 0.5) {\(\omega\)};
            \vertex at (-3.2, 0.5) {\(-\omega\)};
            \diagram* {
                (i1) -- (v1),
                (i2) -- (v1),
                (v1) -- [plain] (v3),
                (v3) -- (i3),
                (v3) -- (i4),
            };
        \end{feynman}
    \end{tikzpicture}
        \caption{Exchange diagram for a pair of cubic abelian vertices with one opposite helicity.}
    \label{figure_abelian2}
\end{figure}

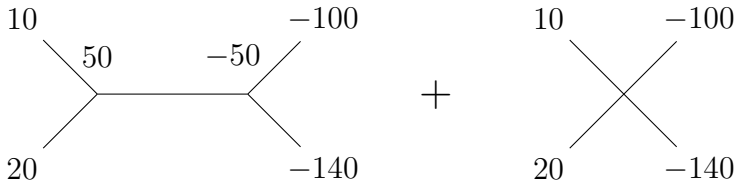
\begin{figure}[H]
    \centering
    \begin{tikzpicture}
        \begin{feynman}
            \vertex (i1) at (-6, 1) {\(10\)};
            \vertex (i2) at (-6,-1) {\(20\)};
            \vertex (i3) at (-2, 1) {\(-100\)};
            \vertex (i4) at (-2,-1) {\(-140\)};

            \vertex (ii1) at (1, 1) {\(10\)};
            \vertex (ii2) at (1,-1) {\(20\)};
            \vertex (ii3) at (3, 1) {\(-100\)};
            \vertex (ii4) at (3,-1) {\(-140\)};

            \vertex (v1) at (-5, 0);
            \vertex (v3) at (-3, 0);

            \node at (-0.5,0) {\Large $+$};
            
            \vertex at (-5, 0.5) {\(50\)};
            \vertex at (-3.2, 0.5) {\(-50\)};
            
            \diagram* {
                (i1) -- (v1),
                (i2) -- (v1),
                (v1) -- [plain] (v3),
                (v3) -- (i3),
                (v3) -- (i4),
            };
            \diagram* {
                (ii1) -- (ii4),
                (ii2) -- (ii3),
            };

        \end{feynman}
    \end{tikzpicture}
        \caption{Example of abelian cubic vertices that satisfy the quartic constraint.}
    \label{figure_twoabelian}
\end{figure}

Let us now consider the case of two abelian couplings that do not form a parity-related pair. Let the vertices be $C^{\lambda_1,\lambda_2,\omega}$ and $\bar{C}^{-\omega,-\lambda_3,-\lambda_4}$, with $\lambda_{1,2,3,4}>0$ and $\omega>0$. In this situation, the relevant diagram is the one in Figure \ref{figure_abelian2}.
\noindent
The conditions that must be satisfied are those in \eqref{kinda_trinagular}, and they are less restrictive than the triangular inequalities mentioned before. An example is the one in Figure \ref{figure_twoabelian}.
\noindent
These cubic vertices, which surpass the triangular inequalities, cannot be expressed using linearised gauge curvatures. 

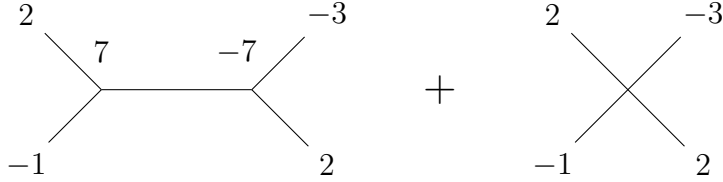
\begin{figure}[H]
    \centering
    \begin{tikzpicture}
        \begin{feynman}
            \vertex (i1) at (-6, 1) {\(2\)};
            \vertex (i2) at (-6,-1) {\(-1\)};
            \vertex (i3) at (-2, 1) {\(-3\)};
            \vertex (i4) at (-2,-1) {\(2\)};

            \vertex (ii1) at (1, 1) {\(2\)};
            \vertex (ii2) at (1,-1) {\(-1\)};
            \vertex (ii3) at (3, 1) {\(-3\)};
            \vertex (ii4) at (3,-1) {\(2\)};

            \vertex (v1) at (-5, 0);
            \vertex (v3) at (-3, 0);

            \node at (-0.5,0) {\Large $+$};
            
            \vertex at (-5, 0.5) {\(7\)};
            \vertex at (-3.2, 0.5) {\(-7\)};
            
            \diagram* {
                (i1) -- (v1),
                (i2) -- (v1),
                (v1) -- [plain] (v3),
                (v3) -- (i3),
                (v3) -- (i4),
            };
            \diagram* {
                (ii1) -- (ii4),
                (ii2) -- (ii3),
            };

        \end{feynman}
    \end{tikzpicture}
        \caption{Example of non-abelian cubic vertices that satisfy the quartic constraint.}
    \label{figure_nonabelian}
\end{figure}

\paragraph{Non-abelian couplings.} Non-abelian couplings are subject to stronger constraints from Lorentz invariance. Nonetheless, it is possible to identify specific cubic vertices that satisfy the quartic consistency conditions. Let us take two generic vertices, $C^{\lambda_1,\lambda_2,\omega}$ and $\bar{C}^{-\omega,\lambda_3,\lambda_4}$. The constraints  \eqref{kinda_trinagular} must still be respected. Such conditions can also be satisfied in the non-Abelian case; an example is in Figure \ref{figure_nonabelian}.

\subsection*{Abelian vs Non-abelian cubic vertices I}
Because we have referred to both abelian and non-abelian cubic vertices, let us briefly clarify this distinction. In Figure \ref{figure_nonabelian}, we have in fact slightly abused the terminology. At first sight, the couplings $C^{-1,2,7}$ and $\bar{C}^{-3,2,-7}$ might appear to be non-abelian, since they are of the (++$-$) and ($--$+) type, rather than the more familiar abelian (+++) and ($---$) vertices. This, however, is not the case.

The criterion distinguishing abelian from non-abelian cubic vertices was explained and found in \cite{Bekaert:2010hp}. We briefly review the relevant results for us here.

First of all, it should be emphasised that the distinction between abelian and non-abelian cubic vertices is naturally formulated in a covariant framework. Indeed, this terminology refers to the way a cubic interaction deforms the gauge algebra. Since in the light-cone formalism the gauge has already been fixed, there is no deformation to discuss.

We start by recalling that not all cubic vertices can be written in the standard (Fronsdal like) covariant formulation. If we fix $s_1\leq s_2\leq s_3$ with $s_{1,2,3}\geq 0$, covariant vertices exist only for the following helicity configurations:
\begin{align}
    &(+s_1,+s_2,+s_3)\oplus(-s_1,-s_2,-s_3) \,,&
    &(-s_1,+s_2,+s_3)\oplus (s_1,-s_2,-s_3)\,.
\end{align}
Note that in the chiral formulation, all cubic vertices can, in principle, be constructed \cite{Krasnov:2021nsq,Skvortsov:2022syz,Sharapov:2022faa}. However, here we restrict our attention to the Fronsdal formulation, since the results discussed below were derived within this framework.

The results of \cite{Bekaert:2010hp} can be summarised as follows. Vertices of the type $(s_1,s_2,s_3)$ and $(-s_1,-s_2,-s_3)$ provide the simplest examples of abelian vertices, since they do not require any deformation of the gauge transformations. There are also vertices of the type $(-s_1,s_2,s_3)$, together with their parity conjugates, satisfying $s_1+s_2\leq s_3$. These do require a deformation of the gauge transformations; however, the commutator of the deformed gauge transformations still vanishes. They are therefore also referred to as abelian, since the gauge algebra remains commutative. Non-abelian vertices are of the type $(-s_1,s_2,s_3)$ with $s_1+s_2> s_3$, together with their parity conjugates. In this case, the gauge transformations must again be deformed, but now the commutator of the deformed gauge transformations is non-vanishing. Equivalently, the gauge algebra itself is deformed into a non-abelian one.

Therefore, the cubic couplings displayed in Figure \ref{figure_nonabelian} should more properly be referred to as abelian vertices according to the definition given above. As can already be inferred from the condition in \eqref{surpass_trinagular}, there also exist local quartic vertices that solve the quartic light-cone constraint while involving non-abelian vertices. We will explore this point in more detail at the end of Section \ref{section7}, where we introduce the notions of single-channel, YM-like, and GR-like amplitudes, which will provide a clearer picture and connections with amplitudes.

\subsection{Unitary local higher-spin theories in flat space}
In this section, we find all possible unitary local higher-spin theories in flat space. Starting from a cubic vertex $C^{\lambda_1,\lambda_2,\lambda_3}$, unitarity forces us to include its parity-related counterpart $\bar{C}^{-\lambda_1,-\lambda_2,-\lambda_3}$. These two vertices form a $C\bar{C}$ pair, which in turn generates quartic constraints that must be solved. 
\begin{figure}[H]
    \centering
    \begin{tikzpicture}
        \begin{feynman}
            \vertex (i1) at (-6, 1) {\(\lambda_2\)};
            \vertex (i2) at (-6,-1) {\(\lambda_1\)};
            \vertex (i3) at (-2, 1) {\(-\lambda_1\)};
            \vertex (i4) at (-2,-1) {\(-\lambda_2\)};

            \vertex (v1) at (-5, 0);
            \vertex (v3) at (-3, 0);

            \vertex at (-5, 0.5) {\(\lambda_3\)};
            \vertex at (-3.2, 0.5) {\(-\lambda_3\)};
            
            \diagram* {
                (i1) -- (v1),
                (i2) -- (v1),
                (v1) -- [plain] (v3),
                (v3) -- (i3),
                (v3) -- (i4),
            };

        \end{feynman}
    \end{tikzpicture}
        \caption{Generic $C\bar{C}$ unitary exchange.}
    \label{figure_unitaryexchange}
\end{figure}
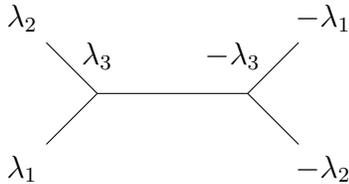
\noindent
Considering that in this case, each of the helicities can end up on the external legs, we obtain the following conditions:
\begin{align}\label{unitary_condition}
&\lambda_1 \leq \lambda_2+\lambda_3\,,&
&\lambda_2 \leq \lambda_1+\lambda_3\,,&
&\lambda_3 \leq \lambda_1+\lambda_2\,.&
&\lambda_1+\lambda_2+\lambda_3>0\,.
\end{align}
These conditions correspond to the triangle inequalities. As discussed above, a theory containing an arbitrary number of pairs of parity-related cubic abelian couplings that satisfy them forms a consistent and unitary theory admitting higher-spin fields.

We now start to study the cases when at least one of the external fields is allowed to be maximal. We will proceed as follows:
\begin{itemize}
    \item Firstly, we start from a generic $C\bar{C}$ unitary exchange diagram, as shown in Figure~\ref{figure_unitaryexchange}, and consider the case where some of the external fields are maximal. There are three possible cases, which we will analyse separately.

    \item Secondly, we examine the constraints imposed by \eqref{kinda_trinagular} and \eqref{surpass_trinagular} and construct additional diagrams that are required to satisfy the quartic constraint.

    \item Thirdly, we notice that these new diagrams generate new cubic vertices. For each of them, we must include the corresponding parity-related pair and ensure that the triangular inequality \eqref{unitary_condition} is satisfied. This will provide additional constraints which we use to restrict the helicities of the external fields in the original diagram.
\end{itemize}

\paragraph{First case.} By imposing that all three external fields have maximal helicities, we find
\begin{align}
\begin{split}
\lambda_1 = \lambda_{23}+n_1\,,\quad
\lambda_2 = \lambda_{13}+n_2\,,&\quad
\lambda_3 = \lambda_{12}+n_3\,,\quad
\lambda_{123}>0\\
&\implies\quad
\lambda_{123}=-(n_1+n_2+n_3)\,.
\end{split}
\end{align}
Recalling that $n_{1,2,3}=1,2$, we see that this system is inconsistent.

\begin{figure}[H]
    \centering
    \begin{tikzpicture}
        \begin{feynman}
            \vertex (i1) at (-10, 1) {\(\lambda_1+\frac{n_2-n_1}{2}\)};
            \vertex (i2) at (-10,-1) {\(\lambda_1\)};
            \vertex (i3) at (-6, 1) {\(-\lambda_1\)};
            \vertex (i4) at (-6,-1) {\(-\lambda_1-\frac{n_2-n_1}{2}\)};

            \vertex (arrow) at (-3,0) {\(\implies\)};

            \vertex (ii1) at (0, 1) {\(\lambda_1\)};
            \vertex (ii2) at (0,-1) {\(-\lambda_1\)};
            \vertex (ii3) at (4, 1) {\(\lambda_1+\frac{n_2-n_1}{2}\)};
            \vertex (ii4) at (4,-1) {\(-\lambda_1-\frac{n_2-n_1}{2}\)};

            \vertex (v1) at (-9, 0);
            \vertex (v3) at (-7, 0);

            \vertex (vv1) at (1, 0);
            \vertex (vv3) at (3, 0);
            
            \vertex at (-9, 0.5);
            \vertex at (-8, 0.5) {\(-\frac{n_1+n_2}{2}\)};
            \vertex at (-7.2, 0.5);

            \vertex at (1, 0.5);
            \vertex at (2, 0.5) {\(2\lambda_1-n_1\)};
            \vertex at (2.8, 0.5);

            \diagram* {
                (i1) -- (v1),
                (i2) -- (v1),
                (v1) -- [plain] (v3),
                (v3) -- (i3),
                (v3) -- (i4),
            };

            \diagram* {
                (ii1) -- (vv1),
                (ii2) -- (vv1),
                (vv1) -- [plain] (vv3),
                (vv3) -- (ii3),
                (vv3) -- (ii4),
            };

        \end{feynman}
    \end{tikzpicture}
    \caption{The left diagram, by quartic consistency, implies the presence of the right one.}
    \label{figure_imply1}
\end{figure}
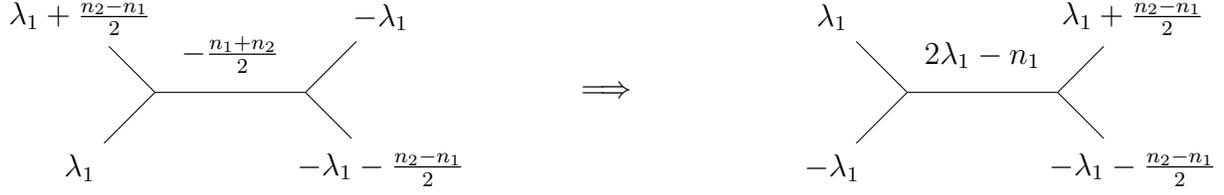

\paragraph{Second case.} By imposing that two external fields have maximal helicities, we find
\begin{align}\label{consistent_1}
\begin{split}
\lambda_1 = \lambda_{23}+n_1\,,&\quad
\lambda_2 = \lambda_{13}+n_2\,\quad
\lambda_3 \leq \lambda_{12}\,,\quad
\lambda_{123}>0\\
&\implies\quad
\lambda_3=-\frac{n_1+n_2}{2}\,,\quad
\lambda_2=\lambda_1+\frac{n_2-n_1}{2}\,,\quad
\lambda_1>\frac{n_1}{2}\,.
\end{split}
\end{align}
The existence of this diagram implies the presence of additional diagrams (see Figure \ref{figure_imply1}).\footnote{When space is limited, only $\omega$ is shown, omitting $-\omega$.} This new diagram, obtained from the original one by a permutation of the external helicities, respects the conditions \eqref{kinda_trinagular} with the possibility of \eqref{surpass_trinagular}. In particular, it satisfies
\begin{align}
\begin{split}
\lambda_1 \leq \lambda_2+\omega\,,&\quad
\lambda_2 = \lambda_1+\omega+n_1\,,\quad
\lambda_3=\lambda_4+\omega+n_2\,,\quad
\lambda_4 \leq \lambda_3+\omega\,,\\
&\lambda_{12}+\omega>0\,,\quad
-\lambda_{34}+\omega>0\,,\quad
\lambda_{12}=\lambda_{34}\,,
\end{split}
\end{align}
where $\lambda_{1,2,3,4}$ indicates the helicities of the diagram on the right in Figure \ref{figure_imply1}, in the same order as in Figure \ref{figure_generic}.
\noindent
As a consequence of unitarity, each newly generated cubic vertex, for instance the vertex $C^{\lambda_1,-\lambda_1,2\lambda_1-n_1}$, must satisfy the condition \eqref{unitary_condition}, up to the possibility of being of maximal helicity. This additional constraint leads to
\begin{equation}
    2\lambda_1-n_1\leq 0\quad
    \implies\quad
    \lambda_1\leq \frac{n_1}{2}\,.
\end{equation}
This is inconsistent with \eqref{consistent_1}. The only remaining possibility is to take this vertex to be maximal, which implies
\begin{equation}
2\lambda_1-n_1=n_3\quad
\implies\quad 
\lambda_1=\frac{n_1+n_3}{2}\,.
\end{equation}
At this point, we can list the cubic vertices that satisfy all the conditions derived. They are given by
\begin{equation}
(\lambda_1,\lambda_2,\lambda_3)=\left(\frac{n_1+n_3}{2},\frac{n_2+n_3}{2},-\frac{n_1+n_2}{2}\right)\,.
\end{equation}
Setting $n_{1,2,3}=1,2$, there are $2^3$ possible combinations. Among all solutions, only two contain exclusively integer helicities. Including the parity-related vertices as well, we obtain
\begin{equation}\label{nonabelian1}
    \Big((1,1,-1) \oplus (-1,-1,1)\Big) \oplus \Big((2,2,-2) \oplus (-2,-2,2)\Big)\,.
\end{equation}
These correspond precisely to the cubic vertices of Yang-Mills theory and gravity. 

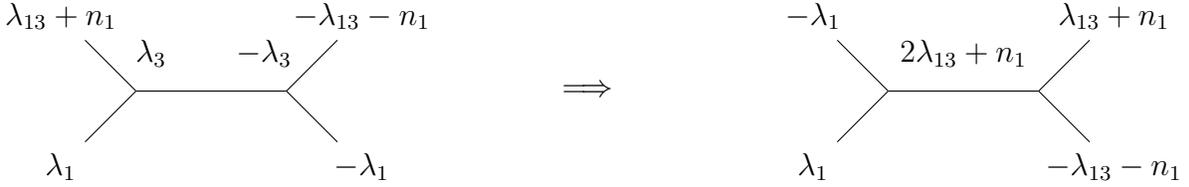
\begin{figure}[H]
    \centering
    \begin{tikzpicture}
        \begin{feynman}
            \vertex (i1) at (-10, 1) {\(\lambda_{13}+n_1\)};
            \vertex (i2) at (-10,-1) {\(\lambda_1\)};
            \vertex (i3) at (-6, 1) {\(-\lambda_{13}-n_1\)};
            \vertex (i4) at (-6,-1) {\(-\lambda_1\)};

            \vertex (arrow) at (-3,0) {\(\implies\)};

            \vertex (ii1) at (0, 1) {\(-\lambda_1\)};
            \vertex (ii2) at (0,-1) {\(\lambda_1\)};
            \vertex (ii3) at (4, 1) {\(\lambda_{13}+n_1\)};
            \vertex (ii4) at (4,-1) {\(-\lambda_{13}-n_1\)};

            \vertex (v1) at (-9, 0);
            \vertex (v3) at (-7, 0);

            \vertex (vv1) at (1, 0);
            \vertex (vv3) at (3, 0);
            
            \vertex at (-9, 0.5);
            \vertex at (-8.8, 0.5) {\(\lambda_3\)};
            \vertex at (-7.3, 0.5) {\(-\lambda_3\)};
            \vertex at (-7.2, 0.5);

            \vertex at (1, 0.5);
            \vertex at (2, 0.5) {\(2\lambda_{13}+n_1\)};
            \vertex at (2.8, 0.5);

            \diagram* {
                (i1) -- (v1),
                (i2) -- (v1),
                (v1) -- [plain] (v3),
                (v3) -- (i3),
                (v3) -- (i4),
            };

            \diagram* {
                (ii1) -- (vv1),
                (ii2) -- (vv1),
                (vv1) -- [plain] (vv3),
                (vv3) -- (ii3),
                (vv3) -- (ii4),
            };

        \end{feynman}
    \end{tikzpicture}
    \caption{The left diagram, by quartic consistency, implies the presence of the right one.}
    \label{figure_imply2}
\end{figure}

\paragraph{Third case.} By imposing that one external field has maximal helicity, we find
\begin{align}\label{unitarycase2}
\begin{split}
\lambda_1 \leq \lambda_{23}\,,\quad
\lambda_2 =& \lambda_{13}+n_1\,,\quad
\lambda_3 \leq \lambda_{12}\,,\quad
\lambda_{123}>0\\
&\implies\quad
\lambda_1\geq-\frac{n_1}{2}\,,\quad
\lambda_3\geq-\frac{n_1}{2}\,,\quad
\lambda_{13}>-\frac{n_1}{2}\,.
\end{split}
\end{align}
This diagram implies the presence of an additional diagram (see Figure \ref{figure_imply2}). This respects the condition \eqref{kinda_trinagular} with the possibility of \eqref{surpass_trinagular}. In particular, it satisfies
\begin{align}
\begin{split}
\lambda_1 \leq \lambda_2+\omega\,,&\quad
\lambda_2 \leq \lambda_1+\omega\,,\quad
\lambda_3=\lambda_4+\omega+n_1\,,\quad
\lambda_4\leq\lambda_3+\omega\,,\\
&\lambda_{12}+\omega>0\,,\quad
-\lambda_{34}+\omega>0\,,\quad
\lambda_{12}=\lambda_{34}\,,
\end{split}
\end{align}
where again by $\lambda_{1,2,3,4}$ we indicate the helicities of the diagram on the right in Figure \ref{figure_imply2}, in the same order as in Figure \ref{figure_generic}.
\noindent
As a consequence of unitarity, each newly generated cubic vertex, for instance, the vertex $C^{\lambda_1,-\lambda_1,2\lambda_{13}+n_1}$, must satisfy the condition \eqref{unitary_condition}, up to the possibility of being of maximal helicity. This additional constraint leads to
\begin{equation}
    2\lambda_{13}+n_1\leq 0\quad
    \implies\quad
    \lambda_{13}\leq -\frac{n_1}{2}\,.
\end{equation}
This is inconsistent with \eqref{unitarycase2}. The only remaining possibility is to take this vertex to be maximal, which implies
\begin{equation}\label{maximal2}
2\lambda_{13}+n_1=n_2\quad
\implies\quad 
\lambda_{13}=\frac{n_2-n_1}{2}\,.
\end{equation}
The constraint \eqref{maximal2}, together with \eqref{unitarycase2}, gives
\begin{equation}
    \lambda_{13}=\frac{n_2-n_1}{2}\,,\qquad
    \lambda_3\geq -\frac{n_1}{2}\,,\quad
    \lambda_1\geq -\frac{n_1}{2}\quad
    \implies\quad
    \lambda_1\leq \frac{n_2}{2}\,,\quad
    \lambda_3\leq \frac{n_2}{2}\,.
\end{equation}
At this point, we can list the cubic vertices that satisfy all the conditions derived. They are given by
\begin{equation}
(\lambda_1,\lambda_{13}+n_1,\lambda_3)\,,\quad
\lambda_3=\frac{n_2-n_1}{2}-\lambda_1\,,\quad
-\frac{n_1}{2}\leq \lambda_1\leq \frac{n_2}{2}\,.
\end{equation}
This gives 
\begin{equation}
\left(\lambda_1,\frac{n_2+n_1}{2},\frac{n_2-n_1}{2}-\lambda_1\right)\,,\quad
-\frac{n_1}{2}\leq \lambda_1\leq \frac{n_2}{2}\,.
\end{equation}
Setting $n_{1,2}=1,2$, there are $3$ possible combinations. Among all solutions, only two contain exclusively integer helicities. Including also the parity-related vertices, we obtain
\begin{equation}\label{nonabelian2}
    \Big((0,1,0) \oplus (0,-1,0)\Big)\oplus\Big((0,2,0) \oplus (0,-2,0)\Big)\oplus\Big((1,2,-1) \oplus (-1,-2,1)\Big)\,.
\end{equation}
These correspond to the non-abelian cubic interactions between fields of helicities $0,1,2$; scalars, Yang-Mills fields, and gravitons.

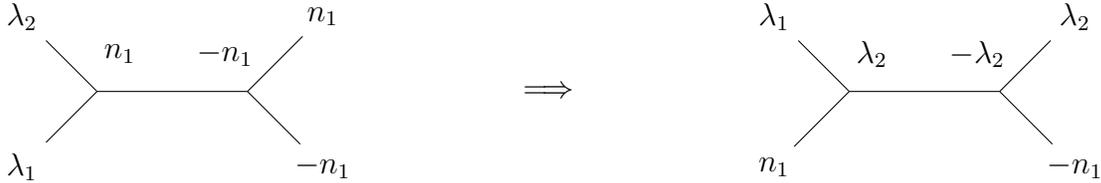
\begin{figure}[H]
    \centering
    \begin{tikzpicture}
        \begin{feynman}
            \vertex (i1) at (-10, 1) {\(\lambda_2\)};
            \vertex (i2) at (-10,-1) {\(\lambda_1\)};
            \vertex (i3) at (-6, 1) {\(n_1\)};
            \vertex (i4) at (-6,-1) {\(-n_1\)};

            \vertex (arrow) at (-3,0) {\(\implies\)};

            \vertex (ii1) at (0, 1) {\(\lambda_1\)};
            \vertex (ii2) at (0,-1) {\(n_1\)};
            \vertex (ii3) at (4, 1) {\(\lambda_2\)};
            \vertex (ii4) at (4,-1) {\(-n_1\)};

            \vertex (v1) at (-9, 0);
            \vertex (v3) at (-7, 0);

            \vertex (vv1) at (1, 0);
            \vertex (vv3) at (3, 0);
            
            \vertex at (-9, 0.5);
            \vertex at (-8.7, 0.5) {\(n_1\)};
            \vertex at (-7.3, 0.5) {\(-n_1\)};
            \vertex at (-7.2, 0.5);

            \vertex at (1, 0.5);
            \vertex at (1.3, 0.5) {\(\lambda_2\)};
            \vertex at (2.7, 0.5) {\(-\lambda_2\)};
            \vertex at (2.8, 0.5);

            \diagram* {
                (i1) -- (v1),
                (i2) -- (v1),
                (v1) -- [plain] (v3),
                (v3) -- (i3),
                (v3) -- (i4),
            };

            \diagram* {
                (ii1) -- (vv1),
                (ii2) -- (vv1),
                (vv1) -- [plain] (vv3),
                (vv3) -- (ii3),
                (vv3) -- (ii4),
            };

        \end{feynman}
    \end{tikzpicture}
    \caption{The left diagram, by quartic consistency, implies the presence of the right one.}
    \label{figure_imply3}
\end{figure}

By applying a procedure analogous to that discussed above, one can show that imposing the presence of any of the non-abelian vertex pairs in \eqref{nonabelian1} and \eqref{nonabelian2} requires all higher-derivative abelian vertices to be set to zero, leaving only the lower-spin abelian interactions. We begin by observing that, due to the holomorphic constraint, as discussed in \cite{Serrani:2025owx}, the presence of the self-dual couplings $C^{0,2,0}$ and $C^{1,2,-1}$ necessarily requires the inclusion of the self-dual gravitational coupling $C^{2,2,-2}$. Similarly, the self-dual coupling $C^{0,1,0}$ requires the presence of the self-dual Yang-Mills coupling $C^{1,1,-1}$. Therefore, it is sufficient to show that the presence of either of the coupling pairs $C^{1,1,-1}\oplus \bar{C}^{-1,-1,1}$ or $C^{2,2,-2}\oplus \bar{C}^{-2,-2,2}$ implies that no additional higher-spin couplings can be consistently introduced. To this end, we start from the most general cubic vertex that forms an exchange, the one on the left of Figure \ref{figure_imply3}, with the $\bar{C}^{n_1,-n_1,-n_1}$ cubic vertex, and require $\lambda_1$ and $\lambda_2$ to be higher-spin fields; then $\lambda_1,\lambda_2>2$. 
\noindent
This diagram implies the presence of an additional diagram (see Figure \ref{figure_imply3}). This respects the condition \eqref{kinda_trinagular} with the possibility of \eqref{surpass_trinagular}, as can be easily checked.

\begin{figure}[H]
    \centering
    \begin{tikzpicture}
        \begin{feynman}
            \vertex (i1) at (-6, 1) {\(\lambda_2\)};
            \vertex (i2) at (-6,-1) {\(\lambda_1\)};
            \vertex (i3) at (-2, 1) {\(\lambda_2\)};
            \vertex (i4) at (-2,-1) {\(-\lambda_2\)};

            \vertex (v1) at (-5, 0);
            \vertex (v3) at (-3, 0);

            \vertex at (-5, 0.5) {\(n_1\)};
            \vertex at (-3.2, 0.5) {\(-n_1\)};
            
            \diagram* {
                (i1) -- (v1),
                (i2) -- (v1),
                (v1) -- [plain] (v3),
                (v3) -- (i3),
                (v3) -- (i4),
            };

        \end{feynman}
    \end{tikzpicture}
        \caption{New exchange diagram between $C^{\lambda_1,\lambda_2,n_1}$ and $\bar{C}^{\lambda_2,-\lambda_2,-n_1}$.}
    \label{figureNoAbelianHS}
\end{figure}
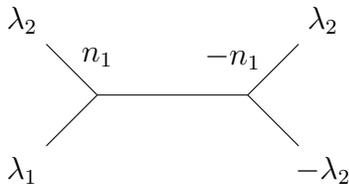

This new diagram generates two new cubic vertices: $C^{\lambda_1,\lambda_2,n_1}$ and $\bar{C}^{\lambda_2,-\lambda_2,-n_1}$. These can form another exchange (see Figure \ref{figureNoAbelianHS}) that must satisfy its own non-holomorphic constraint. The constraints in \eqref{kinda_trinagular} imply
\begin{equation}
    \lambda_2\leq n_1-\lambda_2\,,\quad
    \implies
    \lambda_2\leq\frac{n_2}{2}\,.
\end{equation}
Repeating the same steps with the roles of $\lambda_1$ and $\lambda_2$ reversed gives the constraint
\begin{equation}
    \lambda_1\leq n_1-\lambda_1\,,\quad
    \implies
    \lambda_1\leq\frac{n_1}{2}\,.
\end{equation}
Finally, we have
\begin{align}
    &\lambda_1\leq\frac{n_1}{2}\,,&
    &\lambda_2\leq\frac{n_1}{2}\,,&
    &\lambda_1+\lambda_2+n_1>0\,.
\end{align}
Therefore, no higher-spin fields are allowed.
\noindent
In the case that the external fields are maximal \eqref{surpass_trinagular}, we find
\begin{align}
    &\lambda_1= \frac{n_1+n_2}{2}\,,&
    &\lambda_2= \frac{n_1+n_3}{2}\,.
\end{align}
In any case, no higher-spin fields are allowed.

Let us note that we have discarded possible solutions involving half-integer helicities. This is because we are considering only bosonic fields here. 
However, it seems that allowing for half-integer helicities, we would obtain additional solutions for cubic vertices involving both $\lambda=\pm\frac{1}{2}$ and $\lambda=\pm\frac{3}{2}$. This suggests that, even when massless fermions are included, an analogous structure should emerge. Moreover, it indicates that one could potentially recover --- through a similar consistency analysis at the quartic order --- the existence of interacting and consistent supergravity theories, such as $\mathcal{N}=1$ supergravity in $4d$. 
Investigating solutions to the quartic constraints in the presence of massless fermions is a natural next step, which we plan to pursue in the near future.

In conclusion, we have demonstrated that no unitary and local higher-spin theory in flat space can consistently coexist with the unitary cubic self-interactions of gravity and those of Yang–Mills theory \eqref{nonabelian1} or with the non-abelian couplings \eqref{nonabelian2}. The sole assumption of our analysis is that higher-spin interactions begin at the cubic order.

A potential loophole, however, is the following. One may retain the standard consistent cubic vertices for the lower-spin fields while allowing the higher-spin sector to start directly at quartic order and then attempt to close the algebra from that point onwards. In the light-cone formalism, this strategy would require a complete analysis of the quintic constraint. We leave this investigation for future work.
\subsection{On the possibility of ``quasi-chiral'' higher-spin theories}
As we have shown above, although abelian vertices are consistent among themselves, they become inconsistent once we require the presence of either of the pairs of cubic vertices in \eqref{nonabelian1} or \eqref{nonabelian2}. The interpretation is as follows: higher-spin abelian couplings can be consistently defined on a fixed background geometry, but they cannot interact with the geometry itself, which is represented by the cubic vertices of gravity.

A slightly weaker question is whether it is possible to couple higher-spin vertices to the self-dual sector of gravity or Yang–Mills theory in a way that includes both holomorphic and anti-holomorphic vertices, i.e. beyond the chiral theories described in \cite{Serrani:2025owx}, while satisfying both the holomorphic and non-holomorphic quartic constraints.

Here, we demonstrate the existence of such theories and describe their structure. These constructions allow coupling to one chiral sector, but not to both simultaneously. We refer to these theories as ``quasi-chiral''\footnote{This terminology was introduced in \cite{Adamo:2022lah}, where a related theory featuring a slightly different spectrum and some non-local interactions was constructed via a deformation of the Chalmers-Siegel action of self-dual Yang–Mills theory \cite{Chalmers:1996rq}.} higher-spin theories. By ``quasi-chiral'' we mean a theory that is not parity-invariant yet contains both holomorphic and anti-holomorphic quartic vertices, together with additional quartic interactions required to ensure consistency, at least up to quartic order.

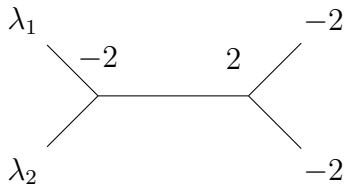
\begin{figure}[H]
    \centering
    \begin{tikzpicture}
        \begin{feynman}
            \vertex (i1) at (-6, 1) {\(\lambda_1\)};
            \vertex (i2) at (-6,-1) {\(\lambda_2\)};
            \vertex (i3) at (-2, 1) {\(-2\)};
            \vertex (i4) at (-2,-1) {\(-2\)};

            \vertex (v1) at (-5, 0);
            \vertex (v3) at (-3, 0);

            \vertex at (-5, 0.5) {\(-2\)};
            \vertex at (-3.2, 0.5) {\(2\)};
            
            \diagram* {
                (i1) -- (v1),
                (i2) -- (v1),
                (v1) -- [plain] (v3),
                (v3) -- (i3),
                (v3) -- (i4),
            };

        \end{feynman}
    \end{tikzpicture}
        \caption{Exchange diagram involving a generic vertex and the anti-MHV cubic vertex of GR.}
    \label{figureHSGR}
\end{figure}

\begin{figure}[H]
    \centering
    \begin{tikzpicture}
        \begin{feynman}
            \vertex (i1) at (-10, 1) {\(\lambda\)};
            \vertex (i2) at (-10,-1) {\(\lambda\)};
            \vertex (i3) at (-6, 1) {\(-2\)};
            \vertex (i4) at (-6,-1) {\(-2\)};

            \vertex (arrow) at (-3,0) {\(\implies\)};

            \vertex (ii1) at (0, 1) {\(\lambda\)};
            \vertex (ii2) at (0,-1) {\(-2\)};
            \vertex (ii3) at (4, 1) {\(\lambda\)};
            \vertex (ii4) at (4,-1) {\(-2\)};

            \vertex (v1) at (-9, 0);
            \vertex (v3) at (-7, 0);

            \vertex (vv1) at (1, 0);
            \vertex (vv3) at (3, 0);
            
            \vertex at (-9, 0.5){\(-2\)};
            \vertex at (-7.2, 0.5){\(2\)};

            \vertex at (1, 0.5){\(\lambda\)};
            \vertex at (2.8, 0.5){\(-\lambda\)};

            \diagram* {
                (i1) -- (v1),
                (i2) -- (v1),
                (v1) -- [plain] (v3),
                (v3) -- (i3),
                (v3) -- (i4),
            };

            \diagram* {
                (ii1) -- (vv1),
                (ii2) -- (vv1),
                (vv1) -- [plain] (vv3),
                (vv3) -- (ii3),
                (vv3) -- (ii4),
            };

        \end{feynman}
    \end{tikzpicture}
        \caption{The left diagram, by quartic consistency, implies the presence of the right one.}
    \label{figureHSGR2}
\end{figure}
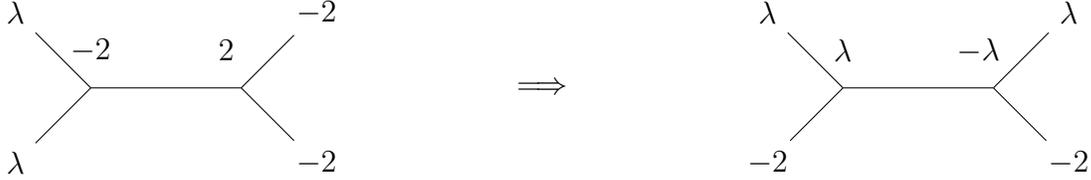

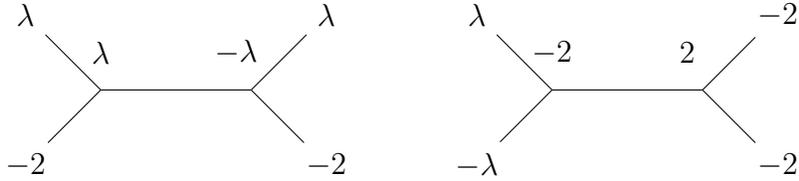
\begin{figure}[H]
    \centering
    \begin{tikzpicture}
        \begin{feynman}
            \vertex (i1) at (-6, 1) {\(\lambda\)};
            \vertex (i2) at (-6,-1) {\(-2\)};
            \vertex (i3) at (-2, 1) {\(\lambda\)};
            \vertex (i4) at (-2,-1) {\(-2\)};

            \vertex (ii1) at (0, 1) {\(\lambda\)};
            \vertex (ii2) at (0,-1) {\(-\lambda\)};
            \vertex (ii3) at (4, 1) {\(-2\)};
            \vertex (ii4) at (4,-1) {\(-2\)};

            \vertex (v1) at (-5, 0);
            \vertex (v3) at (-3, 0);

            \vertex (vv1) at (1, 0);
            \vertex (vv3) at (3, 0);

            \vertex at (-5, 0.5) {\(\lambda\)};
            \vertex at (-3.2, 0.5) {\(-\lambda\)};

            \vertex at (1, 0.5) {\(-2\)};
            \vertex at (2.8, 0.5) {\(2\)};

            \diagram* {
                (i1) -- (v1),
                (i2) -- (v1),
                (v1) -- [plain] (v3),
                (v3) -- (i3),
                (v3) -- (i4),
            };
            \diagram* {
                (ii1) -- (vv1),
                (ii2) -- (vv1),
                (vv1) -- [plain] (vv3),
                (vv3) -- (ii3),
                (vv3) -- (ii4),
            };
        \end{feynman}
    \end{tikzpicture}
        \caption{Holomorphic constraints between $C^{-2,-2,2}$ and $C^{\lambda,-\lambda,-2}$.}
    \label{figureHSGR3}
\end{figure}

We begin by constructing a theory that couples to self-dual gravity. We start by studying the corresponding exchange diagram in Figure \ref{figureHSGR}.
\noindent
This diagram cannot respect \eqref{kinda_trinagular}. We have to allow the external fields to have maximal helicity \eqref{surpass_trinagular}. The existence of quartic vertices requires
\begin{align}
    &\lambda_1=\lambda_2-2+n_1\,,&
    &\lambda_2=\lambda_1-2+n_2\,.&
    &\implies&
    &n_1=n_2=2
    &\implies&
    &\lambda_1=\lambda_2\,.
\end{align}
The existence of this diagram implies the presence of an additional diagram, as shown in Figure \ref{figureHSGR2}.
\noindent
The newly generated cubic vertices are also required to satisfy the holomorphic constraints. In particular, one can construct the diagrams shown in Figure \ref{figureHSGR3}.
\noindent
The holomorphic constraints fix the coefficients of the two anti-holomorphic cubic vertices $C^{-2,-2,2}$ and $C^{\lambda,-\lambda,-2}$ to be equal, implying $C^{-2,-2,2}=C^{\lambda,-\lambda,-2}$. As discussed in \cite{Serrani:2025owx}, these two anti-holomorphic cubic couplings are consistent even in the absence of additional higher-spin vertices, since they form a consistent truncation of the full chiral higher-spin gravity.

We have thus identified a fully consistent quasi-chiral higher-spin theory that we call quasi-chiral HS-GR, at least up to the quartic order, comprising a specific set of cubic and quartic vertices. The theory contains the following interactions:
\begin{equation}
\Big\{C^{-2,-2,2}=C^{\lambda,-\lambda,-2}, C^{\lambda,\lambda,-2}, h^{(2\lambda-2,2)}_{(\lambda,\lambda,-2,-2)}\Big\}\,.
\end{equation}
Notice that summing over all values of $\lambda$ gives a theory that remains consistent. Indeed, no new types of exchange diagrams are generated.

\begin{figure}[H]
    \centering
    \begin{tikzpicture}
        \begin{feynman}
            \vertex (i1) at (-6, 1) {\(\lambda\)};
            \vertex (i2) at (-6,-1) {\(-\lambda\)};
            \vertex (i3) at (-2, 1) {\(2\)};
            \vertex (i4) at (-2,-1) {\(-2\)};

            \vertex (v1) at (-5, 0);
            \vertex (v3) at (-3, 0);

            \vertex at (-5, 0.5) {\(-2\)};
            \vertex at (-3.2, 0.5) {\(2\)};

            \diagram* {
                (i1) -- (v1),
                (i2) -- (v1),
                (v1) -- [plain] (v3),
                (v3) -- (i3),
                (v3) -- (i4),
            };
        \end{feynman}
    \end{tikzpicture}
        \caption{$(2,2)$ $\bar{C}C$ exchange diagram.}
    \label{figure_fullGR}
\end{figure}
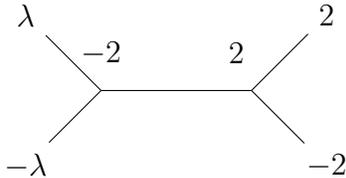

If we now attempt to include the holomorphic cubic vertex of gravity, $C^{2,2,-2}$, we can form the two-derivative exchange diagram in Figure \ref{figure_fullGR}.
\noindent
This exchange diagram does not admit a quartic vertex that solves the quartic constraint.

Interestingly, the same phenomenon also appears in the case of Yang-Mills-like interactions. In this case, we obtain the quasi-chiral HS-YM theory --- an extension of self-dual Yang-Mills theory --- comprising the couplings
\begin{equation}\label{quasi_chiral_YM}
    \{C^{-1,-1,1}=C^{\lambda,-\lambda,-1},C^{\lambda,\lambda,-1},h^{(2\lambda-1,1)}_{[\lambda,\lambda,-1,-1]},h^{(2\lambda-1,1)}_{[\lambda,-1,\lambda,-1]}\}\,.
\end{equation}
As before, summing over all values of $\lambda$ gives a theory that remains consistent. Attempting to include the holomorphic cubic vertex $C^{1,1,-1}$ leads to inconsistencies, similar to those encountered in the case of quasi-chiral HS-GR.

We expect that additional quasi-chiral higher-spin theories may exist that do not include the SDYM ($C^{-1,-1,1}$) or SDGR ($C^{-2,-2,2}$) cubic vertices. It would be interesting to classify all possible quasi-chiral higher-spin theories.

\paragraph{Comparison with the literature.} To our knowledge, the only other work that discusses quasi-chiral higher-spin theories is \cite{Adamo:2022lah}, in which the authors consider the possibility of extending higher-spin self-dual Yang–Mills theory (HS-SDYM) \cite{Ponomarev:2017nrr,Krasnov:2021nsq} to a quasi-chiral framework by allowing the presence of both MHV and anti-MHV cubic vertices.

The theory described in \cite{Adamo:2022lah} features, in our notation, the following spectrum of cubic vertices:
\begin{equation}
\{C^{-s,1,1},C^{-s_1,-s_2,1}\}\,,\qquad
s,s_1,s_2>0\,,
\end{equation}
where, as emphasised in \cite{Adamo:2022lah}, the positive-helicity external fields are restricted to spin-$1$ in order to preserve gauge invariance. The authors also computed four-point color-ordered amplitudes (which we will reproduce later) given by
\begin{align}
    &\tilde{\mA}(1_{s}^-2_{s}^-3_{1}^+4_{1}^+)=\frac{\langle 12\rangle^{2s+1}}{\langle 23\rangle\langle 34\rangle\langle 41\rangle}\,,&
    &\tilde{\mA}(1_{s}^-2_{1}^+3_{s}^-4_{1}^+)=\frac{\langle 13\rangle^{2s+2}}{\langle 12\rangle\langle 23\rangle\langle 34\rangle\langle 41\rangle}\,.
\end{align}
They further discuss an analogue of the Parke–Taylor formula for the $n$-point scattering of this theory. It is also noted that allowing the positive-helicity legs to carry arbitrary spin, rather than being fixed to spin-$1$, would make the amplitudes non-local.

We argue that the theory in \cite{Adamo:2022lah}, named HS-YM, does not define a local theory; rather, it defines a non-local one. As we have proven above, the simultaneous presence of both Yang–Mills cubic couplings $C^{-1,1,1}$ and $C^{-1,-1,1}$ cannot be locally consistent with any cubic higher-spin coupling. In particular, from \eqref{quasi_chiral_YM}, the light-cone analysis indicates that the consistency of the theory requires the presence of the cubic vertex $C^{s,-s,1}$. However, together with $C^{-1,-1,1}$, this inevitably leads to non-locality.

Nevertheless, the theory remains of considerable interest, both for the way it evades various no-go results and for the peculiar structure of its non-localities, which merit further investigation. In particular, it admits a simple Lorentz-invariant formulation.

\section{Four-point higher-spin amplitudes}\label{section7}

In the previous section, we demonstrated that a local and unitary higher-spin theory cannot consistently include parity invariant pairs of non-abelian cubic interactions. Nevertheless, we also observed the presence of many non-trivial local quartic vertices that solve the quartic constraint \eqref{quartic_system}. Whenever such a local quartic vertex exists, one can construct a well-defined local four-point amplitude by combining the exchange contributions with the quartic contact term. Even though the local quartic vertices appear quite complicated,\footnote{While the light-cone formalism has proven extremely efficient in the self-dual sector, where only cubic vertices are present and amplitudes can be computed with remarkable simplicity \cite{Skvortsov:2018jea,Skvortsov:2020wtf,Skvortsov:2020gpn}, it is likely not the most suitable framework for computing amplitudes in the presence of higher-point vertices.} the amplitudes take an especially simple form when expressed in spinor-helicity variables.

As we will now see, contrary to the self-dual sector,\footnote{In the self-dual sector, amplitudes vanish identically \cite{Serrani:2025owx,Serrani:2025oaw}. Moreover, amplitudes can only have helicity configurations (++++) and (+++$-$), and are therefore necessarily non-MHV.} none of these local four-point amplitudes vanish, providing an explicit counterexample to the standard expectation that local higher-spin amplitudes should vanish. These amplitudes evade Weinberg's soft theorem, as shown in \cite{Tran:2022amg}. To determine the local four-point amplitudes for generic spins, one could proceed following the same strategy adopted for Yang–Mills theory and gravity in Section \ref{subsection5.2}.

A more systematic approach is provided by the spinor-helicity formalism, which naturally incorporates the little-group scaling of massless amplitudes while implementing locality through consistent factorisation. In \cite{Ponomarev:2016cwi}, it was shown that the light-cone deformation procedure for constructing interacting massless theories in $4d$ flat space is equivalent to searching directly for spinor-helicity amplitudes satisfying the appropriate little-group scaling properties and locality constraints, the latter being reflected in the presence of only simple poles. It was further argued that this reformulation provides a more efficient framework for the search for consistent higher-spin interactions. Here we follow precisely this strategy.
In several cases, we explicitly verify the resulting amplitudes by summing the contributions from the local quartic vertices and exchange diagrams derived in the light-cone gauge.

Using this approach, we provide a complete classification and explicit expressions for all local higher-spin four-point amplitudes. To the best of our knowledge, this is the first systematic and exhaustive determination of this class of amplitudes. Partial results and specific examples had previously been obtained in \cite{Benincasa:2007xk,Benincasa:2011kn,Benincasa:2011pg,Benincasa:2012wt,McGady:2013sga,Taronna:2017wbx,Roiban:2017iqg,Ananth:2023qrf}. In addition, the quasi-chiral HS--YM amplitudes were previously derived in a covariant formulation in \cite{Adamo:2022lah}.

The recipe for determining the local amplitude $\mA(1_{\lambda_1}2_{\lambda_2}3_{\lambda_3}4_{\lambda_4})$ for generic helicities is as follows:
\begin{enumerate}
    \item We have to match the little group scaling:
    \begin{align}
        &2\lambda_i=\mathbb{N}_{|i]}-\mathbb{N}_{\langle i|}\,,&
        &\mathbb{N}_{|i]}=|i]\frac{\partial}{\partial|i]}\,,&
        &\mathbb{N}_{\langle i|}=\langle i|\frac{\partial}{\partial\langle i|}\,,&
        &i=1,2,3,4\,,
    \end{align}
    where $\mathbb{N}_{|i]}$ and $\mathbb{N}_{\langle i|}$ are the powers of $|i]$ and $\langle i|$ in the amplitude, respectively. 
    
    \item We notice that every local four-point amplitude can be written as follows: 
    \begin{equation}\label{amplitude_base}
        \mA(1_{\lambda_1}2_{\lambda_2}3_{\lambda_3}4_{\lambda_4})=p(s,t)[12]^{x_1}\langle 34\rangle^{x_2}[13]^{x_3}[14]^{x_4}\,,
   \end{equation}
    where $x_1,x_2,x_3,x_4\in \mathbb{Z}$, and $p(s,t)$ is a meromorphic function of the Mandelstam variables $s$ and $t$,\footnote{If we require the locality of the amplitude, $p(s,t)$ must contain at most single poles. Then we can also write $p(s,t)=\frac{1}{s^at^bu^c}P(s,t)$, where $P(s,t)$ is a polynomial in $s$ and $t$, and $a,b,c=0,1$.} as $u$ can always be recovered via $s+t+u=0$. To demonstrate that every spinor-helicity expression for a four-point amplitude can be cast into this form, one needs to apply momentum conservation:
    \begin{subequations}\label{momentum_conserv}
    \begin{align}
        &u=-s-t\,,&
        &\langle 12\rangle=-\frac{s}{[12]}\,,&
        &\langle 13\rangle=-\frac{u}{[13]}\,,&
        &\langle 14\rangle=-\frac{t}{[14]}\,,\\
        &\langle 24\rangle=-\frac{[13]\langle 34\rangle}{[12]}\,,&
        &\langle 23\rangle=\frac{[14]\langle 34\rangle}{[12]}\,,&
        &[23]=\frac{-t[12]}{[14]\langle 34\rangle}\,,&
        &[24]=\frac{u[12]}{[13]\langle 34\rangle}\,.
    \end{align}
    \end{subequations}
   In general, there are other possible structures besides \eqref{amplitude_base}. For instance, one may replace $[13]$ with $[24]$, or $[14]$ with $[23]$. More generally, one is allowed to permute the labels $1,2,3,4$ on the right-hand side of \eqref{amplitude_base}, as well as to exchange square and angle brackets, $[ij]\leftrightarrow \langle ij\rangle$. However, as we will explain below, configurations such as $[12]^{x_1}[34]^{x_2}$ and $\langle 12 \rangle^{x_1}\langle 34\rangle^{x_2}$ are excluded.

    \item The expression above can be constructed for arbitrary choices of the external helicities. We now impose the stronger requirement that the amplitude is generated by the exchange of two cubic vertices, as depicted in Figure \ref{figure_generic}. This requirement enforces the factorisation properties of the amplitude and, consequently, provides the constraints necessary to ensure locality.
    
    The form of the amplitude in \eqref{amplitude_base} requires a minimum of $D$ derivatives carried by the quartic vertex. Therefore\footnote{Here, $D$ can equivalently be understood as the momentum degree (mass dimension) of the
four-point exchange amplitude generated by two cubic vertices $A_4^{\rm exch}\sim V_3^{(n)}\frac{1}{p^2}V_3^{(m)}
\sim p^{n+m-2}\sim [M]^{n+m-2}\,.$} $D\equiv n+m-2=\lambda_{12}-\lambda_{34}+2\omega-2$,\footnote{Here, $\omega$ denotes the helicity of the exchanged field; when only a quartic vertex is present, $\omega$ is just a way to assign a number of derivatives to the specific quartic vertex.} must exceed the minimal number required by the little-group scaling $d\equiv x_1+x_2+x_3+x_4$ in \eqref{amplitude_base}. Taking into account possible poles, then $D-d+2k\geq 0$. The integer $k$ takes the values $k=0$ for a pure quartic vertex, $k=1$ in the single-channel case ($\mA_4\sim 1/s$), $k=2$ in the YM-like case ($\mA_4\sim 1/st$), and $k=3$ in the GR-like case ($\mA_4\sim 1/stu$). By applying the constraint $D-d+2k\geq 0$ to any possible $d$ coming from all local four-point amplitude expressions such as in \eqref{amplitude_base}, we reproduce the same algebraic constraints discussed in \eqref{kinda_trinagular} and \eqref{surpass_trinagular}.
\end{enumerate}
Before studying the constraints imposed on four-point amplitudes, let us justify the statement made above and prove that the following procedure is equivalent to standard factorisation. The advantage of this approach is that it can be applied systematically, given any four external helicities, and it does not require checking factorisation channel by channel.

\paragraph{``\textit{Fast factorisation}''.}
We now explain why the above procedure (which we shall refer to as \emph{fast factorisation}) is equivalent to the usual factorisation. The first step is to show that the general form \eqref{amplitude_base} of a four-point amplitude is unique, up to permutations of the external legs $1,2,3,4$ and the exchange of square ($[\;]$) and angle ($\langle\;\rangle$) spinor brackets. This follows from an explicit construction.

Since the amplitude is gauge invariant, it can only depend on the external spinors through Lorentz-invariant spinor contractions. Without loss of generality, we first fix the little-group weight of particle $2_{\lambda_2}$ by starting with either $\mA_4\sim [12]^{2\lambda_2}$ or $\mA_4\sim \langle 12\rangle^{-2\lambda_2}$. After this choice, the remaining little-group weights must be fixed using spinor contractions involving the other external legs. There are only three independent spinor brackets left, namely $(13)$, $(14)$, and $(34)$, where $(ij)$ denotes either a square bracket $[ij]$ or an angle bracket $\langle ij\rangle$. The most general ansatz therefore takes the form
\begin{equation}
\mA(1_{\lambda_1}2_{\lambda_2}3_{\lambda_3}4_{\lambda_4})
=p(s,t)
(12)^{x_1}
(34)^{x_2}
(13)^{x_3}
(14)^{x_4}\,.
\end{equation}
Since the Mandelstam variables carry zero little-group weight, they can appear with arbitrary powers, $p(s,t)$, without affecting the little-group scaling. 
All other possible ansätze are then found by arbitrary permutations of the external legs. For example, replacing $(13)$ by $(24)$ is equivalent to the permutation $(1,2)\leftrightarrow(4,3)$.

For each factor $(ij)$ we may choose either a square or an angle bracket. However, the two configurations $[12]^{x_1}[34]^{x_2}$ and $\langle 12 \rangle^{x_1}\langle 34\rangle^{x_2}$ are excluded. There are two reasons for this. First, if the amplitude is required to arise from a factorisation channel, such structures can never be generated by gluing two three-point amplitudes. To see this explicitly, consider a factorisation channel in which the internal momentum is $k_{\omega}$:
\begin{equation}
    \mA_4\to\mA_3^{(n)}\frac{1}{s}\mA_3^{(m)}=[12]^{\lambda_{12}-\omega}[2k]^{\lambda_2+\omega-\lambda_1}[k1]^{\lambda_1+\omega-\lambda_2}\frac{1}{s}\langle 34\rangle^{-\lambda_{34}-\omega}\langle 4k\rangle^{\lambda_3+\omega-\lambda_4}\langle k3\rangle^{\lambda_4+\omega-\lambda_3}\,.
\end{equation}
The resulting four-point expression can never reduce to a structure containing $\frac{1}{s}[12]^{x_1}[34]^{x_2}$ with both $x_1,x_2\neq 0$; the same applies to $\frac{1}{s}\langle 12 \rangle^{x_1}\langle 34\rangle^{x_2}$.

There is another, more fundamental reason for excluding such configurations. Allowing them would destroy the predictive power of the pole structure, which is essential for analysing factorisation. Indeed, one would have to enlarge the class of admissible functions $p(s,t)$ by allowing arbitrary dimensionless cross-ratios of the Mandelstam variables, such as powers of ($s/t$), ($t/u$), and ($u/s$). Although these factors preserve both the little-group scaling and the total number of derivatives $D$, they can generate higher-order poles. In fact, such poles can always be removed by trading them for different spinor structures. For example
\begin{align}
&\frac{u^2}{s^2}[12]^2[34]^2=[13]^2[24]^2\,,&
&\frac{t^2}{s^2}[12]^2[34]^2=[14]^2[23]^2\,.
\end{align}
Consequently, the pole structure is no longer uniquely determined by the chosen spinor ansatz. In contrast, for the ansätze considered here, such as \eqref{amplitude_base}, the poles contained in $p(s,t)$ cannot be eliminated by the accompanying spinor factor $[12]^{x_1}\langle34\rangle^{x_2}[13]^{x_3}[14]^{x_4}$ even after using momentum conservation. This uniqueness of the pole structure is what makes this method powerful to classify the most general four-point amplitudes satisfying factorisation.

\paragraph{\textit{Fast factorisation} = Factorisation.}

We now prove that \textit{fast factorisation} is equivalent to the usual notion of factorisation.

\begin{itemize}

\item \textbf{Factorisation $\implies$ \textit{Fast factorisation}.}

Ordinary factorisation implies that, when an internal particle goes on-shell, the four-point amplitude factorises into a product of cubic amplitudes as $\mA_4\to \mA_3^{(n)}\frac{1}{s}\mA_3^{(m)}$ (and analogously in the $t$- and $u$-channels whenever the corresponding poles are present). Since each cubic amplitude has the correct little-group scaling, the resulting four-point amplitude automatically inherits the appropriate little-group weights. It can therefore always be expressed in the form \eqref{amplitude_base}, or, more generally, as a sum of three terms of this type.\footnote{As we will see in the single-channel case $\mA_4$ can be decomposed as $\mA_4\sim \mA_s+\mA_t+\mA_u$, where $\mA_s\sim \frac{1}{s}$, $\mA_t\sim \frac{1}{t}$, and $\mA_u\sim \frac{1}{u}$.} Moreover, by construction $D\ge d-2k$. The special case $D=d-2k$ corresponds to a constant polynomial numerator in the Mandelstam variables, for instance $P(s,t)=1$.

\item \textbf{\textit{Fast factorisation} $\implies$ Factorisation.}

Conversely, \textit{fast factorisation} states that the amplitude can be written in the form \eqref{amplitude_base} with $D\ge d-2k$. If the amplitude contains one ($1/s$), two ($1/st$), or three ($1/stu$) physical poles, then it can be factorised in the corresponding channels.
For the $s$-channel, one may write $\mA_4\to \mA_3^{(n)}\frac{1}{s}\mA_3^{(m)}$, where $n+m-2=\lambda_{12}-\lambda_{34}+2\omega_s-2=D$, which fixes $\omega_s=\frac{1}{2}\left(D+2+\lambda_{34}-\lambda_{12}\right)$. 
Similarly, if a $t$-channel pole is present, $\mA_4\to \mA_3^{(n)}\frac{1}{t}\mA_3^{(m)}$,
with $n+m-2=\lambda_{14}-\lambda_{23}+2\omega_t-2=D$, so that $\omega_t=\frac{1}{2}\left(D+2+\lambda_{23}-\lambda_{14}\right).$
Likewise, if a $u$-channel pole is present, $\mA_4\to \mA_3^{(n)}\frac{1}{u}\mA_3^{(m)}$, with $n+m-2=\lambda_{13}-\lambda_{24}+2\omega_u-2=D$,
which gives $\omega_u=\frac{1}{2}\left(D+2+\lambda_{24}-\lambda_{13}\right)$. Therefore, every amplitude satisfying the \textit{fast factorisation} criterion admits a factorisation into appropriate three-point amplitudes in every physical channel in which a pole is present.

\end{itemize}
Let us make a few remarks. First, the degrees $n=\lambda_{12}+\omega$ (of the holomorphic cubic amplitude) and $m=-\lambda_{34}+\omega$ (of the anti-holomorphic cubic amplitude) must be identical in all three channels. Otherwise, the three factorisation channels cannot form the same four-point amplitude. This follows simply from counting angle and square brackets: momentum conservation can reshuffle spinor products ($\langle 1k\rangle[k4]=\langle 12\rangle[24]=-\langle 13\rangle[34]$) but never changes their number. This is precisely the same condition encountered in the light-cone formalism, where the powers of $\PPb$ and $\PP$ must coincide in all channels for the three exchanges to belong to the same quartic constraint. 

As a consequence, specifying the exchanged helicity $\omega$ in the $s$-channel uniquely determines the corresponding exchanges in the $t$- and $u$-channels. Indeed analysing a single exchange diagram is sufficient to determine whether an amplitude can factorise in one, two, or all three channels.

We also emphasise that there is always a representation free of spurious poles. In particular, the representation with maximal $d$ is manifestly local, with the only poles arising from the physical propagators. This follows from the fact that any denominator in \eqref{amplitude_base} involving spinor brackets can always be converted into a numerator by exchanging angle and square brackets. 

In the following, we derive the constraints imposed by \textit{fast factorisation} and illustrate them through several examples, highlighting the properties discussed above.

Applying the first and second rules described above gives the following form of the amplitude
\begin{equation}\label{amplitude1234}
    \mA(1_{\lambda_1}2_{\lambda_2}3_{\lambda_3}4_{\lambda_4})=p(s,t)[12]^{2\lambda_2}\langle 34\rangle^{\lambda_1-\lambda_{234}}[13]^{\lambda_{13}-\lambda_{24}}[14]^{\lambda_{14}-\lambda_{23}}\,,
\end{equation}
where $d=3\lambda_1-\lambda_{234}$. By considering the constraint $D-d+2k\geq 0$, for \eqref{amplitude1234} we get
\begin{equation}\label{condition1}
\lambda_1\leq \lambda_2+\omega+k-1\,.
\end{equation}
If we start from a different ansatz for the amplitude, for instance, exchanging $[ij]$ with $\langle ij\rangle$ in \eqref{amplitude1234}, then starting from
\begin{equation}\label{amplitude1234_inverted}
    \mA(1_{\lambda_1}2_{\lambda_2}3_{\lambda_3}4_{\lambda_4})=p(s,t)\langle 12\rangle^{-2\lambda_2}[34]^{-\lambda_1+\lambda_{234}}\langle 13\rangle^{-\lambda_{13}+\lambda_{24}}\langle 14\rangle^{-\lambda_{14}+\lambda_{23}}\,,
\end{equation}
then $d=-3\lambda_1+\lambda_{234}$; by considering again the constraint $D-d+2k\geq 0$, we obtain
\begin{equation}\label{condition2}
\lambda_1\geq -\frac{1}{2}(-\lambda_{34}+\omega+k-1)=-\frac{1}{2}(m+k-1)\,.
\end{equation}
If, instead, we permute the RHS labels $(1234)$ in the ansatz \eqref{amplitude1234} with $(3214)$, we have
\begin{equation}\label{amplitude3214}
    \mA(1_{\lambda_1}2_{\lambda_2}3_{\lambda_3}4_{\lambda_4})=p(s,t)[32]^{2\lambda_2}\langle 14\rangle^{\lambda_3-\lambda_{214}}[31]^{\lambda_{31}-\lambda_{24}}[34]^{\lambda_{34}-\lambda_{21}}\,,
\end{equation}
then $d=3\lambda_3-\lambda_{214}$. The constraint does give
\begin{equation}\label{condition3}
\lambda_3\leq \frac{1}{2}(\lambda_{12}+\omega+k-1)=\frac{1}{2}(n+k-1)\,.
\end{equation}
By further exchanging $[ij]$ with $\langle ij\rangle$ and then having
\begin{equation}\label{amplitude3214_inverted}
    \mA(1_{\lambda_1}2_{\lambda_2}3_{\lambda_3}4_{\lambda_4})=p(s,t)\langle 32\rangle^{-2\lambda_2}[14]^{-\lambda_3+\lambda_{214}}\langle 31\rangle^{-\lambda_{31}+\lambda_{24}}\langle 34\rangle^{-\lambda_{34}+\lambda_{21}}\,,
\end{equation}
then $d=-3\lambda_3+\lambda_{214}$, we would obtain
\begin{equation}\label{condition4}
    \lambda_4\leq \lambda_3+\omega+k-1\,.
\end{equation}
We now derive the constraints arising from all possible forms of \eqref{amplitude_base}. It is straightforward to verify that the corresponding ansätze exhaust all possible values of $d = 3\lambda_i - \lambda_{jk\ell}$ for every permutation of $i,j,k,\ell=1,2,3,4$. Evaluating each case, we obtain the following minimal set of independent constraints:
\begin{subequations}\label{total_conditions}
\begin{align}\label{part12}
&\lambda_1 \leq \lambda_2+\omega+k-1\,,&
&\lambda_2 \leq \lambda_1+\omega+k-1\,,\\\label{part34}
&\lambda_3 \leq \lambda_4+\omega+k-1\,,&
&\lambda_4 \leq \lambda_3+\omega+k-1\,,\\
&\lambda_{12} \geq \lambda_{34}\,,&
&k=0,1,2,3.
\end{align}
\end{subequations}
Notice that the condition $\lambda_{12} \geq \lambda_{34}$ is equivalent to \eqref{condition2} and also to \eqref{condition3} once we use \eqref{part12} and \eqref{part34}. These are a rewriting of the conditions found above: \eqref{kinda_trinagular}, \eqref{surpass_trinagular}, and \eqref{Homo_constraints}. The additional conditions $n=\lambda_{12}+\omega> 0$ and $m=-\lambda_{34}+\omega> 0$ must be added by hand to ensure we are looking at non-holomorphic amplitude. 

The conditions \eqref{total_conditions} can be used in two complementary ways. If we are only interested in constructing amplitudes, then for a fixed value of $k$, every choice of helicities $(\lambda_1,\lambda_2,\lambda_3,\lambda_4,\omega)$ satisfying \eqref{total_conditions} defines an amplitude $ \mA(1_{\lambda_1}2_{\lambda_2}3_{\lambda_3}4_{\lambda_4})$ that factorises correctly into $k$ channels, whose explicit form is given by \eqref{amplitude_base}.

Alternatively, if the goal is to recover the information encoded in the light-cone tables, one can proceed as follows. Whenever \eqref{total_conditions} are satisfied, the exchange $(1234)$ is allowed. As a direct consequence, the exchanges $(4123)$, and $(1324)$ are always allowed as well\footnote{This differs slightly from the various tables shown because the helicities are ordered here so that $d$ is maximal.},  whereas $(3412)$ is allowed only if $\lambda_{12}=\lambda_{34}$, $(2341)$ only if $\lambda_{14}=\lambda_{23}$, and $(2413)$ only if $\lambda_{13}=\lambda_{24}$. These conditions follows directly from \eqref{total_conditions}. For instance, if $(1234)$ is allowed, then $\lambda_{12}\geq \lambda_{34}$; demanding that $(3412)$ be allowed as well implies $\lambda_{34}\geq \lambda_{12}$, and hence $\lambda_{12}=\lambda_{34}$. The remaining cases follow analogously. Moreover, the relations among the various couplings can also be determined completely. We will illustrate this procedure below.

We can now begin to present the explicit form of all local four-point higher-spin amplitudes. Even though, as pointed out above, a given amplitude generally admits several equivalent representations, we choose a canonical one by maximising the quantity $d$, namely 
\begin{equation}\label{max_d}
    d=\max\limits_{i\neq j\neq k\neq \ell}\{3\lambda_i-\lambda_{jk\ell},-3\lambda_i+\lambda_{jk\ell}\}\,,\qquad
    i,j,k,\ell=1,2,3,4\,.
\end{equation}
As discussed above, the representation with maximal $d$ makes locality manifest, with the only poles arising from the physical propagators.

\paragraph{Quartic amplitudes from quartic vertices.}
When the conditions \eqref{total_conditions} are satisfied for $k=0$, local quartic vertices are allowed. These correspond to the solution of the homogeneous quartic constraint studied above and lead to the following amplitudes: 
\begin{equation}\label{fourAmplitude_homo}
    \mA^{(D)}_{\text{homo}}(1_{\lambda_1}2_{\lambda_2}3_{\lambda_3}4_{\lambda_4})=\sum_{i=0}^{\frac{D-d}{2}}\left(c_i\,s^it^{\frac{D-d}{2}-i}\right)[12]^{2\lambda_2}\langle 34\rangle^{\lambda_1-\lambda_{234}}[13]^{\lambda_{13}-\lambda_{24}}[14]^{\lambda_{14}-\lambda_{23}}\,,
\end{equation}
where the $c_i$ are $\frac{D-d}{2}+1$ free coefficients. 

\paragraph{Single-channel amplitudes.}
When the conditions \eqref{total_conditions} are satisfied for $k=1$, single-channel amplitudes are allowed, and the most general four-point amplitude is
\begin{equation}\label{single-channel_stu}
    \mA^{(D)}(1_{\lambda_1}2_{\lambda_2}3_{\lambda_3}4_{\lambda_4})=\mA^{(D)}_{s}+\mA^{(D)}_{t}+\mA^{(D)}_{u}\,,
\end{equation}
where
\begin{subequations}\label{single-channel_amplitudes}
\begin{align}
    \mA^{(D)}_{s}(1_{\lambda_1}2_{\lambda_2}3_{\lambda_3}4_{\lambda_4})&=k_s\frac{t^{\frac{D-d+2}{2}}}{s}[12]^{2\lambda_2}\langle 34\rangle^{\lambda_1-\lambda_{234}}[13]^{\lambda_{13}-\lambda_{24}}[14]^{\lambda_{14}-\lambda_{23}}+\mA^{(D)}_{\text{homo}}\,,\\
    \mA^{(D)}_{t}(1_{\lambda_1}2_{\lambda_2}3_{\lambda_3}4_{\lambda_4})&=k_t\frac{u^{\frac{D-d+2}{2}}}{t}[12]^{2\lambda_2}\langle 34\rangle^{\lambda_1-\lambda_{234}}[13]^{\lambda_{13}-\lambda_{24}}[14]^{\lambda_{14}-\lambda_{23}}+\mA^{(D)}_{\text{homo}}\,,\\
    \mA^{(D)}_{u}(1_{\lambda_1}2_{\lambda_2}3_{\lambda_3}4_{\lambda_4})&=k_u\frac{s^{\frac{D-d+2}{2}}}{u}[12]^{2\lambda_2}\langle 34\rangle^{\lambda_1-\lambda_{234}}[13]^{\lambda_{13}-\lambda_{24}}[14]^{\lambda_{14}-\lambda_{23}}+\mA^{(D)}_{\text{homo}}\,,
\end{align}
\end{subequations}
where we selected a representative four-point amplitude, noting that the addition of homogeneous solutions can always modify the explicit form of the term with the pole. The coefficients $k_{\bullet}\sim C\bar{C}$ are the products of cubic couplings in the $s$-, $t$-, and $u$-channel. In particular, we have\footnote{Here by $k_{1,2,3,4,5,6}$ we mean the products of coupling that were used in the various tables presented.} 
\begin{align}
    &k_s=k_1=k_3\,,&
    &k_t=k_2=k_4\,,&
    &k_u=k_5=k_6\,,
\end{align}
where, as discussed above for quartic vertices, coefficients such as $k_1$ and $k_3$ are equal only when both are nonzero. It may nevertheless happen that consistency requires one of them to vanish. This can be seen explicitly by checking whether a given exchange contribution satisfies the conditions in \eqref{total_conditions}. The same reasoning applies to the pairs $(k_2, k_4)$ and $(k_5, k_6)$.

For the special case when $D=d-2$, we have a unique solution corresponding to
\begin{subequations}
\begin{align}
    \mA^{(d-2)}_{s}(1_{\lambda_1}2_{\lambda_2}3_{\lambda_3}4_{\lambda_4})&=\frac{k_s}{s}[12]^{2\lambda_2}\langle 34\rangle^{\lambda_1-\lambda_{234}}[13]^{\lambda_{13}-\lambda_{24}}[14]^{\lambda_{14}-\lambda_{23}}\,,\\
    \mA^{(d-2)}_{t}(1_{\lambda_1}2_{\lambda_2}3_{\lambda_3}4_{\lambda_4})&=\frac{k_t}{t}[12]^{2\lambda_2}\langle 34\rangle^{\lambda_1-\lambda_{234}}[13]^{\lambda_{13}-\lambda_{24}}[14]^{\lambda_{14}-\lambda_{23}}\,,\\
    \mA^{(d-2)}_{u}(1_{\lambda_1}2_{\lambda_2}3_{\lambda_3}4_{\lambda_4})&=\frac{k_u}{u}[12]^{2\lambda_2}\langle 34\rangle^{\lambda_1-\lambda_{234}}[13]^{\lambda_{13}-\lambda_{24}}[14]^{\lambda_{14}-\lambda_{23}}\,.
\end{align}
\end{subequations}
Note that if we sum single-channel amplitudes as in \eqref{single-channel_stu}, they will admit factorisation into three channels if $k_s,k_t,k_u\neq 0$ or two channels (for example, $s$ and $t$) if we take $k_u=0$.

\paragraph{YM-like amplitudes.}
When the conditions \eqref{total_conditions} are satisfied only for $k=2$, and we have $D=d-4$, YM-like amplitudes are allowed, and we obtain the following unique amplitudes
\begin{subequations}\label{YM-like_amplitudes}
\begin{align}
    \mA^{(d-4)}_{st}(1_{\lambda_1}2_{\lambda_2}3_{\lambda_3}4_{\lambda_4})&=\frac{k_{st}}{st}[12]^{2\lambda_2}\langle 34\rangle^{\lambda_1-\lambda_{234}}[13]^{\lambda_{13}-\lambda_{24}}[14]^{\lambda_{14}-\lambda_{23}}\,,\\
    \mA^{(d-4)}_{us}(1_{\lambda_1}2_{\lambda_2}3_{\lambda_3}4_{\lambda_4})&=\frac{k_{us}}{us}[12]^{2\lambda_2}\langle 34\rangle^{\lambda_1-\lambda_{234}}[13]^{\lambda_{13}-\lambda_{24}}[14]^{\lambda_{14}-\lambda_{23}}\,,\\
    \mA^{(d-4)}_{tu}(1_{\lambda_1}2_{\lambda_2}3_{\lambda_3}4_{\lambda_4})&=\frac{k_{tu}}{tu}[12]^{2\lambda_2}\langle 34\rangle^{\lambda_1-\lambda_{234}}[13]^{\lambda_{13}-\lambda_{24}}[14]^{\lambda_{14}-\lambda_{23}}\,.
\end{align}
\end{subequations}
Here, by factorisation, we get the following relations among the various couplings $k_{st}=k_s=k_t$, $k_{us}=k_u=k_s$, and $k_{tu}=k_t=k_u$, then $k_{st}=k_{us}=k_{tu}=k_s=k_t=k_u=k$. There is a unique coupling. Moreover, if we sum over all of them and look for an amplitude that has three poles, we get
\begin{align}
    \begin{split}
    \mA^{(d-4)}(1_{\lambda_1}2_{\lambda_2}3_{\lambda_3}4_{\lambda_4})&=\mA^{(d-4)}_{st}+\mA^{(d-4)}_{us}+\mA^{(d-4)}_{tu}\\
    &=\frac{k(s+t+u)}{stu}[12]^{2\lambda_2}\langle 34\rangle^{\lambda_1-\lambda_{234}}[13]^{\lambda_{13}-\lambda_{24}}[14]^{\lambda_{14}-\lambda_{23}}=0\,.
    \end{split}
\end{align}
This tells us that a GR-like amplitude cannot exist. For instance, Yang–Mills theory does not admit a GR-like amplitude. More generally, however, the statement is stronger: no YM-like amplitude can ever generate a GR-like one. The same conclusion can also be reached in a more conventional way by searching for an amplitude of the form
\begin{equation}
    \mA^{(d-4)}(1_{\lambda_1}2_{\lambda_2}3_{\lambda_3}4_{\lambda_4})
    =\left(\frac{k_{st}}{st}+\frac{k_{us}}{us}+\frac{k_{tu}}{tu}\right)[12]^{2\lambda_2}\langle 34\rangle^{\lambda_1-\lambda_{234}}[13]^{\lambda_{13}-\lambda_{24}}[14]^{\lambda_{14}-\lambda_{23}}\,.
\end{equation}
Consistent factorisation (by matching the $s$, $t$, and $u$ residues) implies:
\begin{equation}\label{system_YMHS}
    R_s=\left(\frac{k_{st}}{t}-\frac{k_{us}}{t}\right)=\frac{k_s}{t}\,,\quad
    R_t=\left(\frac{k_{tu}}{u}-\frac{k_{st}}{u}\right)=-\frac{k_t}{u}\,,\quad
    R_u=\left(\frac{k_{us}}{s}-\frac{k_{tu}}{s}\right)=-\frac{k_u}{s}\,,
\end{equation}
where $R_s$ denoted the $s$-channel residue of $\mA^{(d-4)}$,\footnote{We recall that the $s$-channel residue is obtained in the limit $\langle 12\rangle\to 0$ and $[34]\to 0$. More generally, consider an amplitude $\mA(1_{\lambda_1}2_{\lambda_2}3_{\lambda_3}4_{\lambda_4})$. If $\lambda_{12}+\omega>0$ and $\lambda_{34}-\omega<0$, the $p^2_{12}\to 0$ limit is reached via $\langle 12\rangle\to 0$ and $[34]\to 0$. Conversely, if $\lambda_{12}+\omega<0$ and $\lambda_{34}-\omega>0$, the same limit is reached via $[12]\to 0$ and $\langle 34\rangle\to 0$.} and similarly for the $t$- and $u$-channels. The system of equations \eqref{system_YMHS}, in order to admit a solution \cite{Arkani-Hamed:2017jhn}, does require
\begin{equation}
    k_s-k_t=k_u\,.
\end{equation}
This is the modified Jacobi identity, analogous to the one encountered in Yang–Mills theory in \eqref{JI_YM}. This observation suggests the possibility of extending color–kinematics duality and the double-copy construction to higher-spin amplitudes in the non-holomorphic sector. Color–kinematics duality and the double-copy have already been explored in the self-dual sector for higher-spin theories, where the underlying kinematic algebra is particularly tractable \cite{Ponomarev:2017nrr,Ponomarev:2024jyg}. 

In the following we will show that, at the level of four-point amplitudes, color–kinematics duality and the double-copy construction can indeed be extended beyond the self-dual sector. At higher multiplicity, one could attempt to reorganise the amplitudes into a representation better suited for the analysis of color–kinematics duality \cite{Bern:2019prr,Bern:2022wqg}.

\paragraph{GR-like amplitudes.}
When the conditions \eqref{total_conditions} are satisfied only for $k=3$, and we have $D=d-6$, GR-like amplitudes are allowed, and we obtain the following unique amplitude
\begin{equation}\label{GR-like_amplitude}
    \mA^{(d-6)}_{stu}(1_{\lambda_1}2_{\lambda_2}3_{\lambda_3}4_{\lambda_4})=\frac{k_{stu}}{stu}[12]^{2\lambda_2}\langle 34\rangle^{\lambda_1-\lambda_{234}}[13]^{\lambda_{13}-\lambda_{24}}[14]^{\lambda_{14}-\lambda_{23}}\,,
\end{equation}
where consistent factorisation fixes $k_{stu}=k_s=k_t=k_u=k$.

From the expression of the YM-like amplitudes in \eqref{YM-like_amplitudes} and the GR-like amplitudes in \eqref{GR-like_amplitude}, it is simple to see 
\begin{equation}\label{YM-like-BCJ}
    \mA^{(d-4)}_{st}=\frac{u}{t}\mA^{(d-4)}_{us}=\frac{u}{s}\mA^{(d-4)}_{tu}\,,
\end{equation}
and 
\begin{equation}\label{GR-like-KLT}
    \mA^{(2d-6)}_{stu}(1_{2\lambda_1}2_{2\lambda_2}3_{2\lambda_3}4_{2\lambda_4})=\frac{st}{u}\big(\mA^{(d-4)}_{st}(1_{\lambda_1}2_{\lambda_2}3_{\lambda_3}4_{\lambda_4})\big)^2\,.
\end{equation}
At four points, these structures imply the existence of color-ordered amplitudes, the BCJ amplitude relations for YM-like amplitudes, and the double-copy construction of GR-like amplitudes from YM-like ones, via KLT relations \cite{Kawai:1985xq}. We will investigate these relations in greater detail in the following, providing several explicit examples.

We stress again that all the amplitudes described above are local amplitudes. We now present some explicit four-point amplitudes for several representative examples. 

\paragraph{External scalar fields.} We start with amplitudes involving only external scalar fields, then $d=0$ and $k=0$. From \eqref{fourAmplitude_homo}, we find
\begin{equation}
    \mA^{(D)}_{\text{homo}}(1_02_03_04_0)=\sum_{i=0}^{D/2}\left(c_i\,s^it^{D/2-i}\right)\,.
\end{equation}
In the case of $\phi^4$ theory, then for $D=0$, we have $\mA^{(D)}_{\text{homo}}(1_02_03_04_0)=c_0$, where $c_0$ corresponds to the $\phi^4$ coupling constant. 

If we now consider the single-channel amplitudes, then $d=0$, $k=1$, and $D=2\omega-2$. From \eqref{single-channel_amplitudes} we find 
\begin{equation}
    \mA_s^{(2\omega-2)}(1_02_03_04_0)=k_s\frac{t^{\omega}}{s}+\mA^{(2\omega-2)}_{\text{homo}}(1_02_03_04_0)\,,
\end{equation}
where $k_s=C^{0,0,\omega} \bar{C}^{-\omega,0,0}$ and if we consider all three channels \eqref{single-channel_stu} we get
\begin{equation}
    \mA^{(2\omega-2)}(1_02_03_04_0)=k_s\frac{t^{\omega}}{s}+k_t\frac{u^{\omega}}{t}+k_u\frac{s^{\omega}}{u}+\mA^{(2\omega-2)}_{\text{homo}}(1_02_03_04_0)\,.
\end{equation}
In the case of $\phi^3$ theory, then for $\omega=0$ and $k_s=k_t=k_u=k$, we get
\begin{equation}
    \mA^{(2\omega-2)}(1_02_03_04_0)=k\Big(\frac{1}{s}+\frac{1}{t}+\frac{1}{u}\Big)\,,
\end{equation}
where $k=(C^{0,0,0})^2$ is the square of the $\phi^3$ coupling constant.

It is worth noting that the four-point amplitudes constructed remain valid for the $\phi^3$ theory, even though the case $n,m=0$ was excluded from our analysis.

We also note that four-point amplitudes with external scalar fields and higher-spin exchanges were previously computed in \cite{Ponomarev:2016lrm}, where their explicit expressions in the light-cone gauge were derived.

\paragraph{External helicities $\lambda_{1,2,3,4}=+1$.}
We have again $d=0$ and $k=0$. From \eqref{fourAmplitude_homo}, we find
\begin{equation}
    \mA^{(D)}_{\text{homo}}(1^+2^+3^+4^+)=\sum_{i=0}^{D/2}\left(c_i\,s^it^{D/2-i}\right)\frac{[12]^2}{\langle 34\rangle^2}\,.
\end{equation}
For a local amplitude we require $D\geq 4$. For $D=4$, we find 
\begin{align}
\begin{split}
    \mA^{(4)}_{\text{homo}}(1^+2^+3^+4^+)&=c_2[12]^2[34]^2+c_1[12][34][14][23]+c_0[14]^2[23]^2\\
    &=c'_2[12]^2[34]^2+c'_1[13]^2[24]^2+c'_0[14]^2[23]^2\\
    &=c'_2[12]^2[34]^2+c'_1\frac{u^2}{s^2}[12]^2[34]+c'_0\frac{t^2}{s^2}[12]^2[34]^2\,,
\end{split}
\end{align}
where we used the Schouten identity, $[12][34]+[14][23]=[13][24]$ and redefined the coefficients according to $c'_2=c_2-\frac{c_1}{2}$, $c'_1=\frac{c_1}{2}$, and $c'_0=c_0-\frac{c_1}{2}$. As already remarked, this illustrates that an ansatz containing the factor $[12][34]$ would require the function $p(s,t)$ to include contributions involving cross-ratios, such as $u/s$ and $t/s$. Consequently, the locality of the amplitude would no longer be manifest in this parametrization.

If we now consider the single-channel amplitudes, then $d=0$, $k=1$. Moreover, by the requirement that $m>0$, we get $\omega\geq 3$, then $D\geq 4$. From \eqref{single-channel_amplitudes} we find 
\begin{equation}
    \mA_s^{(2\omega-2)}(1^+2^+3^+4^+)=k_s\frac{t^{\omega}}{s}\frac{[12]^2}{\langle 34\rangle^2}+\mA^{(2\omega-2)}_{\text{homo}}(1^+2^+3^+4^+)\,,
\end{equation}
where $k_s\sim C^{1,1,\omega} \bar{C}^{-\omega,1,1}$ and if we consider all three channels \eqref{single-channel_stu}, we get
\begin{equation}
    \mA^{(2\omega-2)}(1^+2^+3^+4^+)=\Big(k_s\frac{t^{\omega}}{s}+k_t\frac{u^{\omega}}{t}+k_u\frac{s^{\omega}}{u}\Big)\frac{[12]^2}{\langle 34\rangle^2}+\mA_s^{(2\omega-2)}(1^+2^+3^+4^+)\,.
\end{equation}

\paragraph{Yang-Mills theory.} For Yang-Mills theory, from \eqref{YM-like_amplitudes}, we find
\begin{subequations}\label{YM-amplitudes}
\begin{align}
      \mA^{(0)}_{st}(1^+2^+3^-4^-)=\frac{k_{\text{YM}}}{st}[12]^2\langle 34\rangle^2&=-k_{\text{YM}}\frac{[12]^3}{[23][34][41]}\,,\\
    \mA^{(0)}_{us}(1^+2^+3^-4^-)=\frac{k_{\text{YM}}}{us}[12]^2\langle 34\rangle^2&=-k_{\text{YM}}\frac{[12]^3}{[13][34][24]}\,,\\
    \mA^{(0)}_{tu}(1^+2^+3^-4^-)=\frac{k_{\text{YM}}}{tu}[12]^2\langle 34\rangle^2&=-k_{\text{YM}}\frac{[12]^4}{[23][24][13][14]}\,,
\end{align}    
\end{subequations}
where $k_{\text{YM}}\sim C^{1,1,-1}C^{1,-1,-1}$. From this representation, it is immediate to see $\mA^{(0)}_{us}(1^+2^+3^-4^-)=\frac{t}{u}\mA^{(0)}_{st}(1^+2^+3^-4^-)=\mA^{(0)}_{st}(1^+2^+4^-3^-)$. The remaining two amplitudes correspond precisely to the two standard color-ordered amplitudes:
\begin{align}
    \tilde{\mA}(1^+2^+3^-4^-)=&-\frac{1}{k_{\text{YM}}}\mA^{(0)}_{st}(1^+2^+3^-4^-)=\frac{[12]^{3}}{[23][34][41]}\,,\\
    \tilde{\mA}(1^+3^-2^+4^-)=&-\frac{1}{k_{\text{YM}}}\mA^{(0)}_{tu}(1^+2^+3^-4^-)=\frac{[12]^{4}}{[13][32][24][41]}\,.
\end{align}
These are again related, as a consequence of $s+t+u=0$, that implies $\mA^{(0)}_{st}+\mA^{(0)}_{us}+\mA^{(0)}_{tu}=0$. Moreover, just by looking at \eqref{YM-amplitudes} we find $\mA^{(0)}_{tu}(1^+_12^+_13^-_14^-_1)=\frac{s}{u}\mA^{(0)}_{st}(1^+_12^+_13^-_14^-_1)$. This is the simplest example of BCJ amplitude relation \cite{Bern:2019prr}.

\paragraph{Gravity.} For gravity, from \eqref{GR-like_amplitude}, we find
\begin{equation}
    \mA^{(2)}_{stu}(1^+_22^+_23^-_24^-_2)=k_{\text{GR}}\frac{[12]^4\langle 34\rangle^4}{stu}\,.
\end{equation}
where $k_{\text{GR}}\sim C^{2,2,-2}C^{2,-2,-2}$. This can also be obtained by double-copy from the Yang-Mills quartic amplitude via the simplest of the KLT relations \cite{Bern:2019prr}
\begin{equation}
    \mA^{(2)}_{stu}(1^+_22^+_23^-_24^-_2)=\frac{st}{u}\big(\mA^{(0)}_{st}(1^+2^+3^-4^-)\big)^2\,,
\end{equation}
after the identification of $k_{\text{GR}}=k_{\text{YM}}^2$.

\paragraph{Quasi-chiral HS-YM.}
For the quasi-chiral HS-YM theories found above, four-point amplitudes are a simple generalisation of the YM ones
\begin{align}
    &\mA^{(2s-2)}_{st}(1^+_s2^+_s3^-_14^-_1)=\frac{k^{\text{HS}}_{\text{YM}}}{st}[12]^{2s}\langle 34\rangle^2=-k^{\text{HS}}_{\text{YM}}\frac{[12]^{2s+1}}{[23][34][41]}\,,\\
    &\mA^{(2s-2)}_{us}(1^+_s2^+_s3^-_14^-_1)=\frac{k^{\text{HS}}_{\text{YM}}}{us}[12]^{2s}\langle 34\rangle^2=-k^{\text{HS}}_{\text{YM}}\frac{[12]^{2s+1}}{[13][34][24]}\,,\\
    &\mA^{(2s-2)}_{tu}(1^+_s2^+_s3^-_14^-_1)=\frac{k^{\text{HS}}_{\text{YM}}}{tu}[12]^{2s}\langle 34\rangle^2=-k^{\text{HS}}_{\text{YM}}\frac{[12]^{2s+2}}{[23][24][13][14]}\,,
\end{align}
where $k^{\text{HS}}_{\text{YM}}\sim C^{s,s,-1}C^{1,-1,-1}$.
As before, we have $\mA^{(2s-2)}_{us}(1^+_s2^+_s3^-_14^-_1)=\frac{t}{u}\mA^{(2s-2)}_{st}(1^+_s2^+_s3^-_14^-_1)=\mA^{(2s-2)}_{st}(1^+_s2^+_s4^-_13^-_1)$. The remaining two amplitudes correspond to the two color-ordered amplitudes:
\begin{subequations}\label{color-ordered-HSYM}
\begin{align}
    \tilde{\mA}(1_{s}^+2_{s}^+3_{1}^-4_{1}^-)=&-\frac{1}{k^{\text{HS}}_{\text{YM}}}\mA^{(2s-2)}_{st}(1^+_s2^+_s3^-_14^-_1)=\frac{[12]^{2s+1}}{[23][34][41]}\,,\\
    \tilde{\mA}(1_{s}^+3_{1}^-2_s^+4_{1}^-)=&-\frac{1}{k^{\text{HS}}_{\text{YM}}}\mA^{(2s-2)}_{tu}(1^+_s2^+_s3^-_14^-_1)=\frac{[12]^{2s+2}}{[13][32][24][41]}\,.
\end{align}
\end{subequations}
These amplitudes are also related as a consequence of $s+t+u=0$, that implies $\mA^{(2s-2)}_{st}+\mA^{(2s-2)}_{us}+\mA^{(2s-2)}_{tu}=0$. Moreover, just by looking at \eqref{YM-amplitudes}, we find $\mA^{(2s-2)}_{tu}(1^+_s2^+_s3^-_14^-_1)=\frac{s}{u}\mA^{(2s-2)}_{st}(1^+_s2^+_s3^-_14^-_1)$. This is a BCJ amplitude relation for higher-spin amplitudes in the non-holomorphic sector. More generally, the same applies to all YM-like amplitudes \eqref{YM-like-BCJ}.

The expressions \eqref{color-ordered-HSYM} coincide with those found in \cite{Adamo:2022lah}, although we have already emphasized that the theory described there exhibits a certain degree of non-locality. Setting $s=1$, we recover the standard Yang-Mills four-point amplitudes.

\paragraph{Quasi-chiral HS-GR.} For the quasi-chiral HS-GR theories found above, the four-point amplitudes are a simple generalisation of the GR ones
\begin{equation}
    \mA^{(2s-2)}(1_{s}^+2_{s}^+3_{2}^-4_{2}^-)=-k^{\text{HS}}_{\text{GR}}\frac{[12]^{2s}\langle 34 \rangle^4 }{stu}\,,
\end{equation}
where $k^{\text{HS}}_{\text{GR}}\sim C^{s,s,-2}C^{2,-2,-2}$.
Setting $s=2$, we recover the standard MHV four-point graviton amplitude. 
Some of them can also be obtained by a ''double-copy'' for higher-spin amplitudes from the four-point amplitude of HS-YM via the simplest of the KLT relation
\begin{equation}
    \mA^{(4s-2)}_{stu}(1^+_{2s}2^+_{2s}3^-_24^-_2)=\frac{st}{u}\big(\mA^{(2s-2)}_{st}(1^+_s2^+_s3^-_14^-_1)\big)^2\,,
\end{equation}
after the identification of $k^{\text{HS}}_{\text{GR}}=(k^{\text{HS}}_{\text{YM}})^2$.
More generally, the same applies to all GR-like amplitude \eqref{GR-like-KLT}.

We emphasise once again that the conditions in \eqref{total_conditions} precisely coincide with those obtained in the light-cone Hamiltonian formulation, namely \eqref{kinda_trinagular}, \eqref{surpass_trinagular}, and \eqref{Homo_constraints}. This agreement is expected, as it was shown in \cite{Ponomarev:2016cwi} that the light-cone deformation procedure can be reformulated entirely in terms of the spinor-helicity formalism at all orders in the interaction.

Let us stress, however, that the existence of local four-point amplitudes does not by itself guarantee the existence of a local Lorentz-invariant theory. Indeed, while a given four-point amplitude may be local, other amplitudes in the same theory may still be non-local. Establishing Lorentz invariance of a theory requires verifying the full set of amplitudes, a task that we carry out explicitly in Section \ref{section6}. From the light-cone perspective, this corresponds to determining whether, for a given set of cubic vertices, the quartic light-cone consistency condition admits a local solution. In some cases, although local quartic vertices solving some of the constraints do exist, additional non-local quartic vertices are nevertheless required to solve all of them.

The strategy pursued here, as already suggested in \cite{Ponomarev:2016cwi}, may provide a viable route towards solving the light-cone constraints at all orders in perturbation theory. This would require a systematic investigation of the existence and form of general $n$-point higher-spin amplitudes. Such a program would be analogous to the search for compact expressions for known classes of amplitudes, such as the Parke–Taylor formula for tree-level MHV scattering amplitudes of $n$ gluons \cite{Parke:1986gb,Berends:1987me}, the tree-level MHV graviton amplitudes \cite{Bedford:2005yy,Hodges:2012ym}, and also that of the quasi chiral HS-YM amplitudes \cite{Adamo:2022lah}.

The same ideas are likely applicable to holomorphic amplitudes, potentially reproducing the results of \cite{Ponomarev:2016lrm,Serrani:2025owx}. Furthermore, from the amplitude perspective, the extension to fermions and massive fields seems more direct \cite{Arkani-Hamed:2017jhn}.

We conclude by observing that the YM-like and GR-like amplitudes constructed above do not admit homogeneous quartic contributions. This absence suggests the possibility that a suitable complex momentum deformation, analogous to the BCFW shift, may exist such that the boundary contribution at $z\to\infty$ vanishes. Establishing the existence of such a shift would allow for the construction of higher-point amplitudes through on-shell recursion relations.

\subsection*{Abelian vs Non-abelian cubic vertices II}

Here we show that the conditions in \eqref{total_conditions} contain information about the nature, abelian or non-abelian, of the cubic vertices entering the amplitude. It is straightforward to see that, if we require an amplitude to contain at least one cubic vertex saturating the bound in \eqref{total_conditions} for $k=1$, then at least one of the cubic vertices must satisfy $(-s_1,s_2,s_3)$ with $s_1+s_2=s_3$, or a parity-related equivalent condition. Similarly, requiring saturation of the bound for $k=2$ implies that at least one cubic vertex satisfy $(-s_1,s_2,s_3)$ with $s_1+s_2+1=s_3$, or a parity-related equivalent condition. Finally, saturation of the bound for $k=3$ requires the presence of at least one cubic vertex that satisfy $(-s_1,s_2,s_3)$ with $s_1+s_2+2=s_3$, or a parity-related equivalent condition. These three condition are associated with the three classes of four-point amplitudes: respectively, single-channel amplitudes ($k=1$, then $s_1+s_2=s_3$), YM-like ($k=2$, then $s_1+s_2+1=s_3$) and GR-like ($k=3$, then $s_1+s_2+2=s_3$) amplitudes. 

This is consistent with the interpretation of cubic vertices as abelian or non-abelian given in Section \ref{section6}, following \cite{Bekaert:2010hp}. Our analysis provides an on-shell characterisation of this distinction, making its relation with the light-cone formulation more explicit. Furthermore, it reveals a stronger constraint: while all abelian deformations are compatible with locality, the only non-abelian deformations allowed by locality are those satisfying $s_3=s_1+s_2+1$ corresponding to YM-like deformations, or $s_3=s_1+s_2+2$ corresponding to GR-like deformations. More generally, deformations with $s_3=s_1+s_2+\ell$, $\ell>2$ are excluded by locality.

\section{Comments on non-local quartic vertices}\label{section8}

From the analysis above, it is clear that although the landscape of local higher-spin interactions is richer than previously thought, admitting some non-trivial higher-spin quartic vertices and, as a consequence, certain quasi-chiral higher-spin theories, it also presents a clear obstruction to achieving parity invariance and unitarity without having to abandon locality. This also helps explain why, in the covariant framework, where parity is sometimes implicit, no parity-invariant higher-spin theory has ever been found.

This inevitably leads us, if we aim to construct a parity-invariant theory, to extend our search to non-local higher-spin interactions. The difficulty with exploring non-local vertices in a perturbative Noether-like approach is that, once we allow for non-localities of the form $\frac{1}{H_2}$, the deformation procedure becomes trivial. In the light-cone, allowing for such non-localities trivialises the quartic constraint, as shown in \eqref{maineqB_full}.

We then follow the guidance of the light-cone deformation procedure. In doing so, we should allow for certain non-localities, with the aim of constructing a unitary theory, while explicitly excluding terms of the form $\frac{1}{H_2}$. This can be referred to as \textit{mild non-localities}. Although the resulting quartic Hamiltonian will contain transverse momenta in the denominator --- and thus be effectively non-local --- it does not have to trivialise the scattering or make the amplitudes ill-defined. In general, mild non-localities extend the space of admissible quartic vertices, opening the possibility for a unitary higher-spin theory that accommodates both chiral and anti-chiral vertices.

A systematic classification and analysis of all possible mild non-localities is a highly non-trivial task. For instance, a mild non-local version of the $(1,1)$ quartic vertex could take the form
\begin{align}\label{(n,n)-(1,1)_vertex}
\begin{split}
h^{(1,1)}_4(x,y)=&\;\frac{\PP^n_{12}}{\PP^n_{34}}f_1(x,y)+\frac{\PP^{n-1}_{12}}{\PP^{n-1}_{34}}f_2(x,y)+\cdots+\frac{\PP_{12}}{\PP_{34}}f_n(x,y)+f_{n+1}(x,y)\\
&+\frac{\PPb_{12}}{\PPb_{34}}f_{n+2}(x,y)+\cdots+\frac{\PPb^{n-1}_{12}}{\PPb^{n-1}_{34}}f_{2n}(x,y)+\frac{\PPb^n_{12}}{\PPb^n_{34}}f_{2n+1}(x,y)\,.
\end{split}
\end{align}
However, other types of mild non-localities may also arise, such as
\begin{align}
    &\frac{c_1(\beta_i)\PP_{12}}{c_2(\beta_i)\PP_{12}+c_3(\beta_i)\PP_{34}}\,,&
    &\frac{c_1(\beta_i)\PP^2_{12}}{c_2(\beta_i)\PP^2_{12}+c_3(\beta_i)\PP_{12}\PP_{34}+c_4(\beta_i)\PP^2_{34}}\,.
\end{align}
More generally, one may encounter expressions of the form
\begin{equation}
    \frac{p_1(\PP_{12},\PP_{34},\beta_i)}{p_2(\PP_{12},\PP_{34},\beta_i)}+\frac{p_1(\PPb_{12},\PPb_{34},\beta_i)}{p_2(\PPb_{12},\PPb_{34},\beta_i)}\,,
\end{equation}
where the complex conjugate has been added to ensure unitarity. Such mild non-local structures would allow us to solve the quartic constraint \eqref{quartic_system}. However, even without trivialising the perturbative procedure, they may still give rise to a very large number of possibilities.

We can try to follow a different line of thinking. A method to address the potential non-localities was proposed in \cite{Metsaev:1991nb}. Instead of searching for a local solution to the system \eqref{quartic_system}, the quartic Hamiltonian density is split into two parts:
\begin{equation}
h_4 = -h_4^{\text{exch}} + h_4^{\text{homo}}\,,
\end{equation}
where $h_4^{\text{exch}}$ is a non-local quartic Hamiltonian density (constructed ad hoc, as we review below) with exchange-type non-localities (these are not of the form $\sim\frac{1}{H_2}$) that solves the system \eqref{quartic_system}, and $h_4^{\text{homo}}$ is a solution of the corresponding homogeneous system.

As emphasised in \cite{Metsaev:1991nb}, a solution with exchange-type non-localities to \eqref{quartic_system} always exists, taking the form
\begin{equation}\label{exchange_solution}
h_4^{\text{exch}} = \frac{9}{2s} (-)^{\omega} C^{\lambda_1,\lambda_2,\omega} \bar{C}^{-\omega,\lambda_3,\lambda_4} \frac{\PPb_{12}^{\lambda_{12}+\omega} \PP_{34}^{-\lambda_{34}+\omega} }{(\beta_1+\beta_2)^{2\omega}}\frac{\beta_3^{\lambda_3} \beta_4^{\lambda_4}}{\beta_1^{\lambda_1} \beta_2^{\lambda_2}}\,,
\end{equation}
to cancel the $(1234)$ exchange diagram \eqref{exchange_diagram}. The same form takes the solution for the other exchanges upon permuting the helicities in \eqref{exchange_solution} and substituting $s$ with either $t$ or $u$.

Once $h_4^{\text{exch}}$ is constructed, one can then search for the most general solution of the homogeneous system, which Metsaev also provided in terms of Jacobi polynomials. The next step is to combine the homogeneous solution $h_4^{\text{homo}}$ with the non-local $h_4^{\text{exch}}$ and determine under which conditions a local solution to \eqref{quartic_system} exists. This is the crucial question, which, however, is not solved in \cite{Metsaev:1991nb}.

This is the question we addressed above. We have shown that a local solution to \eqref{quartic_system} exists if and only if $D-d+2k\geq 0$, where $D \equiv \lambda_{12} - \lambda_{34} + 2\omega -2$ is the total number of derivatives carried by the quartic Hamiltonian density, $d=\max\limits_{i\neq j\neq k\neq \ell}\{\lambda_i-\lambda_{jk\ell},-\lambda_i+\lambda_{jk\ell}\}$ and $k=1,2,3$ for single-channel, YM-like, and GR-like amplitudes, respectively. This condition, however, leads to the impossibility of a consistent local higher-spin theory, except for the case containing only abelian cubic interactions. 

Therefore, the only possible way to construct an interacting unitary higher-spin theory in $4d$ is to allow for a certain degree of non-locality (while avoiding singularities of the type $\frac{1}{H_2}$). Such non-localities would necessarily modify the pole structure of the resulting amplitudes. For instance, they may introduce higher-order poles, or conversely remove poles altogether. As an example, by choosing the quartic vertex to precisely cancel the exchange-type contributions as in \eqref{exchange_solution}, the resulting four-point amplitude can be made to vanish.

Then a minimal example of a non-local theory at quartic order, consistent with the quartic light-cone constraint, would consist of turning on all the local quartic vertices described above and then reproducing all local four-point higher-spin amplitudes, while cancelling all remaining non-local four-point amplitudes through the exchange-type non-local contributions in \eqref{exchange_solution}.

It would be interesting to investigate the structure of all possible exchange-type non-localities, at least at quartic order, and determine whether they exhibit any special or universal properties.

Concerning higher-order light-cone constraints \eqref{LFNoether}, we expect them to behave in a way analogous to the quartic one.\footnote{For example, in the covariant approach, it is possible to extend the triangular inequalities for abelian vertices to quadrilateral inequalities and, more generally, to $n$-point inequalities that ensure the consistency of higher-point abelian vertices, following the same line of reasoning used for cubic interactions. We can therefore expect that analogous constraints to those found in the light-cone formulation can be extended to higher-point vertices as well; where spin-$1$ and spin-$2$ will maintain a special role.}

\section{Conclusions and discussion}\label{section9}
 
In the paper, we carried out a detailed study of the full quartic constraint \eqref{maineqB}. We began by developing a method to consistently address the problem of determining both the local Hamiltonian quartic density $h_4$ and the corresponding boost densities $j^{z-}_4$ and $j^{\zb-}_4$. We then introduced two complementary approaches: one to construct explicit solutions and another for systematically checking the existence of consistent quartic vertices (i.e. those that solve the associated system of quartic constraints \eqref{quartic_system}), via Frobenius integrability. The latter approach allowed us to compile several tables listing all quartic vertices that solve \eqref{quartic_system}. We then conjectured all such $(n,m)$ quartic vertices, revealing the special role taken by the spin-$1$ and spin-$2$ fields.

Our analysis reproduced known no-go and yes-go results for lower-spin interactions. In particular, chiral higher-spin gravity \cite{Metsaev:1991mt,Metsaev:1991nb,Ponomarev:2016lrm,Sharapov:2022faa,Sharapov:2022wpz,Sharapov:2022awp,Sharapov:2022nps,Sharapov:2023erv} does not have a local completion in flat space; see also \cite{Ponomarev:2017nrr}. An important novelty of our analysis is the ability to extract quartic vertices involving higher-spin fields. Indeed, the method does not rely on any assumptions about the high-energy behaviour but comes as a solution to the quartic constraint \eqref{quartic}, which is always well-defined. The same analysis can be carried out for fermions, as well as for massive fields, starting from the work \cite{Metsaev:2005ar,Metsaev:2007rn}.

Using the conjectured local quartic vertices, we determined all possible local unitary higher-spin theories. We find that a consistent local higher-spin theory can contain only abelian cubic vertices that satisfy the triangular inequality. Conversely, attempting to include nontrivial lower-spin cubic vertices, such as the Yang–Mills or gravitational cubic couplings, leads to a no-go result for local higher-spin interactions.

Even though the impossibility of having a local unitary higher-spin theory is confirmed, many local higher derivative higher-spin quartic vertices have been discovered. This, in turn, suggests the existence of certain consistent local ``quasi-chiral'' sectors. We have studied and commented on the existence of a quasi-chiral HS-YM theory and a quasi-chiral HS-GR theory. Quasi-chiral theories have both MHV and anti-MHV vertices, but they enter in an asymmetric way with respect to the spins involved (the symmetric combinations of type $(\lambda,\lambda,-\lambda)\oplus (-\lambda,-\lambda,\lambda)$ can exist only for $\lambda=1,2$). 

Using the spinor-helicity formalism and imposing locality in the form of factorisation, we determine all local higher-spin four-point amplitudes and identify four distinct classes: amplitudes arising from self-consistent quartic vertices, amplitudes requiring a single exchange channel, amplitudes requiring two exchange channels (YM-like), and amplitudes requiring all three channels (GR-like). The latter two classes are closely related to the standard Yang–Mills and gravity MHV four-point amplitudes. Furthermore, the YM-like amplitudes exhibit color–kinematics duality \cite{Bern:2008qj}, while the GR-like amplitudes can be obtained through a double-copy construction \cite{Bern:2010ue}, extending these ideas to higher-spin non-holomorphic amplitudes. Indeed, the holomorphic case has already been analysed in \cite{Ponomarev:2017nrr,Ponomarev:2024jyg}. See also \cite{Serrani:2026azw}, where color–kinematics duality, and in particular the self-dual higher-spin kinematic algebra, is employed to construct all $n$-point half-collinear self-dual higher-spin amplitudes as generalisations of half-collinear self-dual Yang–Mills amplitudes \cite{Guevara:2026qzd}.

A very interesting question to explore is whether colour–kinematics duality and the double-copy construction in the form of the KLT relations we have found for higher-spin non-holomorphic four-points amplitudes, extend in some way to all local $n$-point tree-level amplitudes.

One important remark is that, although it was long believed that higher-spin theories require an infinite spectrum of fields in order to be consistent, this is not necessarily the case. In fact, it was already observed in the chiral sector in \cite{Serrani:2025owx}, and also hinted at in \cite{Ponomarev:2016lrm}, that such an infinite spectrum is not strictly required. Moreover, the study of the full quartic constraint does not appear to indicate any preference for an infinite spectrum over a finite one. As we have seen, the obstruction arising from locality is the same in both cases, whether the spectrum is infinite or finite. Moreover, the ``quasi-chiral'' theories we have found seems not to require an infinite spectrum of higher-spin fields.

An interesting future direction would be to extend the idea of mild non-locality to $AdS$ and see if it improves the locality properties of the quartic vertices found via the AdS/CFT correspondence in \cite{Bekaert:2015tva}. Note that for cubic vertices, the flat-space limit --- when the cosmological constant goes to zero --- is smooth, and the two classifications (cubic vertices in flat and AdS spaces) coincide \cite{Metsaev:2018xip}. Also, there does not seem to be any mechanism that would prefer a non-vanishing cosmological constant over flat space, as far as the higher-spin problem is concerned, and we expect the same yes-go/no-go results to extend to AdS if one follows the light-cone gauge path \cite{Metsaev:2018xip}.

Perhaps most importantly, there is the search for the optimal definition of mild non-locality for higher-spin quartic vertices. Such a definition would ideally allow for the existence of a unitary higher-spin theory while preserving the desirable features of perturbative field theory. In this work, we have proposed a concrete notion of mild non-locality. The analysis of possible non-localities has also been explored from the CFT side \cite{Ponomarev:2017qab, Sleight:2017pcz, Sleight:2016xqq,Neiman:2023orj}, and it would be interesting to compare the two approaches to determine whether they are related in some way.

Finally, it would be interesting to interpret the results obtained for the local quartic vertices in the context of a putative dual 2d celestial CFT. Indeed, in \cite{Serrani:2025oaw}, building on the ideas of \cite{Ren:2022sws,Monteiro:2022lwm,Monteiro:2022xwq}, it was shown that the  holomorphic light-cone constraint is equivalent to the associativity of the celestial OPE. We expect that this relation can be extended to the full quartic light-cone constraint and beyond. Achieving this would require a study of the “all-order” celestial OPE \cite{Adamo:2022wjo, Ren:2023trv} and likely also the multi-particle celestial OPE \cite{Calkins:2026hpg}. Such a connection could provide valuable insight into the relation between the existence of a consistent gravity theory in the bulk and a well-defined celestial CFT with an associative OPE on the boundary.

\section*{Acknowledgments}
\label{sec:Aknowledgements1}
 I am grateful to Evgeny Skvortsov for suggesting the project and for many insightful discussions, particularly for drawing my attention to the role of triangular inequalities for abelian vertices. I would also like to thank Dmitry Ponomarev for pointing out the relevance of both the $J^{z-}$ and $J^{\bar{z}-}$  quartic constraints, as well as for many useful and insightful comments. I am further indebted to Dario Francia and Evgeny Skvortsov for encouraging me to provide more detailed explanations of the various tables. Finally, I would like to thank Dmitry Ponomarev once again for his valuable comments on the first version of this paper, and for pointing out the subtle distinction between abelian and non-abelian vertices. This project has received funding from the European Research Council (ERC) under the European Union’s Horizon 2020 research and innovation (grant agreement No 101002551).

\appendix

\section{Unitarity and parity in the light-cone}\label{AppendixA}
Parity transformations are defined as
\begin{align}
    &P:\phi^{\lambda}\rightarrow\phi^{-\lambda}\,,&
    &P:C^{\lambda_i}\rightarrow \bar{C}^{-\lambda_i}\,,&    &P:\bar{q}\rightarrow q\implies P:\PPb\rightarrow \PP\,.
\end{align}
At the cubic level, parity invariance requires $C^{\lambda_1,\lambda_2,\lambda_3}=\bar{C}^{-\lambda_1,-\lambda_2,-\lambda_3}$. On the other hand, unitarity (i.e. a Hermitian Lagrangian) requires $C^{\lambda_1,\lambda_2,\lambda_3}=(\bar{C}^{-\lambda_1,-\lambda_2,-\lambda_3})^*$.\footnote{Note that if we take even-helicity fields to be Hermitian matrices and odd-helicity fields to be anti-Hermitian matrices, the conditions become $C^{\lambda_1,\lambda_2,\lambda_3}=(-)^{\lambda_{123}}\bar{C}^{-\lambda_1,-\lambda_2,-\lambda_3}$ for parity invariance and $C^{\lambda_1,\lambda_2,\lambda_3}=(-)^{\lambda_{123}}(\bar{C}^{-\lambda_1,-\lambda_2,-\lambda_3})^*$ for unitarity.} When both $C$ and $\bar{C}$ are real, the two conditions coincide, so the theory is simultaneously parity-invariant and unitary. However, if the couplings are complex, one can, in principle, obtain a theory that is unitary but not parity-invariant, or vice versa.

\paragraph{Imposing parity to $\mathbf{h_4}$.} We begin by recalling that, for a three-point vertex, parity invariance requires the presence of both its holomorphic and anti-holomorphic components. In fact, parity exchanges the two as
\begin{equation}
    C^{\lambda_1,\lambda_2,\lambda_3} \frac{\PPb^{\lambda_{123}}}{\beta_1^{\lambda_1}\beta_2^{\lambda_2}\beta_3^{\lambda_3}}\quad
    \overset{P}{\longleftrightarrow}\quad
    \bar{C}^{-\lambda_1,-\lambda_2,-\lambda_3} \frac{\PP^{\lambda_{123}}}{\beta_1^{\lambda_1}\beta_2^{\lambda_2}\beta_3^{\lambda_3}}\,.
\end{equation}
For the quartic vertices, parity acts as follows:
\begin{equation}
 H_4\sim h_{\lambda_1,\lambda_2,\lambda_3,\lambda_4}^{q_1,q_2,q_3,q_4}\, \phi^{\lambda_1}_{q_1}\phi^{\lambda_2}_{q_2}\phi^{\lambda_3}_{q_3}\phi^{\lambda_4}_{q_4}\quad
    \overset{P}{\longleftrightarrow}\quad
    \bar{H}_4\sim \bar{h}_{-\lambda_1,-\lambda_2,-\lambda_3,-\lambda_4}^{\bar{q}_1,\bar{q}_2,\bar{q}_3,\bar{q}_4}\, \phi^{-\lambda_1}_{q_1}\phi^{-\lambda_2}_{q_2}\phi^{-\lambda_3}_{q_3}\phi^{-\lambda_4}_{q_4}\,,
\end{equation}
where by $\bar{h}$ we mean the same expression as $h$, with the replacement $q \leftrightarrow \bar q$, which effectively corresponds to $\mathbb{P} \leftrightarrow \bar{\mathbb{P}}$.
We can distinguish two cases: 
\begin{itemize}
    \item If the helicities of $\bar{h}_4$ are not a permutation of those of $h_4$, no additional condition needs to be imposed on the quartic vertex. In that case, achieving parity invariance proceeds exactly as in the cubic case: we simply add $\bar{h}_4$ to the spectrum of quartic vertices. However, the presence of both vertices can introduce new constraints, possibly rendering previously consistent configurations inconsistent, exactly as it happens in the cubic case when attempting to parity-complete a chiral theory.
    
     Let us also note that for the quartic Hamiltonian density of type $(n,m)$, with $n\neq m$, that has the following dependence on transverse momenta
    \begin{equation}
    h_4\sim \PPb^{n-1}\PP^{m-1}\quad
    \overset{P}{\longleftrightarrow}\quad
    \bar{h}_4\sim \PPb^{m-1}\PP^{n-1}\,,
    \end{equation}
    parity exchanges an $(n,m)$ quartic vertex with an $(m,n)$ one, in direct analogy with the cubic case. Therefore, to obtain a parity-invariant quartic vertex, both the $(n,m)$ and $(m,n)$ must be present together. If one of them exists, it can always be ``parity completed'' by adding the other.

    \item A distinctive feature of quartic vertices is that, for an $(n,n)$ vertex, the helicities of $\bar{h}_4$ may coincide with those of $h_4$ up to a permutation. In such cases, one must impose non-trivial symmetry conditions on the quartic vertex. Specifically, any permutation symmetry acting on the helicities must be satisfied by the transverse momenta $q$. In addition, we must recall that parity exchanges $\PPb$ and $\PP$.
\end{itemize}
Some useful transformation properties of $f_i(x,y)$, used in the main text, under permutations of the external legs, are listed below. When $x$ and $y$ are defined as in \eqref{xy_1234}, one finds
\begin{subequations}\label{f_parity_GR}
\begin{align}
    &f_i(x,y)=f_i\left(-x,y\right)&
    &(1234)\leftrightarrow (2134)\,,\\
    &f_i(x,y)=f_i\left(x,-y\right)&
    &(1234)\leftrightarrow (1243)\,,\\
    &f_i(x,y)=f_i(y,x)&
    &(1234)\leftrightarrow (1243)\,.
\end{align}
\end{subequations}
When $x$ and $y$ are defined as in \eqref{xy_1324}, one finds
\begin{subequations}
\begin{align}
    &f_i(x,y)=f_i(y,-x)&
    &(1234)\leftrightarrow (2341)\,,\\
    &f_i(x,y)=f_i(-x,-y)&
    &(1234)\leftrightarrow (3412)\,,\\
    &f_i(x,y)=f_i(-y,x)&
    &(1234)\leftrightarrow (4123)\,.
\end{align}
\end{subequations}
\section{Useful formulas and relations}\label{AppendixB}
For the variables $x,y$ defined in \eqref{xy_1234}, we have the following relations
\begin{subequations}\label{xy_1234tranf}
\begin{align}
    &\frac{\beta_1-\beta_3}{\beta_1+\beta_3}=\frac{2+x+y}{x-y}\,,&
    &\frac{\beta_1-\beta_4}{\beta_1+\beta_4}=\frac{2+x-y}{x+y}\,,\\
    &\frac{\beta_2-\beta_3}{\beta_2+\beta_3}=\frac{-2+x-y}{x+y}\,,&
    &\frac{\beta_2-\beta_4}{\beta_2+\beta_4}=\frac{-2+x+y}{x-y}\,.
\end{align}
\end{subequations}
While for the variables $x,y$ defined in \eqref{xy_1324}, we have the following relations
\begin{subequations}\label{xy_1324tranf}
\begin{align}
    &\frac{\beta_1-\beta_2}{\beta_1+\beta_2}=\frac{2+x+y}{x-y}\,,&
    &\frac{\beta_1-\beta_4}{\beta_1+\beta_4}=\frac{2+x-y}{x+y}\,,\\
    &\frac{\beta_2-\beta_3}{\beta_2+\beta_3}=\frac{2-x+y}{x+y}\,,&
    &\frac{\beta_3-\beta_4}{\beta_3+\beta_4}=\frac{-2+x+y}{x-y}\,.
\end{align}
\end{subequations}
For four-point scattering, we have $2$ independent $\PPb$ variables, which, for example, can be chosen to be $\PPb_{12}$ and $\PPb_{34}$. Additionally, there are three independent $\beta$'s. All other $\PPb_{ij}$ can be expressed as
\begin{subequations}\label{PP_relations}
\begin{align}
    \PPb_{13}&=\frac{\beta_3 \PPb_{12}+\beta_1 \PPb_{34}}{\beta_1+\beta_2}\,,
    &&\PPb_{14}=-\frac{\PPb_{12} (\beta_1+\beta_2+\beta_3)+\beta_1 \PPb_{34}}{\beta_1+\beta_2}\,,\\
    \PPb_{23}&=\frac{\beta_2 \PPb_{34}-\beta_3 \PPb_{12}}{\beta_1+\beta_2}\,,
    &&\PPb_{24}=\frac{\PPb_{12} (\beta_1+\beta_2+\beta_3)-\beta_2 \PPb_{34}}{\beta_1+\beta_2}\,.
\end{align}
\end{subequations}
The same applies to $\PP$. On-shell, the following relations hold:
\begin{align}
    \begin{split}
   &\langle ij\rangle [ij]=-\frac{2}{\beta_i\beta_j}\PP_{ij}\PPb_{ij}=2\,q_i \cdot q_j=(q_i+q_j)^2\,,\\
   &\sum^{n}_{j=1}q^{\mu}_j=0\quad\Rightarrow\quad\sum^{n}_{j=1}\langle ij\rangle[jk]=\sum^{n}_{j=1}\frac{\PP_{ij}\PPb_{jk}}{\beta_j}=0\,.
   \end{split}
\end{align}
In particular, using the relation above for the four-point scattering, we find
\begin{align}
    &\frac{\PPb_{12}\PPb_{34}}{(q_1+q_2)^2}=\frac{\PPb_{31}\PPb_{24}}{(q_1+q_3)^2}=\frac{\PPb_{14}\PPb_{23}}{(q_1+q_4)^2}\,,&
    &s_{ij}=-(q_i+q_j)^2\,,
\end{align}
where $s\equiv s_{12}$, $t\equiv s_{14}$, and $u\equiv s_{13}$ are the standard Mandelstam variables for massless four-point scattering. Off-shell, the Mandelstam variables are defined as
\begin{align}
    &s\equiv\frac{\PPb_{12}\PP_{12}}{\beta_1\beta_2}+\frac{\PPb_{34}\PP_{34}}{\beta_3\beta_4}\,,&
    &t\equiv\frac{\PPb_{14}\PP_{14}}{\beta_1\beta_4}+\frac{\PPb_{23}\PP_{23}}{\beta_2\beta_3}\,,&
    &u\equiv\frac{\PPb_{13}\PP_{13}}{\beta_1\beta_3}+\frac{\PPb_{24}\PP_{24}}{\beta_2\beta_4}\,.
\end{align}

\section{Cubic scalar vertex}\label{AppendixC}
The holomorphic constraint originally analysed in \cite{Metsaev:1991mt,Metsaev:1991nb,Ponomarev:2016lrm} was further studied in \cite{Serrani:2025owx}, whose notation we follow. Here, we examine the holomorphic constraint for the cubic scalar vertex. Although this case is not explicitly discussed in \cite{Serrani:2025owx}, it highlights several interesting features and provides a useful testing ground for the light-cone deformation procedure. As we will see, it reproduces all the correct predictions for lower-spin vertices while also providing answers for the higher-spin ones. The holomorphic constraint for generic vertices is
\begin{align}\label{holo_constraint}
    \begin{split}
    [H_3,J_3^{z-}]=&\sum_{\lambda_i,\omega}\int d^{12}q\;\delta \left(\sum_i q_i\right)\frac{9}{2}\Big[(-)^{\omega}\frac{(\lambda_1+\omega-\lambda_2)\beta_1-(\lambda_2+\omega-\lambda_1)\beta_2}{(\beta_1+\beta_2)\beta_1^{\lambda_1}\beta_2^{\lambda_2}\beta_3^{\lambda_3}\beta_4^{\lambda_4}}\,\times\\
    &\mathcal{F}^{1234\omega}\PPb_{12}^{\lambda_{12}+\omega-1}\PPb_{34}^{\lambda_{34}-\omega}\,\,(\phi^{\lambda_1}_{q_1})^{a_1}(\phi^{\lambda_2}_{q_2})^{a_2}(\phi^{\lambda_3}_{q_3})^{a_3}(\phi^{\lambda_4}_{q_4})^{a_4}\Big]=0\,,
    \end{split}
\end{align}
where $\mathcal{F}^{1234\omega}=\fA_{a_1a_2c}\fA^c_{\phantom{c}a_3a_4}C^{\lambda_1,\lambda_2,\omega}C^{-\omega,\lambda_3,\lambda_4}$ is built from a holomorphic pair of cubic couplings (a $CC$ pair).
Following the approach of \cite{Serrani:2025owx}, we analyse the solutions in the case where one of the two couplings is the scalar cubic self-interaction $C^{0,0,0}$. As in the non-holomorphic constraint, six independent kinematical structures contribute:
\begin{equation}
    (1234)+(1324)+(1423)+(3412)+(2413)+(2314)\,.
\end{equation}
Typically, the holomorphic constraint receives contributions from both $(1234)$ and $(3412)$ (and similarly for the $t$ and $u$ channels), because they involve the same product of couplings $C^{\lambda_1,\lambda_2,\omega}C^{-\omega,\lambda_3,\lambda_4}=C^{\lambda_3,\lambda_4,-\omega}C^{\omega,\lambda_1,\lambda_2}$. However, when one of the couplings is the scalar self-coupling $C^{0,0,0}$, we notice from \eqref{holo_constraint} that while the power of $\PPb$ is zero, the power of $\PP$ is negative, giving a non-local contribution. The resolution to this apparent issue is that Eq.~\eqref{holo_constraint} is no longer valid in this special case. Before inserting the explicit expressions for the densities, the commutator reads 
\begin{align}
\begin{split}
[H_3,J_3^{z-}]= &\sum_{\lambda_i,\alpha_j}\int d^9p\;d^9q\; \delta\left(\sum_i q_i\right) \delta\left(\sum_j p_j\right)9\,\delta^{\lambda_3,-\alpha_3}\frac{\delta(q_3+p_3)}{2q_3^+}\,\phi^{\lambda_1}_{q_1}\phi^{\lambda_2}_{q_2}\phi^{\alpha_1}_{p_1}\phi^{\alpha_2}_{p_2}\,\times\\
&\left(j_3^{\lambda_i}(q_i)+\sum_{k\neq 3}\frac{\partial}{\partial \bar{q}_{k}}\frac{h_3^{\lambda_i}(q_i)}{3}\right)h_3^{\alpha_j}(p_j)\,.
\end{split}
\end{align}
Here, only $h_3$ can generate the $C^{0,0,0}$ contribution, since $j_3$ would vanish.\footnote{In particular, the holomorphic constraint for the pair of couplings $C^{0,0,0}C^{0,0,0}$ gives identically zero.}. The solution is then \eqref{holo_constraint} once all contributions that would have produced non-local terms are set to zero. 

Let us start by looking at the holomorphic constraint for the pair of coupling $C^{\lambda_1,\lambda_2,0}C^{0,0,0}$. In this case, we just get one contribution
\begin{align}
    &[H_3,J_3^{z-}]\sim C^{\lambda_1,\lambda_2,0}C^{0,0,0}(\lambda_1-\lambda_2)\PP_{12}^{\lambda_{12}-1}=0&
    &\implies&
    &\lambda_1=\lambda_2\,.
\end{align}
For the pair of couplings $C^{\lambda,0,0}C^{0,0,0}$, we have three independent contributions coming from $(1234)$, $(1324)$, and $(1423)$, then we get
\begin{align}
    &[H_3,J_3^{z-}]\sim C^{\lambda,0,0}C^{0,0,0}\left(\PP_{12}^{\lambda-1}+\PP_{13}^{\lambda-1}+\PP_{14}^{\lambda-1}\right)=0&
    &\implies&
    &\lambda=2\,.
\end{align}
Lorentz consistency between $C^{1,0,0}$ and $C^{0,0,0}$ is possible in the case of a color scalar vertex; thus, we need to consider only the cyclic contribution $[1234]$, and two terms survive
\begin{align}
    &[H_3,J_3^{z-}]\sim C^{\lambda,0,0}C^{0,0,0}\left(\PP_{12}^{\lambda-1}-\PP_{41}^{\lambda-1}\right)=0&
    &\implies&
    &\lambda=1\,.
\end{align}
Interestingly, we can also make $C^{1,0,0}$ consistent with $C^{0,0,0}$ if the cubic scalar is made out of three different scalars, then we find
\begin{align}
    \begin{split}
    [H_3,J_3^{z-}]\sim \;&C^{\lambda,0_1,0_1}C^{0_1,0_2,0_3}\PP_{12}^{\lambda-1}+C^{\lambda,0_2,0_2}C^{0_2,0_3,0_1}\PP_{13}^{\lambda-1}+C^{\lambda,0_3,0_3}C^{0_3,0_1,0_2}\PP_{14}^{\lambda-1}=0\\
    &\implies\quad
    \lambda=1\; \wedge\; C^{1,0_1,0_1}+C^{1,0_2,0_2}+C^{1,0_3,0_3}=0\,,\\
    &\implies\quad
    \lambda=2\; \wedge\; C^{2,0_1,0_1}=C^{2,0_2,0_2}=C^{2,0_3,0_3}\,,
    \end{split}
\end{align}
where $0_1$, $0_2$, and $0_3$ denote three distinct scalar fields. Notice that the solution for $\lambda=1$ implies that the total charge must vanish, while the case $\lambda=2$ confirms the universality of gravitational interactions.

As we have seen, the consistency of the scalar cubic couplings $C^{0,0,0}$ with higher-spin cubic couplings is restricted to interactions of the form $C^{\lambda,\lambda,0}$.
Consequently, for example, in the full chiral higher-spin theory, where all holomorphic cubic couplings are turned on, the cubic scalar coupling is not allowed.

The above results for lower-spin couplings are consistent with what is known in the covariant formulation. Recall that the Lorentz-invariance constraint in the light-cone formalism corresponds to gauge invariance in the covariant language. Indeed, the minimal coupling of a scalar field to gravity remains consistent even in the presence of a $\phi^3$ interaction term.

\section{Explicit form of lower-derivative quartic vertices}\label{AppendixD}

We present explicit examples of lower-derivative quartic vertices. Free coefficients arising from solutions of the homogeneous system of quartic constraints are denoted by $c_i$ when present. Throughout, we use the commutator in the form given in \eqref{commutator_general}.

\paragraph{(1,1) quartic vertices.}
\begin{align}
    &h^{(1,1)}_{[0,0,0,0]}:&
    &f(x,y)=k_5\frac{x y}{4}+k_1\frac{x y-1}{(x-y)^2}-k_2\frac{1+x y}{(x+y)^2}+c_1\,,\\
    &h^{(1,1)}_{[1,-1,0,0]}:&
    &f(x,y)=k_1\frac{1-x y}{(x-y)^2}\,,\\
    &h^{(1,1)}_{[1,0,-1,0]}:&
    &f(x,y)=-\frac{k_1}{2}\,.
\end{align}
In the case of $h^{(1,1)}_{[0,0,0,0]}$, the quartic constraint allows for single-channel, YM-like, and GR-like vertices.\footnote{This notation is explained both in section \ref{section6} and section \ref{section7}.} In particular, we have
\begin{align}
    h^{(1,1)}_{[0,0,0,0]}:\qquad\text{single-channel:}\qquad f(x,y)&=k_1\frac{x y-1}{(x-y)^2}+c_1\,,\\
    \text{YM-like:}\qquad f(x,y)&=k_1\left(\frac{x y-1}{(x-y)^2}-\frac{1+x y}{(x+y)^2}\right)+c_1\,,\\
    \text{GR-like:}\qquad f(x,y)&=k_1\left(\frac{x y}{4}+\frac{x y-1}{(x-y)^2}-\frac{1+x y}{(x+y)^2}\right)+c_1\,,
\end{align}
\paragraph{(2,2) quartic vertices.}
\begin{subequations}
\begin{align}
\nonumber
    &h^{(2,2)}_{(-1,1,-1,1)}:\\
    &f_1(x,y)=\frac{(x-1) (y+1) \left(k_1 (x+y)^4-2 k_2 \left(2 x^2+3 x y+x+y (2 y-1)+1\right)\right)}{2 (x+y)^4}\,,\\
    & f_2(x,y)=\frac{\left(x^2-1\right) \left(y^2-1\right) \left(3 k_1 (x+y)^4-8 k_2 \left(2 x^2+3 x y+2 y^2-1\right)\right)}{32 (x+y)^4}\,,\\
    &f_3(x,y)=\frac{(x+1) (y-1) \left(k_1 (x+y)^4-2 k_2 \left(3 x y+x (2 x-1)+2 y^2+y+1\right)\right)}{2 (x+y)^4}\,.
\end{align}
\end{subequations}
\begin{subequations}
\begin{align}
    h^{(2,2)}_{(-2,2,0,0)}:\qquad f_1(x,y)&=\frac{1}{2} k_1 (x-2) y\,,\\
     f_2(x,y)&=\frac{1}{32} k_1 \left(3 \left(x^2-3\right) y^2-x^2+7\right)\,,\\
    f_3(x,y)&=\frac{1}{2} k_1 (x+2) y\,.
\end{align}
\end{subequations}
\begin{subequations}
\begin{align}
    h^{(2,2)}_{(-2,2,-1,1)}:f_1(x,y)&=-\frac{k_1 (y+1) (x+y-1) (x (x+y-1)-2 y+1)}{2 (x+y)^2}\,,\\
    f_2(x,y)&=-\frac{k_1 \left(y^2-1\right) \left(3 x^4+3 \left(x^2-3\right) y^2+6 \left(x^2-1\right) x y-x^2+4\right)}{32 (x+y)^2}\,,\\
    f_3(x,y)&=-\frac{k_1 (y-1) (x+y+1) \left(x^2+(x+2) y+x+1\right)}{2 (x+y)^2}\,.
\end{align}
\end{subequations}

\section{Light-cone and spinor-helicity}\label{AppendixE}
For our spinor-helicity conventions for massless particles, we follow \cite{Elvang:2013cua}:
\begin{align}
    &q_{a\dot{b}}= q_{\mu}(\sigma^{\mu})_{a\dot{b}}\,,&
    &\text{det}(q_{a\dot{b}})=-q^{\mu}q_{\mu}=m^2\,,&
    &q^2=0\;\; \Rightarrow\;\; q_{a\dot{b}}=-|q]_a\langle q|_{\dot{b}}\equiv -\lambda_a\tilde{\lambda}_{\dot{b}}\,,
\end{align}
\begin{align}
        &\langle ij\rangle\definition \langle q_i|_{\dot{a}}|q_j\rangle^{\dot{a}}\equiv\tilde{\lambda}_{i\dot{a}}\tilde{\lambda}_j^{\dot{a}}\,,&
    &[ij]\definition [q_i|^a|q_j]_a\equiv \lambda_i^{a}\lambda_{ja}\,\,,&
    &\lambda^{a}=\epsilon^{ab}\lambda_{b}\,,&
&\tilde{\lambda}^{\dot{a}}=\epsilon^{\dot{a}\dot{b}}\tilde{\lambda}_{\dot{b}}\,,
\end{align}

\begin{align}
&\sigma^0=
    \begin{pmatrix}
        1 & 0\\
        0 & 1
    \end{pmatrix}\,,&
    &\sigma^1=
    \begin{pmatrix}
        0 & 1\\
        1 & 0
    \end{pmatrix}\,,&
    &\sigma^2=
    \begin{pmatrix}
        0 & -i\\
        i & 0
    \end{pmatrix}\,,&
    &\sigma^3=
    \begin{pmatrix}
        1 & 0\\
        0 & -1
    \end{pmatrix}\,,
\end{align}
\begin{align}
    &\epsilon^{ab}=\epsilon_{ab}=
    \begin{pmatrix}
        0 & 1\\
        -1 & 0
    \end{pmatrix}\,,&
    &\epsilon^{\dot{a}\dot{b}}=\epsilon_{\dot{a}\dot{b}}=
    \begin{pmatrix}
        0 & 1\\
        -1 & 0
    \end{pmatrix}\,.
\end{align}
Notice that for complex momenta $q^{\mu}$ the two spinors $(\lambda_a,\tilde{\lambda}_{\dot{b}})$ are independent two-dimensional complex vectors. In Minkowski space and for real momenta $q_{a\dot{b}}$ is hermitian, and the two spinors become complex conjugate $\tilde{\lambda}_{\dot{a}}=\pm(\lambda_a)^*$ (where the sign depends on whether the energy is taken to be positive or negative, then on the convention we use on the background flat metric). Spinor-helicity variables $(\lambda_i,\tilde{\lambda}_i)$ are defined up to little group scaling $(\lambda_i,\tilde{\lambda}_i)\sim (t_i\lambda_i,t_i^{-1}\tilde{\lambda}_i)$ for $t_i\in\mathbb{C}^*$. 

In the context of the light-cone Hamiltonian approach for massless higher-spin fields, we adopt the following notation \cite{Ponomarev:2016cwi}:
\begin{equation}
    q_{a\dot{b}}=q_{\mu}(\sigma^{\mu})_{a\dot{b}}=\sqrt{2}
    \begin{pmatrix}
        q^- & \bar{q}\\
        q & - \beta
    \end{pmatrix}\approx\sqrt{2}
    \begin{pmatrix}
        -\frac{q\bar{q}}{\beta} & \bar{q}\\
        q & - \beta
    \end{pmatrix}= -|q]_a\langle q|_{\dot{b}}=-\lambda_a\tilde{\lambda}_{\dot{b}}\,,
\end{equation}
\begin{align}
&\langle i|=\frac{2^{\frac{1}{4}}}{\sqrt{\beta_i}}\begin{pmatrix}
    q_i & -\beta_i
    \end{pmatrix}\,,&
    &\langle ij\rangle=-\sqrt{\frac{2}{\beta_i\beta_j}}\PP_{ij}\,,&
    &|i]=\frac{2^{\frac{1}{4}}}{\sqrt{\beta_i}}
    \begin{pmatrix}
       \bar{q}_i\\
        -\beta_i
    \end{pmatrix}\,,&
    &[ij]=\sqrt{\frac{2}{\beta_i\beta_j}}\PPb_{ij}\,.
\end{align}
The relation between the cubic amplitude and the cubic Hamiltonian density follows:
\begin{align}\label{AmplitudeandHamiltonian_holo}
    \mathcal{A}_3&=C^{\lambda_1,\lambda_2,\lambda_3}[12]^{d_{12}} [23]^{d_{23}} [31]^{d_{31}}=C^{\lambda_1,\lambda_2,\lambda_3}\frac{\sqrt{2}^{\lambda_{123}}\PPb^{\lambda_{123}}}{\beta_1^{\lambda_1}\beta_2^{\lambda_2}\beta_3^{\lambda_3}}=\sqrt{2}^{\lambda_{123}}h_3\,,\\\label{AmplitudeandHamiltonian_antiholo}
    \bar{\mathcal{A}}_3&=\bar{C}^{\lambda_1,\lambda_2,\lambda_3}\langle 12\rangle^{-d_{12}} \langle 23\rangle^{-d_{23}} \langle 31\rangle^{-d_{31}}=\bar{C}^{\lambda_1,\lambda_2,\lambda_3}\frac{\sqrt{2}^{\lambda_{123}}\PP^{-\lambda_{123}}}{\beta_1^{-\lambda_1}\beta_2^{-\lambda_2}\beta_3^{-\lambda_3}}=\sqrt{2}^{\lambda_{123}}\bar{h}_3\,,
\end{align}
where $\mathcal{A}_3$ is valid for $\lambda_{123}>0$ and $\bar{\mathcal{A}}_3$ for $\lambda_{123}<0$, and we have defined
\begin{align}
    &d_{12}=\lambda_{12}-\lambda_3\,,&
    &d_{23}=\lambda_{23}-\lambda_1\,,&
    &d_{31}=\lambda_{31}-\lambda_2\,.
\end{align}
The relations above, in the light-cone approach, hold only when the external particles are on-shell.  Indeed, one of the main differences between the cubic Hamiltonian density and the amplitudes in \eqref{AmplitudeandHamiltonian_holo}-\eqref{AmplitudeandHamiltonian_antiholo} is that the latter are inherently on-shell objects, while the former contains off-shell information. 

\begin{figure}[H]
    \centering
    \begin{tikzpicture}
        \begin{feynman}
            \vertex (i1) at (-6, 1) {\(\lambda_2\)};
            \vertex (i2) at (-6,-1) {\(\lambda_1\)};
            \vertex (i3) at (-2, 1) {\(\lambda_3\)};
            \vertex (i4) at (-2,-1) {\(\lambda_4\)};

            \vertex (v1) at (-5, 0);
            \vertex (v3) at (-3, 0);

            \vertex at (-5, 0.5) {\(\lambda_I\)};
            \vertex at (-3.2, 0.5) {\(-\lambda_I\)};

            \diagram* {
                (i1) -- (v1),
                (i2) -- (v1),
                (v1) -- [plain] (v3),
                (v3) -- (i3),
                (v3) -- (i4),
            };
        \end{feynman}
    \end{tikzpicture}
    \caption{Generic four-point exchange diagram.}
    \label{gen_exch}
\end{figure}
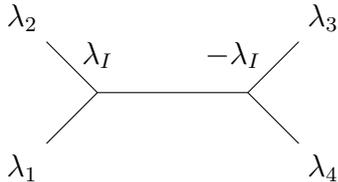
\noindent
Given a four-point scattering, as shown in Figure \ref{gen_exch}, momentum conservation implies
\begin{align}
    \begin{split}
     \langle 1|q_I|4]&=-\langle 1I\rangle[I4]=\langle 1|-q_1-q_2|4]=\langle 1|q_3+q_4|4]\\
     &\implies
     \langle 1I\rangle[I4]=\langle 12\rangle[24]=-\langle 13\rangle[34]\,.
     \end{split}
\end{align}

\section{Tables for higher-derivatives quartic vertices}\label{AppendixF}
Tables \ref{tab(4,1)}--\ref{tab(4,4)} collect quartic vertices that solve the system of quartic constraints~\eqref{quartic_system}. These tables\footnote{We display them only up to $(4,4)$, although our analysis extended up to $(8,8)$.} have been used to conjecture the list of all quartic vertices that solve~\eqref{quartic_system}, as discussed in the main text. Table \ref{tabHomo} collects quartic vertices that solve the homogeneous system of quartic constraints.

\begin{table}[H]
    \centering
    \begin{tabular}{|c|c|c|c|c|c|c|}
\hline
$(\lambda_1,\lambda_2,\omega,\lambda_3,\lambda_4)$ & $k_2=(2341)$ & $k_3=(3412)$ & $k_4=(4123)$ & $k_5=(1324)$ & $k_6=(2413)$ \\ \hline
$(-1,3,2,-1,2)$ & $k_1$ & $0$ & $0$ & $0$ & $k_1$ \\ \hline
$(-1,3,2,0,1)$ & $k_1$ & $0$ & $0$ & $0$ & $k_1$ \\ \hline
$(0,2,2,-1,2)$ & $0$ & $0$ & $k_4$ & $0$ & $k_1-k_4$ \\ \hline
$(0,2,2,0,1)$ & $k_2$ & $0$ & $0$ & $0$ & $k_6$ \\ \hline
$(0,3,1,0,0)$ & $k_1$ & $0$ & $0$ & $0$ & $k_1$ \\ \hline
$(1,1,2,-1,2)$ & $0$ & $0$ & $k_4$ & $0$ & $k_1-k_4$ \\ \hline
$(1,1,2,0,1)$ & $0$ & $0$ & $k_4$ & $0$ & $k_6$ \\ \hline
    \end{tabular}
    \caption{$(4,1)$ quartic vertices satisfying the quartic constraints.}
    \label{tab(4,1)}
\end{table}

\begin{table}[H]
    \centering
    \begin{tabular}{|c|c|c|c|c|c|c|}
\hline
$(\lambda_1,\lambda_2,\omega,\lambda_3,\lambda_4)$ & $k_2=(2341)$ & $k_3=(3412)$ & $k_4=(4123)$ & $k_5=(1324)$ & $k_6=(2413)$ \\ \hline
$(-2,3,3,-2,3)$ & $k_1$ & $k_1$ & $k_1$ & $0$ & $k_1$ \\ \hline
$(-2,3,3,-1,2)$ & $k_1$ & $k_1$ & $0$ & $0$ & $k_1$ \\ \hline
$(-2,3,3,0,1)$ & $k_1$ & $k_1$ & $0$ & $0$ & $k_1$ \\ \hline
$(-1,2,3,-1,2)$ & $k_2$ & $k_1$ & $k_2$ & $0$ & $k_6$ \\ \hline
$(-1,2,3,0,1)$ & $k_2$ & $k_1$ & $0$ & $0$ & $k_6$ \\ \hline
$(-1,3,2,-1,1)$ & $k_1$ & $0$ & $0$ & $0$ & $k_1$ \\ \hline
$(-1,3,2,0,0)$ & $k_1$ & $0$ & $0$ & $0$ & $k_1$ \\ \hline
$(0,1,3,0,1)$ & $k_2$ & $k_1$ & $k_2$ & $0$ & $k_6$ \\ \hline
$(0,2,2,-2,2)$ & $0$ & $0$ & $k_1$ & $0$ & $k_1$ \\ \hline
$(0,2,2,0,0)$ & $k_2$ & $0$ & $0$ & $0$ & $k_6$ \\ \hline
$(1,1,2,-2,2)$ & $0$ & $0$ & $k_1$ & $0$ & $k_1$ \\ \hline
$(1,1,2,-1,1)$ & $0$ & $0$ & $k_4$ & $0$ & $k_6$ \\ \hline
    \end{tabular}
    \caption{$(4,2)$ quartic vertices satisfying the quartic constraints.}
    \label{tab(4,2)}
\end{table}

\begin{table}[H]
    \centering
    \begin{tabular}{|c|c|c|c|c|c|c|}
\hline
$(\lambda_1,\lambda_2,\omega,\lambda_3,\lambda_4)$ & $k_2=(2341)$ & $k_3=(3412)$ & $k_4=(4123)$ & $k_5=(1324)$ & $k_6=(2413)$ \\ \hline
$(-2,3,3,-2,2)$ & $k_1$ & $0$ & $0$ & $0$ & $k_1$ \\ \hline
$(-2,3,3,-1,1)$ & $k_1$ & $0$ & $0$ & $0$ & $k_1$ \\ \hline
$(-2,3,3,0,0)$ & $k_1$ & $0$ & $0$ & $0$ & $k_1$ \\ \hline
$(-1,2,3,-2,2)$ & $0$ & $0$ & $k_4$ & $0$ & $k_1-k_4$ \\ \hline
$(-1,2,3,-1,1)$ & $k_2$ & $0$ & $0$ & $0$ & $k_6$ \\ \hline
$(-1,2,3,0,0)$ & $k_2$ & $0$ & $0$ & $0$ & $k_6$ \\ \hline
$(-1,3,2,-1,0)$ & $k_1$ & $0$ & $0$ & $0$ & $k_1$ \\ \hline
$(0,1,3,-2,2)$ & $0$ & $0$ & $k_4$ & $0$ & $k_1-k_4$ \\ \hline
$(0,1,3,-1,1)$ & $0$ & $0$ & $k_4$ & $0$ & $k_6$ \\ \hline
$(0,1,3,0,0)$ & $k_2$ & $0$ & $0$ & $0$ & $k_6$ \\ \hline
$(1,1,2,-2,1)$ & $0$ & $0$ & $k_4$ & $0$ & $k_1-k_4$ \\ \hline
    \end{tabular}
    \caption{$(4,3)$ quartic vertices satisfying the quartic constraints.}
    \label{tab(4,3)}
\end{table}

\begin{table}[H]
    \centering
    \begin{tabular}{|c|c|c|c|c|c|c|}
\hline
$(\lambda_1,\lambda_2,\omega,\lambda_3,\lambda_4)$ & $k_2=(2341)$ & $k_3=(3412)$ & $k_4=(4123)$ & $k_5=(1324)$ & $k_6=(2413)$ \\ \hline
$(-3,3,4,-3,3)$ & $k_1$ & $k_1$ & $k_1$ & $0$ & $k_1$ \\ \hline
$(-3,3,4,-2,2)$ & $k_1$ & $k_1$ & $0$ & $0$ & $k_1$ \\ \hline
$(-3,3,4,-1,1)$ & $k_1$ & $k_1$ & $0$ & $0$ & $k_1$ \\ \hline
$(-3,3,4,0,0)$& $k_1$ & $k_1$ & $0$ & $0$ & $k_1$ \\ \hline
$(-2,2,4,-2,2)$ & $k_2$ & $k_1$ & $k_2$ & $0$ & $k_6$ \\ \hline
$(-2,2,4,-1,1)$ & $k_2$ & $k_1$ & $0$ & $0$ & $k_6$ \\ \hline
$(-2,2,4,0,0)$ & $k_2$ & $k_1$ & $0$ & $0$ & $k_6$ \\ \hline
$(-2,3,3,-2,1)$ & $k_1$ & $0$ & $0$ & $0$ & $k_1$ \\ \hline
$(-2,3,3,-1,0)$ & $k_1$ & $0$ & $0$ & $0$ & $k_1$ \\ \hline
$(-1,1,4,-1,1)$ & $k_2$ & $k_1$ & $k_2$ & $0$ & $k_6$ \\ \hline
$(-1,1,4,0,0)$ & $k_2$ & $k_1$ & $0$ & $0$ & $k_6$ \\ \hline
$(-1,2,3,-3,2)$ & $0$ & $0$ & $k_1$ & $0$ & $k_1$ \\ \hline
$(-1,2,3,-1,0)$ & $k_2$ & $0$ & $0$ & $0$ & $k_6$ \\ \hline
$(-1,3,2,-1,-1)$ & $k_1$ & $0$ & $0$ & $0$ & $k_1$ \\ \hline
$(0,0,4,0,0)$ & $k_2$ & $k_1$ & $k_2$ & $k_5$ & $k_5$ \\ \hline
$(0,1,3,-3,2)$ & $0$ & $0$ & $k_1$ & $0$ & $k_1$ \\ \hline
$(0,1,3,-2,1)$ & $0$ & $0$ & $k_4$ & $0$ & $k_6$ \\ \hline
$(1,1,2,-3,1)$ & $0$ & $0$ & $k_1$ & $0$ & $k_1$ \\ \hline
    \end{tabular}
    \caption{$(4,4)$ quartic vertices satisfying the quartic constraints.}
    \label{tab(4,4)}
\end{table}

\begin{table}[H]
    \centering
    \begin{tabular}{|c|c|c|c|c|c|c|}
\hline
$(n,m),D$ & $(\lambda_1,\lambda_2,\lambda_3,\lambda_4)$ & $(n,m),D$ & $(\lambda_1,\lambda_2,\lambda_3,\lambda_4)$ & $(n,m),D$ & $(\lambda_1,\lambda_2,\lambda_3,\lambda_4)$ \\ \hline
$(1,1),0$ & $(0,0,0,0)$ &$(4,3),5$ & $(1,0,0,0)$&$(5,3),6$ & $(2,1,0,-1)$\\ \hline
$(2,2),2$ & $(0,0,0,0)$ & $(4,3),5$ & $(1,1,0,-1)$&$(5,3),6$ & $(2,2,-1,-1)$\\ \hline
$(3,1),2$ & $(1,1,0,0)$ & $(5,2),5$ & $(1,1,1,0)$&$(6,2),6$ & $(1,1,1,1)$\\ \hline
$(3,2),3$ & $(1,0,0,0)$ &  $(5,2),5$ & $(2,1,0,0)$&$(6,2),6$ & $(2,1,1,0)$\\ \hline
$(4,1),3$ & $(1,1,1,0)$ &  $(6,1),5$ & $(2,1,1,1)$&$(6,2),6$ & $(2,2,0,0)$\\ \hline
$(3,3),4$ & $(0,0,0,0)$ & $(6,1),5$ & $(2,2,1,0)$&$(7,1),6$ & $(2,2,1,1)$\\ \hline
$(3,3),4$ & $(1,0,0,-1)$ & $(4,4),6$ & $(0,0,0,0)$&$(7,1),6$ & $(3,1,1,1)$\\ \hline
$(3,3),4$ & $(1,1,-1,-1)$ & $(4,4),6$ & $(1,0,0,-1)$&$(7,1),6$ & $(2,2,2,0)$\\ \hline
$(4,2),4$ & $(1,1,0,0)$ & $(4,4),6$ & $(1,1,-1,-1)$&$(7,1),6$ & $(3,2,1,0)$\\\hline
$(5,1),4$ & $(1,1,1,1)$ &$(5,3),6$ & $(1,1,0,0)$&$(7,1),6$ & $(3,3,0,0)$\\\hline
$(5,1),4$ & $(2,1,1,0)$ &$(5,3),6$ & $(2,0,0,0)$&...& ...\\\hline
$(5,1),4$ & $(2,2,0,0)$ &$(5,3),6$ & $(1,1,-1,-1)$& ...&... \\\hline
    \end{tabular}
    \caption{Homogeneous quartic vertices satisfying the quartic constraints.}
        \label{tabHomo}
\end{table}

\footnotesize
\providecommand{\href}[2]{#2}\begingroup\raggedright\endgroup

\end{document}